\documentstyle[12pt,cite,psfig,axodraw]{article}

\makeatletter
\newtoks\@stequation

\def\subequations{\refstepcounter{equation}%
  \edef\@savedequation{\the\c@equation}%
  \@stequation=\expandafter{\theequation}
  \edef\@savedtheequation{\the\@stequation}
  \edef\oldtheequation{\theequation}%
  \setcounter{equation}{0}%
  \def\theequation{\oldtheequation\alph{equation}}}

\def\endsubequations{%
  \ifnum\c@equation < 2 \@warning{Only \the\c@equation\space subequation
    used in equation \@savedequation}\fi
  \setcounter{equation}{\@savedequation}%
  \@stequation=\expandafter{\@savedtheequation}%
  \edef\theequation{\the\@stequation}%
  \global\@ignoretrue}

\def\eqnarray{\stepcounter{equation}\let\@currentlabel\theequation
\global\@eqnswtrue\m@th
\global\@eqcnt\z@\tabskip\@centering\let\\\@eqncr
$$\halign to\displaywidth\bgroup\@eqnsel\hskip\@centering
     $\displaystyle\tabskip\z@{##}$&\global\@eqcnt\@ne
      \hfil$\;{##}\;$\hfil
     &\global\@eqcnt\tw@ $\displaystyle\tabskip\z@{##}$\hfil
   \tabskip\@centering&\llap{##}\tabskip\z@\cr}

\def\dlinepattern#1#2{%
\ifdim#2<#1
   \errmessage{the 1st argument is less than the 2nd argument.}%
\else
   \gdef\dline@solid{#1}\gdef\dline@period{#2}%
\fi}

\def\dline#1{\@dline[#1]}
\def\@dline[#1-#2]{\noalign{\global\@dla#1\relax
\global\advance\@dla\m@ne
\ifnum\@dla>\z@\global\let\@gtempa\@dlinea\else
  \global\let\@gtempa\@dlineb\fi
\global\@dlb#2\relax
\global\advance\@dlb-\@dla}\@gtempa
\noalign{\vskip-\arrayrulewidth}}

\def\@dlinea{\multispan\@dla&\multispan\@dlb
\unskip\cleaders\hbox to \dline@period
{\hss\rule{\dline@solid}{\arrayrulewidth}\hss}\hfill\cr}
 
\newcount\@dla
\newcount\@dlb

\def\@dlineb{\multispan\@dlb
\unskip\cleaders\hbox to \dline@period
{\hss\rule{\dline@solid}{\arrayrulewidth}\hss}\hfill\cr}

\dlinepattern{2pt}{5pt}

\renewcommand{\theequation}{\thesection.\arabic{equation}}

\def\section{{\setcounter{equation}{0}}
\@startsection {section}{1}{\z@}{-3.5ex plus-1ex minus
             -.2ex}{2.3ex plus.2ex}{\reset@font\Large\bf}}

\makeatother

\newcommand\epem{\mbox{$e^+e^-$}}

\newcommand\textfrac[2]{{\textstyle\frac{#1}{#2}}}

\def\simgt{\rlap{\lower 3.5 pt \hbox{$\mathchar \sim$}}%
           \raise 1pt \hbox {$>$}}
\def\simlt{\rlap{\lower 3.5 pt \hbox{$\mathchar \sim$}}%
           \raise 1pt \hbox {$<$}}
\def\mev{{\,\rm MeV}}

\def\gev{{\,\rm GeV}}

\def\tev{{\,\rm TeV}}

\def\msbar{\overline{\rm MS }}  
\def\to{\rightarrow}

\def\mz{m_Z^{}}
\def\mw{m_W^{}}
\def\mt{m_t^{}}
\def\mh{m_H^{}}
\def\mmv{m_V^2}
\def\mmz{m_Z^2}
\def\mmw{m_W^2}

\def\ebar{\bar{e}}
\def\sbar{\bar{s}}
\def\cbar{\bar{c}}
\def\gzbar{\bar{g}_Z}
\def\gwbar{\bar{g}_W}
\def\ehat{\hat{e}}
\def\shat{\hat{s}}
\def\chat{\hat{c}}
\def\gzhat{\hat{g}_Z}
\def\ghat{\hat{g}}

\def\pibar{\overline{\Pi}}

\def\delb{\mbox{$\bar{\delta}_{b}$}}
\def\delg{\bar{\delta}_{G}^{}}
\def\delgbox{\bar{\delta}_{G}^{\rm box}}
\def\delgvertex{\bar{\delta}_{G}^{\rm vertex}}
\def\zbb{Zb_L^{}b_L^{}}
\def\xs{\mbox{$x_s$}}
\def\xt{\mbox{$x_t$}}
\def\xh{\mbox{$x_H^{}$}}

\def\da{\mbox{$\delta_{\alpha}$}}
\def\xa{\mbox{$x_\alpha$}}
 
\def \ibid {{\it ibid}.}
\def \etal {{\it et al}.}

\newcommand {\PLB}[3]{Phys. Lett.~{\bf B#1}, #3 (#2)}

\newcommand {\NPB}[3]{Nucl. Phys.~{\bf B#1}, #3 (#2)}
\newcommand {\ZP}[3] {Z.~Phys.~{\bf#1}, #3 (#2)}
\newcommand {\ZPC}[3]{Z.~Phys.~{\bf C#1}, #3 (#2)}

\newcommand {\PRD}[3]{Phys. Rev.~{\bf D#1}, #3 (#2)}
\newcommand {\PTP}[3]{Prog. Theor. Phys. {\bf#1}, #3 (#2)}
\newcommand {\PRL}[3]{Phys. Rev.~Lett.~{\bf#1}, #3 (#2)}
\newcommand {\MPL}[3]{Mod. Phys.~Lett.~{\bf#1}, #3 (#2)}

\newcommand {\both}[1]  {\relax\ifmmode#1\else$#1$\fi\relax}
\newcommand {\bea}      {\begin{eqnarray}}
\newcommand {\eea}      {\end{eqnarray}}
\newcommand {\be}       {\begin{equation}}
\newcommand {\ee}       {\end{equation}}
\newcommand {\bsub}     {\begin{subequations}}
\newcommand {\esub}     {\end{subequations}}

\def\fitofsb{%
\begin{equation}
   \sbar^2(\mmz ) = 0.23065 \pm 0.00024 
\label{fitofsb96}   
\end{equation}
}%

\def\fitofsbwithleptonuniversality{%
\begin{equation}
   \sbar^2(\mmz ) = 0.23064 \pm 0.00025 
\label{fitofsb96withleptonuniversality}   
\end{equation}
}%

\def\fitofsbfromleptonicasymmetries{%
\begin{equation}
   \sbar^2(\mmz ) = 0.23019 \pm 0.00031 
\label{fitofsb96fromleptonicasymmetries}   
\end{equation}
}%

\def\fitofgzbsbalpspdb{%
  \begin{subequations}
     \label{fitofgzbsbalpspdb}
  \begin{eqnarray}
  &&\!\! \left.
    \begin{array}{c@{\,}r@{\,}r@{\,}r@{\,}r}
       \gzbar^2(\mmz) &=&   0.55557  &\pm&   0.00074
         \\[1mm]
       \sbar^2(\mmz)  &=&   0.23065  &\pm&   0.00025
         \\[1mm]
       \alpha_s'      &=&   0.1218   &\pm&   0.0038
         \\[1mm]
       \delb(\mmz)    &=&  -0.0051   &\pm&   0.0028
    \end{array}
    \right\} \;
   \rho_{\rm corr} = \left(
      \begin{array}{rrrr}
             1.00 &  0.13 & -0.57 &  0.00  \\[1mm]
                  &  1.00 &  0.11 &  0.05  \\[1mm]
                  &       &  1.00 &  0.01  \\[1mm]
                  &       &       &  1.00  \\[1mm]
      \end{array}
      \right), \,\,\,
   \qquad \label{fitofgzbsbalpspdb_mean}
   \\
   & & \chi^2_{\rm min}/({\rm d.o.f.}) =  15.4/(9)\,.
     \label{chisqofgzbsbalpspdb}
  \end{eqnarray}
  \end{subequations}
}%
\def\fitofgzbsb{%
  \begin{subequations}
   \label{fitofgzbsb}
  \begin{eqnarray}
     & &
     \left.
     \begin{array}{ll}
     \gzbar^2(\mmz) &\!\!= 0.55557
      -0.00042\,\textfrac{\alpha_s' -0.1218}{0.0038}
       \pm 0.00061
      \\[1mm]
     \sbar^2(\mmz)  &\!\!= 0.23065
      +0.00003\,\textfrac{\alpha_s' -0.1218}{0.0038}
       \pm 0.00024
     \end{array}
    \right\}\,\,
   \rho_{\rm corr} = 0.24,
   \\ 
   & & \quad
   \chi^2_{\rm min} =  15.4
         +\biggl(\frac{\alpha_s' -0.1218}{0.0038}\biggr)^2
         +\biggl(\frac{\delb+0.0051}{0.0028}\biggr)^2\,.
   \label{fitofgzbsbchisq}
  \end{eqnarray}
  \end{subequations}
}

\def\fitofmw{%
  \begin{equation}
     \label{fitofmw}
	\gwbar^2(0) = 0.4237\pm 0.0013\,,
  \end{equation}
}%

\def\fitofstu{%
  \begin{subequations}
     \label{fitofstu}
  \begin{eqnarray}
  &&\!\! \left.
    \begin{array}{c@{}r@{}r@{}r@{}r}
       S =&   -0.33 &
              -0.056\,\frac{\alpha_s'-0.1227}{0.0037} &
              +0.06\,\frac{\delta_\alpha-0.03}{0.09} &
              \pm 0.13
         \\[1mm]
       T =&    0.61 &
              -0.094\,\frac{\alpha_s'-0.1227}{0.0037} & &
              \pm 0.14
         \\[1mm]
       U =&    0.48 &
              +0.069\,\frac{\alpha_s'-0.1227}{0.0037} &
              +0.02\,\frac{\delta_\alpha-0.03}{0.09} &
              \pm 0.38 \\[1mm]
    \end{array}
    \right\} \;
   \rho_{\rm corr} = \left(
      \begin{array}{rrr}
           1    &  0.86 & -0.11  \\[1mm]
                &  1    & -0.21  \\[1mm]
                &       &  1     \\[1mm]
      \end{array}
      \right), \,\,\,
   \qquad \label{fitofstu_mean}
   \\
   & & \chi^2_{\rm min} =  20.3
         +\biggl(\frac{\alpha_s' -0.1227}{0.0037}\biggr)^2
         +\biggl(\frac{\delb+0.0051}{0.0028}\biggr)^2\,, \quad
   ({\rm d.o.f.} = 21)\,.
     \label{chisqofstu}
  \end{eqnarray}
  \end{subequations}
}%
\def\dataofnqccfrorig{%
\begin{equation}
   K = 0.5626 \pm 0.0025\,\mbox{(stat)} \pm 0.0036\,\mbox{(sys)}
              \pm 0.0028\,\mbox{(model)}
               -0.0029\,\frac{m_c-1.31\gev}{0.24\gev}, 
   \label{dataofnqccfrorig}
\end{equation}
}

\def\fitofnqccfr{%
\begin{eqnarray}
  \sbar^2(0) 
             = 0.2421 + 1.987[\gzbar^2(0)-0.5486] \pm 0.0058\,.
   \label{fitofnqccfr}
\end{eqnarray}
}

\def\fitofnqfh{%
  \begin{subequations}
   \label{fitofnqfh}
  \begin{eqnarray}
     & &
     \left.
     \begin{array}{ll}
     \gzbar^2(0) &= 0.5454^{+0.0076}_{-0.0082}
      \\[1mm]
     \sbar^2(0)  &= 0.2419^{+0.0130}_{-0.0142}
     \end{array}
    \right\}\quad
   \rho_{\rm corr} = 0.916,
   \\ 
   & & \quad
   \chi^2_{\rm min} =  0.13\,.
   \label{fitofnqfhchisq}
  \end{eqnarray}
  \end{subequations}
}

\def\fitofnq{%
  \begin{subequations}
   \label{fitofnq}
  \begin{eqnarray}
     & &
     \left.
     \begin{array}{ll}
     \gzbar^2(0) &= 0.5476^{+0.0070}_{-0.0076}
      \\[1mm]
     \sbar^2(0)  &= 0.2429^{+0.0128}_{-0.0140}
     \end{array}
    \right\}\quad
   \rho_{\rm corr} = 0.955,
   \\ 
   & & \quad
   \chi^2_{\rm min} =  0.7 \quad ({\rm d.o.f.}=3).
   \label{fitofnqchisq}
  \end{eqnarray}
  \end{subequations}
}

\def\fitoflenc{%
  \begin{subequations}
   \label{fitoflenc}
  \begin{eqnarray}
     & &
     \left.
     \begin{array}{ll}
     \gzbar^2(0) &= 0.5441 \pm 0.0029
      \\[1mm]
     \sbar^2(0)  &= 0.2362 \pm 0.0044
     \end{array}
    \right\}\quad
   \rho_{\rm corr} = 0.70,
   \\ 
   & & \quad
   \chi^2_{\rm min} =  2.7 \quad ({\rm d.o.f.}=8)\,.
   \label{fitoflencchisq}
  \end{eqnarray}
  \end{subequations}
}

\def\fitoflencatmz{%
  \begin{subequations}
   \label{fitoflencatmz}
  \begin{eqnarray}
     & &
     \left.
     \begin{array}{ll}
     \gzbar^2(\mmz) &= 0.5512 \pm 0.0030
      \\[1mm]
     \sbar^2(\mmz)  &= 0.2277 \pm 0.0047
     \end{array}
    \right\}\quad
   \rho_{\rm corr} = 0.70,
   \\ 
   & & \quad
   \chi^2_{\rm min} =  2.7 \quad ({\rm d.o.f.}=8)\,.
   \label{fitoflencatmzchisq}
  \end{eqnarray}
  \end{subequations}
}

\def\chisqsm{
  \begin{subequations}
  \label{total_chisqsm}
  \begin{eqnarray}
     \chi^2_{\rm SM}(m_t,\mh,\alpha_s,\delta_\alpha)
     &=& \biggl(\frac{m_t -\langle m_t\rangle}{\Delta m_t} \biggr)^2
         +\chi^2_{H}(\mh,\alpha_s,\delta_\alpha)\,,
     \label{chisqsm}
  \end{eqnarray}
}
\def\fitofmt{%
  \begin{eqnarray}
     \label{fitofmt}
    \langle m_t \rangle &=& 162.4 +13.0\, \log\frac{\mh}{100}
                            +0.8 \, \log^2\frac{\mh}{100}
           -0.85\,\biggl(\frac{\alpha_s-0.118}{0.003}\biggr)
           -4.9\,\biggl(\frac{\delta_\alpha-0.03}{0.09}\biggr),
     \nonumber \\
     && \label{fitofmtbest}
     \\
     \Delta m_t   &=& 5.5 - 0.06\, \log \frac{\mh}{100}
          - \Bigl(0.090 -0.018\,\log \frac{\mh}{100}\Bigr)\,
            \frac{m_t -175}{6}\,,
     \label{fitofmterror}
  \end{eqnarray}
}
\def\chisqhsm{%
  \begin{eqnarray}
     \chi^2_H(\mh,\alpha_s,\delta_\alpha) &=&
      22.1 + \biggl(\frac{\delta_\alpha -0.19}{0.18} \biggr)^2
          + \biggl(\frac{\alpha_s-0.1201+0.0011\,\delta_\alpha}{0.0031}
            \biggr)^2 \quad
     \nonumber \\
       && - \biggl(\frac{\alpha_s-0.1343+0.063\,\delta_\alpha}{0.0071} 
                   \biggr) \log \frac{\mh}{100}
          - \biggl(\frac{\alpha_s-0.1305}{0.0129}\biggr)
                   \log^2 \frac{\mh}{100} \,.
    \nonumber \\ &&
    \label{chisqhsm}
  \end{eqnarray}
  \end{subequations}
}
\def\fitofmtstandard{%
\begin{eqnarray} \label{mtfit_standard}
   m_t = 178 \pm 6 {}^{+19(\mh=1000)}_{-21(\mh=60)}
             \mp 1 (\alpha_s)
             \mp 5 (\delta_\alpha) \,,
\end{eqnarray}
}

\def\fitofmtxhalpsda{%
  \begin{subequations}
     \label{fitofmtxhalpsda}
  \begin{eqnarray}
  &&\!\! \left.
    \begin{array}{c@{\,}r@{\,}l}
       m_t(\gev)     &=&  151    \pm  13     \\[1mm]
       \xh           &=& -0.5    \pm  1.5   \\[1mm]
       \alpha_s      &=&  0.1198 \pm  0.0031 \\[1mm]
       \delta_\alpha &=&  0.13   \pm  0.34
    \end{array}
    \right\} \;
   \rho_{\rm corr} = \left(
      \begin{array}{rrrr}
         1.0 &  0.0 & -0.0 &  0.5\\
             &  1.0 & -0.1 & -0.8\\
             &      &  1.0 &  0.1\\
             &      &      &  1.0
      \end{array}
      \right), \,\,\,
   \qquad \label{fitofmtxhalpsda_mean}
   \\
   & & \chi^2_{\rm min}/({\rm d.o.f.}) =  21.9/(21)\,.
     \label{chisqofmtxhalpsda}
  \end{eqnarray}
  \end{subequations}
}%

\def\fitofmtxhalpsdawithalpsdaej{%
  \begin{subequations}
     \label{fitofmtxhalpsdawithalpsdaej}
  \begin{eqnarray}
  &&\!\! \left.
    \begin{array}{c@{\,}r@{\,}l}
       m_t(\gev)     &=&  153    \pm  10     \\[1mm]
       \xh           &=& -0.8    \pm  0.8   \\[1mm]
       \alpha_s      &=&  0.1190 \pm  0.0022 \\[1mm]
       \delta_\alpha &=&  0.04   \pm  0.09
    \end{array}
    \right\} \;
   \rho_{\rm corr} = \left(
      \begin{array}{rrrr}
         1.0 &  0.6 & -0.1 &  0.3\\
             &  1.0 & -0.1 & -0.3\\
             &      &  1.0 &  0.1\\
             &      &      &  1.0
      \end{array}
      \right), \,\,\,
   \qquad \label{fitofmtxhalpsdawithalpsdaej_mean}
   \\
   & & \chi^2_{\rm min}/({\rm d.o.f.}) =  22.2/(23)\,,
     \label{chisqofmtxhalpsdawithalpsdaej}
  \end{eqnarray}
  \end{subequations}
}%

\def\fitofmtxhalpsdawithalpsdamz{%
  \begin{subequations}
     \label{fitofmtxhalpsdawithalpsdamz}
  \begin{eqnarray}
  &&\!\! \left.
    \begin{array}{c@{\,}r@{\,}l}
       m_t(\gev)     &=&  151    \pm  11  \\[1mm]
       \xh           &=& -0.5    \pm  0.8  \\[1mm]
       \alpha_s      &=&  0.1189 \pm  0.0022  \\[1mm]
       \delta_\alpha &=&  0.12   \pm  0.06
    \end{array}
    \right\} \;
   \rho_{\rm corr} = \left(
      \begin{array}{rrrr}
         1.00 &  0.8 & -0.1 &  0.1\\
              &  1.0 & -0.0 & -0.2\\
              &       & 1.0 &  0.0\\
              &       &     &  1.0
      \end{array}
      \right), \,\,\,
   \qquad \label{fitofmtxhalpsdawithalpsdamz_mean}
   \\
   & & \chi^2_{\rm min}/({\rm d.o.f.}) =  22.1/(23)\,.
     \label{chisqofmtxhalpsdawithalpsdamz}
  \end{eqnarray}
  \end{subequations}
}%

\def\fitofmtxhalpsdawithalpsdaejwithoutrbrc{%
  \begin{subequations}
     \label{fitofmtxhalpsdawithalpsdaejwithoutrbrc}
  \begin{eqnarray}
  &&\!\! \left.
    \begin{array}{c@{\,}r@{\,}l}
       m_t(\gev)     &=&  158   \pm  12   \\[1mm]
       \xh           &=& -0.5   \pm  1.0   \\[1mm]
       \alpha_s      &=&  0.1188\pm  0.0022   \\[1mm]
       \delta_\alpha &=&  0.03  \pm  0.09
    \end{array}
    \right\} \;
   \rho_{\rm corr} = \left(
      \begin{array}{rrrr}
         1.0 &  0.8 & -0.1 &  0.1\\
             &  1.0 & -0.0 & -0.4\\
             &      &  1.0 &  0.0\\
             &      &      &  1.0
      \end{array}
      \right), \,\,\,
   \qquad \label{fitofmtxhalpsdawithalpsdaejwithoutrbrc_mean}
   \\
   & & \chi^2_{\rm min}/({\rm d.o.f.}) =  20.5/(22)\,.
     \label{chisqofmtxhalpsdawithalpsdaejwithoutrbrc}
  \end{eqnarray}
  \end{subequations}
}%

\def\fitofmtxhalpsdawithmt{%
  \begin{subequations}
     \label{fitofmtxhalpsdawithmt}
  \begin{eqnarray}
  &&\!\! \left.
    \begin{array}{c@{\,}r@{\,}l}
       m_t(\gev)     &=&  173    \pm  6      \\[1mm]
       \xh           &=&  1.7    \pm  1.1    \\[1mm]
       \alpha_s      &=&  0.1218 \pm  0.0037 \\[1mm]
       \delta_\alpha &=&  0.30   \pm  0.26
    \end{array}
    \right\} \;
   \rho_{\rm corr} = \left(
      \begin{array}{rrrr}
    1.0  &  0.4  &  0.2  & -0.1 \\
         &  1.0  &  0.5  & -0.9 \\
         &       &  1.0  & -0.4 \\
         &       &       &  1.0
      \end{array}
      \right), \,\,\,
   \qquad \label{fitofmtxhalpsdawithmt_mean}
   \\
   & & \chi^2_{\rm min}/({\rm d.o.f.}) =  23.8/(22)\,.
     \label{chisqofmtxhalpsdawithmt}
  \end{eqnarray}
  \end{subequations}
}%

\def\fitofmtxhalpsdawithmtalpsdaej{
  \begin{subequations}
     \label{fitofmtxhalpsdawithmtalpsdaej}
  \begin{eqnarray}
  &&\!\! \left.
    \begin{array}{c@{\,}r@{\,}l}
       m_t(\gev)     &=&  171    \pm  6      \\[1mm]
       \xh           &=&  0.5    \pm  0.6    \\[1mm]
       \alpha_s      &=&  0.1191 \pm  0.0022 \\[1mm]
       \delta_\alpha &=&  0.05   \pm  0.08
    \end{array}
    \right\} \;
   \rho_{\rm corr} = \left(
      \begin{array}{rrrr}
         1.0  &  0.6  &  0.1  & -0.0\\
              &  1.0  &  0.2  & -0.6\\
              &       &  1.0  & -0.1\\
              &       &       &  1.0
      \end{array}
      \right), \,\,\,
   \qquad \label{fitofmtxhalpsdawithmtalpsdaej_mean}
   \\
   & & \chi^2_{\rm min}/({\rm d.o.f.}) =  24.9/(24) \,.
     \label{chisqofmtxhalpsdawithmtalpsdaej}
  \end{eqnarray}
  \end{subequations}
}%

\def\fitofmtxhalpsdawithmtalpsdamz{%
  \begin{subequations}
     \label{fitofmtxhalpsdawithmtalpsdamz}
  \begin{eqnarray}
  &&\!\! \left.
    \begin{array}{c@{\,}r@{\,}l}
       m_t(\gev)     &=&  172    \pm  6      \\[1mm]
       \xh           &=&  0.9    \pm  0.6    \\[1mm] 
       \alpha_s      &=&  0.1193 \pm  0.0022 \\[1mm]
       \delta_\alpha &=&  0.12   \pm  0.06
    \end{array}
    \right\} \;
   \rho_{\rm corr} = \left(
      \begin{array}{rrrr}
         1.0  &  0.7  &  0.1  & -0.0\\
              &  1.0  &  0.2  & -0.4\\
              &       &  1.0  & -0.1\\
              &       &       &  1.0
      \end{array}
      \right), \,\,\,
   \qquad \label{fitofmtxhalpsdawithmtalpsdamz_mean}
   \\
   & & \chi^2_{\rm min}/({\rm d.o.f.}) =  24.6/(24)\,.
     \label{chisqofmtxhalpsdawithmtalpsdamz}
  \end{eqnarray}
  \end{subequations}
}%

\def\fitofncgzbsbalpspdb{%
  \begin{subequations}
     \label{fitofncgzbsbalpspdb}
  \begin{eqnarray}
  &&\!\! \left.
    \begin{array}{c@{\,}r@{\,}r@{\,}r@{\,}r}
       \gzbar^2(\mmz) &=&   0.55525  &\pm&   0.00070  \\[1mm]
       \sbar^2(\mmz)  &=&   0.23065  &\pm&   0.00024  \\[1mm]
       \alpha_s'      &=&   0.1227   &\pm&   0.0037   \\[1mm]
       \delb(\mmz)    &=&  -0.0051   &\pm&   0.0028
    \end{array}
    \right\} \;
    \rho_{\rm corr} = \left(
      \begin{array}{rrrr}
             1.00 &  0.14 & -0.54 &  0.00  \\[1mm]
                  &  1.00 &  0.11 &  0.05  \\[1mm]
                  &       &  1.00 &  0.01  \\[1mm]
                  &       &       &  1.00  \\[1mm]
      \end{array}
      \right), \,\,\,
   \qquad \label{fitofncgzbsbalpspdb_mean}
   \\
   & & \chi^2_{\rm min}/({\rm d.o.f.}) =  20.4/(19)
     \label{chisqofncgzbsbalpspdb}
  \end{eqnarray}
  \end{subequations}
}%
\def\fitofncgzbsb{%
  \begin{subequations}
   \label{fitofncgzbsb}
  \begin{eqnarray}
     & &
     \left.
     \begin{array}{ll}
      \gzbar^2(\mmz) &\!\!= 0.55525
        -0.00038\,\textfrac{\alpha_s' -0.1227}{0.0037} \pm 0.00059
        \\[1mm]
      \sbar^2(\mmz)  &\!\!= 0.23065
        +0.00003\,\textfrac{\alpha_s' -0.1227}{0.0037} \pm 0.00024
     \end{array}
     \right\}\,\,
   \rho_{\rm corr} = 0.24,
   \\ 
   & & \quad
   \chi^2_{\rm min} =  20.4
         +\biggl(\frac{\alpha_s' -0.1227}{0.0037}\biggr)^2
         +\biggl(\frac{\delb+0.0051}{0.0028}\biggr)^2\,,
   \label{fitofncgzbsbchisq}
  \end{eqnarray}
  \end{subequations}
}

\def\fitofmtxhalpsdafuture{%
  \begin{subequations}
     \label{fitofmtxhalpsdafuture}
  \begin{eqnarray}
  &&\!\! \left.
    \begin{array}{c@{\,}r@{\,}l}
       m_t(\gev)     &=&  161    \pm  5      \\[1mm]
       \xh           &=& -1.24   \pm  0.95   \\[1mm]
       \alpha_s      &=&  0.1204 \pm  0.0035 \\[1mm]
       \delta_\alpha &=& -0.13   \pm  0.16
    \end{array}
    \right\} \;
   \rho_{\rm corr} = \left(
      \begin{array}{rrrr}
         1.00  &  0.35  & -0.10  &  0.18\\
               &  1.00  & -0.47  & -0.80\\
               &        &  1.00  &  0.42\\
               &        &        &  1.00
      \end{array}
      \right), \,\,\,
   \qquad \label{fitofmtxhalpsdafuture_mean}
   \\
   & & \chi^2_{\rm min}/({\rm d.o.f.}) =  24.6/(23)\,.
     \label{chisqofmtxhalpsdafuture}
  \end{eqnarray}
  \end{subequations}
}%

\def\fitofmtxhalpsdafuturewithmtalpsda{%
  \begin{subequations}
     \label{fitofmtxhalpsdafuturewithmtalpsda}
  \begin{eqnarray}
  &&\!\! \left.
    \begin{array}{c@{\,}r@{\,}l}
       m_t(\gev)     &=&  172    \pm  6      \\[1mm]
       \xh           &=&  0.49   \pm  0.60   \\[1mm]
       \alpha_s      &=&  0.1190 \pm  0.0022 \\[1mm]
       \delta_\alpha &=&  0.04   \pm  0.08
    \end{array}
    \right\} \;
   \rho_{\rm corr} = \left(
      \begin{array}{rrrr}
         1.00  &  0.84  &  0.13  & -0.26\\
               &  1.00  &  0.13  & -0.63\\
               &        &  1.00  & -0.06\\
               &        &        &  1.00
      \end{array}
      \right), \,\,\,
   \qquad \label{fitofmtxhalpsdafuturewithmtalpsda_mean}
   \\
   & & \chi^2_{\rm min}/({\rm d.o.f.}) =  24.9/(26)\,.
     \label{chisqofmtxhalpsdafuturewithmtalpsda}
  \end{eqnarray}
  \end{subequations}
}%

\def\fitofmtxhalpsdafuturewithfuturemtalpsda{%
  \begin{subequations}
     \label{fitofmtxhalpsdafuturewithfuturemtalpsda}
  \begin{eqnarray}
  &&\!\! \left.
    \begin{array}{c@{\,}r@{\,}l}
       m_t(\gev)     &=&  175    \pm  2      \\[1mm]
       \xh           &=&  0.75   \pm  0.35   \\[1mm]
       \alpha_s      &=&  0.1192 \pm  0.0021 \\[1mm]
       \delta_\alpha &=&  0.05   \pm  0.08
    \end{array}
    \right\} \;
   \rho_{\rm corr} = \left(
      \begin{array}{rrrr}
         1.00  &  0.48  &  0.05  & -0.09\\
               &  1.00  &  0.07  & -0.73\\
               &        &  1.00  & -0.03\\
               &        &        &  1.00
      \end{array}
      \right), \,\,\,
   \qquad \label{fitofmtxhalpsdafuturewithfuturemtalpsda_mean}
   \\
   & & \chi^2_{\rm min}/({\rm d.o.f.}) =  25.2/(26)\,.
     \label{chisqofmtxhalpsdafuturewithfuturemtalpsda}
  \end{eqnarray}
  \end{subequations}
}%

\def\fitofmtxhalpsdafuturewithfuturemtalpsfutureda{%
  \begin{subequations}
     \label{fitofmtxhalpsdafuturewithfuturemtalpsfutureda}
  \begin{eqnarray}
  &&\!\! \left.
    \begin{array}{c@{\,}r@{\,}l}
       m_t(\gev)     &=&  175    \pm  2      \\[1mm]
       \xh           &=&  0.69   \pm  0.26   \\[1mm]
       \alpha_s      &=&  0.1192 \pm  0.0021 \\[1mm]
       \delta_\alpha &=&  0.03   \pm  0.03
    \end{array}
    \right\} \;
   \rho_{\rm corr} = \left(
      \begin{array}{rrrr}
         1.00  &  0.58  &  0.05  & -0.04\\
               &  1.00  &  0.07  & -0.38\\
               &        &  1.00  & -0.01\\
               &        &        &  1.00
      \end{array}
      \right), \,\,\,
   \qquad \label{fitofmtxhalpsdafuturewithmtalpsfutureda_mean}
   \\
   & & \chi^2_{\rm min}/({\rm d.o.f.}) =  25.3/(26)\,.
     \label{chisqofmtxhalpsdafuturewithfuturemtalpsfutureda}
  \end{eqnarray}
  \end{subequations}
}%

\def\fitofstuwithfuturedata{%
  \begin{subequations}
     \label{fitofstuwithfuturedata}
  \begin{eqnarray}
  &&\!\! \left.
    \begin{array}{c@{}r@{}r@{}r}
       S' =&  -0.32 &
              -0.061\,\frac{\alpha_s'-0.1075}{0.0037} &
              \pm 0.11
         \\[1mm]
       T' =&    0.61 &
              -0.096\,\frac{\alpha_s'-0.1075}{0.0037} & 
              \pm 0.14
         \\[1mm]
       U'' =&    0.47 &
              +0.065\,\frac{\alpha_s'-0.1075}{0.0037} &
              \pm 0.11 \\[1mm]
    \end{array}
    \right\} \;
   \rho_{\rm corr} = \left(
      \begin{array}{rrr}
           1    &  0.92 & -0.60  \\[1mm]
                &  1    & -0.79  \\[1mm]
                &       &  1     \\[1mm]
      \end{array}
      \right), \,\,\,
   \qquad \label{fitofstuwithfuturedata_mean}
   \\
   & & \chi^2_{\rm min} =  20.4
         +\biggl(\frac{\alpha_s' -0.1075}{0.0037}\biggr)^2
         +\biggl(\frac{\delb+0.0051}{0.0028}\biggr)^2\,, \quad
   ({\rm d.o.f.} = 23)\,.
     \label{chisqofstuwithfuturedata}
  \end{eqnarray}
  \end{subequations}
}%
\begin{document}


\begin{center}

\thispagestyle{empty}
\vspace*{-25mm}
\baselineskip10pt
\begin{flushright}
\begin{tabular}{l}
{\bf KEK-TH-512}\\
{\bf DESY 96-192}\\
{\bf hep-ph/9706331}\\
\today
\end{tabular}
\end{flushright}
\baselineskip18pt 
\vglue 12mm 

{\Large\bf
Analysis of Electroweak Precision Data \\
and Prospects for Future Improvements
}
\vspace{5mm}


\def\thefootnote{\alph{footnote}}
\setcounter{footnote}{0}
{\bf
Kaoru Hagiwara$^{1,2,}\footnote{E-mail: kaoru.hagiwara@kek.jp}$,
Dieter Haidt$^{3,}\footnote{E-mail: haidt@dice2.desy.de}$, 
Seiji Matsumoto$^{1,}\footnote{E-mail: seiji.matsumoto@kek.jp}$
}
\vspace{5mm}
\def\thefootnote{\arabic{footnote}}
\setcounter{footnote}{0}


{\it
$^1$ Theory Group, KEK, Tsukuba, Ibaraki 305, Japan
\\
$^2$ ICEPP, University of Tokyo, Hongo, Bunkyo-ku, Tokyo 113, Japan
\\
$^3$ DESY, Notkestrasse 85, D-22603 Hamburg, Germany
}

\end{center}
\vspace{5mm}


\begin{center}
{\bf Abstract}\\[3mm]
\begin{minipage}{12cm}
\baselineskip 12pt
\noindent
We update our previous work on an analysis of the electroweak data 
by including new and partly preliminary data available up to the 1996
summer conferences. 
The new results on the $Z$ partial decay widths into $b$ and $c$ 
hadrons now offer a consistent interpretation of all data in the 
minimal standard model.
The value extracted for the strong interaction coupling constant 
$\alpha_s(\mz)$ 
agrees well with determinations in other areas. 
New constraints on the universal parameters $S$, $T$ and $U$ are obtained 
from the updated measurements.
No signal of new physics is found in the $S$, $T$, $U$ analysis once 
the SM contributions with $m_t \sim 175$GeV and those of not a too heavy 
Higgs boson are accounted for. 
The naive QCD-like technicolor model is now ruled out at the 99\%CL 
even for the minimal model with ${\rm SU(2)_{TC}}$.
In the absence of a significant new physics effect in the electroweak 
observables, constraints on masses of the top quark, $m_t$, 
and Higgs boson, $\mh$, are
derived as a function 
of $\alpha_s$ and the QED effective coupling $\bar{\alpha}(m_Z^2)$.
The preferred range of $\mh$ depends rather strongly on the actual value 
of $m_t$\,: $\mh<360 \gev$ for $\mt=170\gev$, while $\mh>130\gev$ 
for $\mt=180\gev$ at 95~\%CL.   
Prospects due to forthcoming improved measurements of asymmetries, 
the mass of the weak boson $W$ 
$m_W$, $m_t$ and $\bar{\alpha}(m_Z^2)$ are discussed.
Anticipating uncertainties of
0.00020 for $\sbar^2(\mmz )$, 20~MeV for $\mw$, 
and 2~GeV for $\mt$, the new physics contributions to the $S$, $T$, $U$ 
parameters will be constrained more severely, 
and, within the SM, 
the logarithm of the Higgs mass can be constrained to about $\pm 0.35$.
The better constraints on $S$, $T$, $U$ and on $\mh$ within the minimal SM 
should be accompanied with matching precision in
$\bar{\alpha}(m_Z^2)$. 
\end{minipage}
\end{center}

\begin{center}
{\it To be published in Zeitshrift f\"ur Physik C}
\end{center}


\section{Introduction}
 
The physics program of LEP1 is completed and has brought a wealth of precise 
data at the $Z$-resonance. With the presentation of the updated measurements 
at the 1996 summer conferences\cite{lepewwg96} an appropriate moment has come 
to assess the impact of the new data in the context of the theoretical 
framework introduced in Refs.\cite{hhkm,sm95}.
   
The $Z$-shape variables are now quite well measured (see Table~1), 
also the apparent 
discrepancy of the previous $R_b$ and $R_c$ measurements\cite{lepewwg95} with 
their Standard Model (SM) expectations seems to be solved. After combining 
the preliminary data from all LEP experiments and from SLD, the $R_c$ value 
is now in good agreement with the SM, while $R_b$ is less than 2 standard 
deviations away from the SM prediction. These new measurements are of 
importance when  extracting a reliable value for the QCD coupling constant 
$\alpha_s(\mz)$ from the electroweak data. 
 
The paper is organized as follows.
In section 2 
all electroweak measurements from LEP, SLC and Tevatron reported 
to the Warsaw Conference\cite{lepewwg96}, are collected. These data are 
compared with the SM predictions\cite{hhkm} and a few remarkable features 
are pointed out.
In section 3 
a brief review is given of the electroweak radiative corrections in generic 
$\rm{SU(2)_L \times U(1)_Y}$ models following the formalism of Ref.\cite{hhkm}.
In section 4
the impact of the new measurements is discussed, in particular 
the $Z$-shape parameter measurements at LEP/SLC and 
the new neutrino measurement of CCFR. 
A comprehensive fit to all the electroweak data is performed in terms 
of the three parameters\cite{stu} $S$, $T$, $U$, which characterize possible 
new physics contributions through the electroweak gauge-boson propagator 
corrections,  and $\delb$ which characterizes possible new physics 
contributions to the $\zbb$ vertex.
Section 5
is devoted to the interpretation of all electroweak data within the minimal 
SM. Their constraints are shown as functions of $\alpha_s(\mz)$ and 
$\bar{\alpha}(\mmz)$ in the $(m_t,m_H)$-plane. 
A brief discussion on the significance of bosonic radiative corrections 
containing the weak boson self-couplings is also given.
In section 6 
the impact of future improved 
measurements of the $Z$ boson asymmetries, the $W$ and top-quark 
masses and $\bar{\alpha}(\mmz)$ are studied.    
Finally, section 7
gives a summary and outlook.


\section{%
Electroweak Precision Data }
%
Since our first analysis of electroweak data\cite{hhkm} a considerable
improvement occurred in three areas, which is summarized in Table~1. The
LEP Electroweak Working Group\cite{lepewwg96} has updated their results 
by including their preliminary electroweak data 
available up to summer 1996. The table contains also the results from 
SLC\cite{lepewwg96} and new Tevatron data on the $W$ mass\cite{mw96} 
and the neutrino neutral current experiment\cite{ccfr95_kevin}. 
Correlation matrices among the errors of the 
line-shape parameters and the heavy-quark parameters are given in Tables~2 
and 3, respectively. All the numerical results presented in this paper are 
based on the unchanged data in Ref.\cite{hhkm} and the updated data in 
Tables~1--3, unless otherwise stated. Also shown in Table~1 are the SM 
predictions\cite{hhkm} for $m_t=175$~GeV, equal to the present best value 
from CDF and D0\cite{mt96}, $\mh=100$~GeV, $\alpha_s(\mz )=0.118$ and
$1/\bar{\alpha}(\mmz )=128.75$. The sensitivity of the fit results due to
the uncertainties of the QCD and QED running coupling strengths will be
discussed in sections 4, 5 and 6.
The right-most column gives the 
difference between the mean of the data and the corresponding SM prediction 
in units of the experimental error. The data and the SM predictions agree 
fairly well. The previously\cite{lepewwg95} larger values of $R_b$ and 
smaller values of $R_c$ are now close to the SM prediction.
 
\begin{table}[t]
\begin{center}
\caption{\protect\footnotesize\sl
Summary of new electroweak results since our first analysis\cite{hhkm}. 
These data represent the status as of the 1996 summer 
conferences
and contain contributions from LEP and SLC\protect\cite{lepewwg96} and  
Tevatron, $p\bar{p}$\protect\cite{mw96} and CCFR\protect\cite{ccfr95_kevin}. 
The SM predictions\cite{hhkm} are calculated for 
$m_t=175$~GeV, $\mh=100$~GeV, $\alpha_s(\mz)=0.118$, and 
$1/\bar{\alpha}(\mmz)=128.75$; 
see section 3 for the definition of $\bar{\alpha}(\mmz )$ 
and its uncertainty.  
Heavy flavor results are obtained by combining data from 
LEP and SLC\protect\cite{lepewwg96}.
}
\def\afb{A_{\rm FB}}
\def\mzdata    { $ 91.1863\pm 0.0020$}
\def\gammazdata{ $ 2.4946 \pm 0.0027$}
\def\sigmahdata{ $ 41.508 \pm 0.056$ }
\def\rldata    { $ 20.778 \pm 0.029$ }
\def\afbldata  { $ 0.0174 \pm 0.0010$}
\def\ataudata  { $ 0.1401 \pm 0.0067$}
\def\aedata    { $ 0.1382 \pm 0.0076$}
\def\rbdata    { $ 0.2178 \pm 0.0011$}
\def\rcdata    { $ 0.1715 \pm 0.0056$}
\def\afbbdata  { $ 0.0979 \pm 0.0023$}
\def\afbcdata  { $ 0.0735 \pm 0.0048$}
\def\slleptdata{ $ 0.2320 \pm 0.0010$}
\def\alrdata   { $ 0.1542 \pm 0.0037$}
\def\abdata    { $ 0.863  \pm 0.049 $}
\def\acdata    { $ 0.625  \pm 0.084 $}
\def\mwdata    { $ 80.356 \pm 0.125 $}
\def\ccfrdata  { $ 0.5626 \pm 0.0060$}
\def\mzsm    { ---          }
\def\gammazsm{ $  2.4972 $}
\def\sigmahsm{ $ 41.474  $}
\def\rlsm    { $ 20.747  $}
\def\afblsm  { $  0.0168 $}
\def\atausm  { $  0.1485 $}
\def\aesm    { $  0.1486 $}
\def\rbsm    { $  0.2157 $}
\def\rcsm    { $  0.1721 $}
\def\afbbsm  { $  0.1041 $}
\def\afbcsm  { $  0.0747 $}
\def\slleptsm{ $  0.2313 $}
\def\alrsm   { $  0.1485 $}
\def\absm    { $  0.935  $}
\def\acsm    { $  0.668  $}
\def\mwsm    { $  80.400 $}
\def\ccfrsm  { $  0.5669 $}
\def\mzpull    { ---         }
\def\gammazpull{ $ -1.0 $}
\def\sigmahpull{ $  0.6 $}
\def\rlpull    { $  1.1 $}
\def\afblpull  { $  0.6 $}
\def\ataupull  { $ -1.3 $}
\def\aepull    { $ -1.4 $}
\def\rbpull    { $  1.9 $}
\def\rcpull    { $ -0.1 $}
\def\afbbpull  { $ -2.7 $}
\def\afbcpull  { $ -0.2 $}
\def\slleptpull{ $  0.7 $}
\def\alrpull   { $  1.5 $}
\def\abpull    { $ -1.5 $}
\def\acpull    { $ -0.5 $}
\def\mwpull    { $ -0.4 $}
\def\ccfrpull  { $  0.7 $}
{\footnotesize
\begin{tabular}{|lr|c|c|r|} 
\hline
& & data  &  SM  & ${\rm \frac{\langle data\rangle-SM}{(error)}}$
\\
\hline
{\bf LEP} & & & & \\[-2mm]
&\multicolumn{1}{l|}{line shape:} & & &\\
&${ \mz}$(GeV)       & \mzdata     & \mzsm     & \mzpull     \\[0.5mm] 
&${\Gamma _Z}$(GeV)  & \gammazdata & \gammazsm & \gammazpull \\[0.5mm]
&${ \sigma_h^0}$(nb) & \sigmahdata & \sigmahsm & \sigmahpull \\[0.5mm]
&${ R_\ell\equiv}\Gamma_{ h/}\Gamma_{ \ell}$ 
                     & \rldata     & \rlsm     & \rlpull     \\[0.5mm]
&${ \afb^{0,\ell}}$  & \afbldata   & \afblsm   & \afblpull   \\[0.5mm]
&\multicolumn{1}{l|}{$\tau$ polarization:} & & &\\[0.5mm]
&${ A_\tau}$         & \ataudata   & \atausm   & \ataupull \\[0.5mm]
&${ A_e}$            & \aedata     & \aesm     & \aepull   \\[0.5mm]
&\multicolumn{1}{l|}{heavy flavor results:}
 & & &\\[0.5mm]
&${ R_b\equiv}\Gamma_{ b/}\Gamma_{ h}$
                     & \rbdata     & \rbsm     & \rbpull   \\[0.5mm]
&${ R_c\equiv}\Gamma_{ c/}\Gamma_{ h}$
                     & \rcdata     & \rcsm     & \rcpull   \\[0.5mm]
&${ \afb^{0,b}}$     & \afbbdata   & \afbbsm   & \afbbpull \\[0.5mm]
&${ \afb^{0,c}}$     & \afbcdata   & \afbcsm   & \afbcpull \\[1.5mm]
&\multicolumn{1}{l|}{ jet charge asymmetry:}
& & &\\[0.5mm]
&{ $ \sin^2\theta^{\rm lept}_{\rm eff}(\langle Q_{FB} \rangle)$}
                     & \slleptdata & \slleptsm & \slleptpull \\[1.5mm]
{\bf SLC} & & & & \\[-2mm]
&${ A_{LR}^0}$       & \alrdata    & \alrsm    & \alrpull \\[0.5mm]
&${ A_b}$            & \abdata     & \absm     & \abpull  \\[0.5mm]
&${ A_c}$            & \acdata     & \acsm     & \acpull  \\[1.5mm]
\multicolumn{2}{|l|}{\bf Tevatron} & & & \\[-1mm]
{ $ p\bar{p}$}& & & &\\[-2mm]
&${ m_W^{}}$         & \mwdata     & \mwsm     & \mwpull  \\[1.5mm]
{ CCFR}& & & &\\[-2mm]
&$K$                 & \ccfrdata   & \ccfrsm     & \ccfrpull  \\[1.5mm]
\hline
\end{tabular}
}
\end{center}
\end{table}
%
 
\begin{table}[t]
\begin{minipage}[t]{2.65in}
\begin{center}
\caption{\protect\footnotesize\sl
The error correlation matrix for the $Z$ line-shape 
parameters\protect\cite{lepewwg96}.
}
\label{tableofcorrzlineshape}
{\footnotesize
\begin{tabular}{|c|r@{\,\,}r@{\,\,}r@{$\,\,$}r@{$\,\,$}r|}
\hline
&
\multicolumn{1}{c}{$\mz$} & 
\multicolumn{1}{c}{$\Gamma_Z$} &
\multicolumn{1}{c}{$\sigma_{\rm h}^0$} &
\multicolumn{1}{c}{$R_\ell$} &
\multicolumn{1}{c|}{$A_{\rm FB}^{0,\ell}$}
\\ \hline
$\mz$                 &   1.00 &   0.09 &$-$0.01 &$-$0.01 & 0.08\\
$\Gamma_Z$            &   0.09 &   1.00 &$-$0.14 &$-$0.01 & 0.00\\
$\sigma_{\rm h}^0$    &$-$0.01 &$-$0.14 &   1.00 &   0.15 & 0.01\\
$R_\ell$              &$-$0.01 &$-$0.01 &   0.15 &   1.00 & 0.01\\
$A_{\rm FB}^{0,\ell}$ &   0.08 &   0.00 &   0.01 &   0.01 & 1.00\\
\hline
\end{tabular}
}
\end{center}
\end{minipage}
\hfill
\begin{minipage}[t]{3.20in}
\begin{center}
\caption{\protect\footnotesize\sl
The error correlation matrix for the $b$ and $c$ 
quark results\protect\cite{lepewwg96}.
}
\label{tableofcorrbc}
{\footnotesize
\begin{tabular}{|c|r@{\,\,}r@{\,\,}r@{\,\,}r@{\,\,}r@{\,\,}r@{\,\,}|}
\hline
&
\multicolumn{1}{c}{$R_b$} &
\multicolumn{1}{c}{$R_c$} &
\multicolumn{1}{c}{$A_{\rm FB}^{0,b}$} &
\multicolumn{1}{c}{$A_{\rm FB}^{0,c}$} &
\multicolumn{1}{c}{$A_b$} &
\multicolumn{1}{c|}{$A_c$}
\\ \hline
$R_b$             &   1.00 &$-$0.23 &   0.00 &   0.00 &$-$0.03 &   0.01\\
$R_c$             &$-$0.23 &   1.00 &   0.04 &$-$0.06 &   0.05 &$-$0.07\\
$A_{\rm FB}^{0,b}$&   0.00 &   0.04 &   1.00 &   0.10 &   0.04 &   0.02\\
$A_{\rm FB}^{0,c}$&   0.00 &$-$0.06 &   0.10 &   1.00 &   0.01 &   0.10\\
$A_b$             &$-$0.03 &   0.05 &   0.04 &   0.01 &   1.00 &   0.12\\
$A_c$             &   0.01 &$-$0.07 &   0.02 &   0.10 &   0.12 &   1.00\\
\hline
\end{tabular}
}
\end{center}
\end{minipage}
\end{table}
 
\begin{figure}[t]
\begin{center}
  \leavevmode\psfig{file=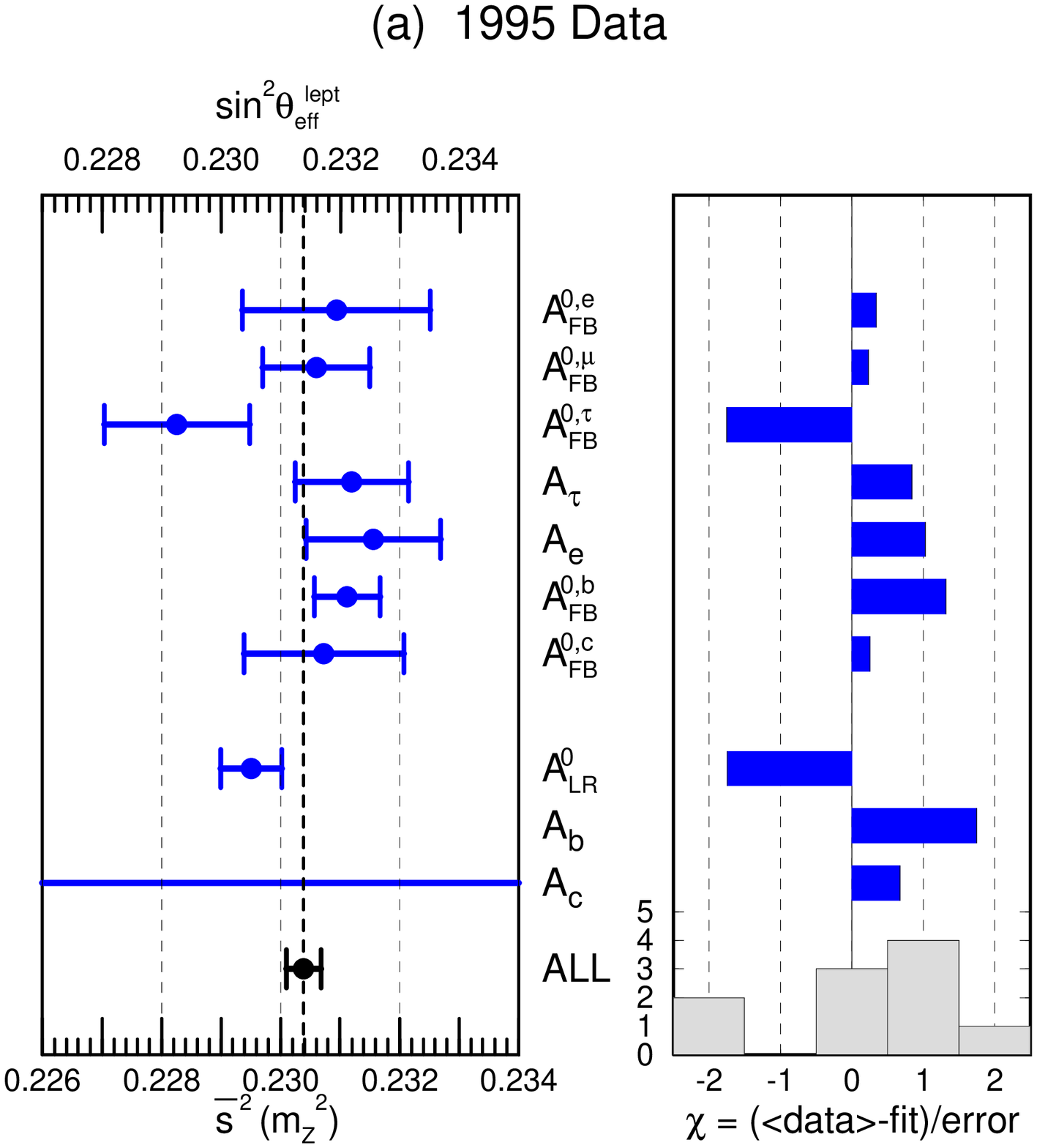,height=8cm,silent=0}
  \hfill
  \leavevmode\psfig{file=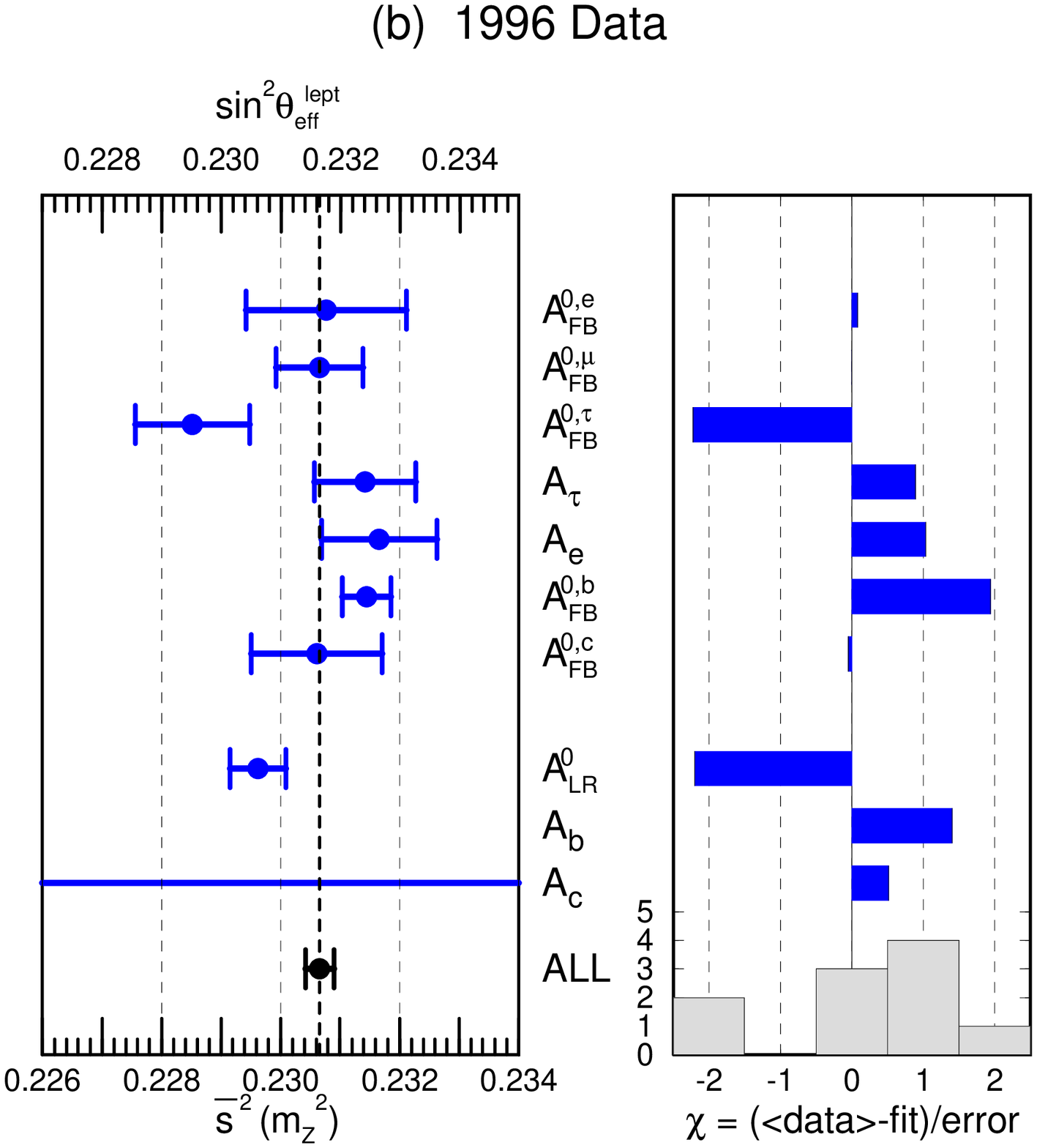,height=8cm,silent=0}
\vspace{3mm}
\caption{\protect\footnotesize\sl
The effective electroweak mixing parameter ${\sbar}^2(\mmz)$ is determined 
from the asymmetry data from LEP and SLC. The data up to 1995 and up to
1996 are displayed separately. The effective parameter 
$\sin^2\theta_{\rm eff}^{\rm lept}$ of the LEP Electroweak Working Group%
\protect\cite{lepewwg95,lepewwg96} is related to 
${\sbar}^2(\mmz)$\protect\cite{hhkm}  
by $\sin^2\theta_{\rm eff}^{\rm lept} 
\approx {\sbar}^2(\mmz)+0.0010$. 
The $A_b$-measurement is off scale.
}
\label{fig:sb2}
\end{center}
\end{figure}
 
All the asymmetry data, including the left-right beam-polarization asymmetry, 
$A_{\rm LR}$, from SLC are compared in Fig.~\ref{fig:sb2}.  
It shows the result of a one-parameter fit to the asymmetry data 
in terms of the effective electroweak mixing angle, 
$\sbar^2(\mmz )$\cite{hhkm}. In the SM 
(for details see sect.4) its numerical value is related to the effective 
parameter $\sin^2\theta_{\rm eff}^{\rm lept}$ adopted by the LEP group
\cite{lepewwg95,lepewwg96} as follows : 
$\sbar^2(\mmz )\approx\sin^2\theta_{\rm eff}^{\rm lept}-0.0010$\cite{hhkm}. 
The lepton forward-backward asymmetry is shown 
separately for each species. The fit to all 10 measurements yields :
\fitofsb
with $\chi^2_{\rm min}/({\rm d.o.f.})=17.3/(9)$. 
The updated measurements of the asymmetries barely agree (4\%CL) 
with the hypothesis of being determined by a universal electroweak 
mixing parameter. The new fit is slightly worse than the corresponding one
to the 1995 data\cite{lepewwg95} which gave\cite{khlp95}
$\sbar^2(\mmz )=0.23039\pm0.00029$ with 
$\chi^2_{\rm min}/({\rm d.o.f.})=13.0/(9)$ or 16\%CL.  
 
In the analysis presented below we use the data of Table~1 and combine,
assuming lepton ($e$--$\mu$--$\tau$) universality, the three forward-backward 
lepton asymmetries into the average forward-backward lepton asymmetry 
$A_{\rm FB}^{\ell,0}$ on the $Z$-pole. Using the data of Table~1
with $A_{\rm FB}^{\ell,0}$ instead of the three separate asymmetry 
measurements (see Fig.~\ref{fig:sb2}) one obtains~:
%
\fitofsbwithleptonuniversality
%
with $\chi^2_{\rm min}/({\rm d.o.f})=14.1/(7)$. 
Both the value and the probability of 
the fit (5\%CL) remain nearly unchanged compared to (\ref{fitofsb96}). 
The somewhat low probability of the fits 
reflects the fact that two of the most accurate 
measurements, $A_{\rm LR}^0$ and $A_{\rm FB}^{b,0}$, are about two
standard deviations from the 
mean to opposite sides as seen in Fig.~\ref{fig:sb2}. 
For instance, ignoring all hadron jet asymmetries and performing the fit
with the lepton asymmetry data alone 
($A_{\rm FB}^{\ell,0},\,A_\tau,\,A_e\,A_{\rm LR}^0$) one obtains
%
\fitofsbfromleptonicasymmetries
%
with $\chi^2/({\rm d.o.f})=6.0/(3)$. 
The fitted mean value decreases by about two standard deviations 
and the probability of the fit improves to 11\%CL.

The quantity $K$ in Table~1 
is a new measurement\cite{ccfr95_kevin} obtained by the 
CCFR Collaboration from their neutrino data.
 
The value of the $W$-mass has been slightly improved\cite{mw96}. 

\vspace*{5mm}

\section{ Theoretical framework --- 
             Brief Review of Electroweak Radiative Corrections 
             in $\rm{SU(2)_L \times U(1)_Y}$ Models }
The formalism introduced in Ref.\cite{hhkm} is used to interpret the
electroweak data. 
We use only those electroweak data that are most model independent, such 
as those listed in Table~1 of this report and those in Table~6 of
Ref.\cite{hhkm}. We then express them in terms of the $S$-matrix
elements of the processes with external quarks and leptons 
(with or without
external QED and QCD corrections, depending on how the electroweak data
are evaluated by experiments). These $S$-matrix elements 
are then evaluated in a
generic $\rm SU(2)\times U(1)$ model with four charge form factors, 
$\ebar^2(q^2)$, $\sbar^2(q^2)$, $\gzbar^2(q^2)$ and $\gwbar^2(q^2)$. 
An additional parameter $\delb(\mmz)$ 
related to the $\zbb$ vertex form factor is also introduced. 
By assuming negligible new physics
contribution to the remaining vertex and box corrections, we 
derive constraints on the $4+1$ form factors from the model-independent
data. By further assuming negligible new physics contribution to the
running of the charge form factors, we derive constraints on $S$, 
$T$, $U$ and $\delb(\mmz)$. Finally, by assuming no new physics
contribution at all, we can constrain $m_t$ and $\mh$.
In this section a brief review of the salient features
are given.
 
The propagator corrections in generic $\rm{SU(2)_L \times U(1)_Y}$ 
models can be conveniently expressed in terms of the following four effective 
charge form factors\cite{hhkm}:

\def\propagator#1#2{%
    \begin{picture}(80,35)(0,18)
       \Text(20,27)[cb]{#1}
       \Text(50,27)[cb]{#2}
       \Line(0,35)(10,20)
       \Line(0, 5)(10,20)
       \Line(70,35)(60,20)
       \Line(70, 5)(60,20)
       \Photon(10,20)(30,20){3}{3}
       \Photon(40,20)(60,20){3}{3}
       \GCirc(35,20){7}{0.5}
    \end{picture}
}
\bsub \label{barcharges}
  \begin{eqnarray}
    \propagator{$\gamma$}{$\gamma$} &\sim& \ebar^2(q^2) =\ehat^2
    \Bigl[\,1-{\rm Re}\pibar_{T,\gamma}^{\gamma\gamma}(q^2)\,\Bigr]\,,
\\[-2mm]
    \propagator{$\gamma$}{$Z$}      &\sim& \sbar^2(q^2)      =\shat^2
    \Bigl[\,1+\frac{\chat}{\shat}\,
    {\rm Re}\,\pibar_{T,\gamma}^{\gamma Z}(q^2)
    \,\Bigr]\,,
\\[-2mm]
    \propagator{$Z$}{$Z$}           &\sim& \gzbar^2(q^2)     =\gzhat^2
    \Bigl[\,1 -{\rm Re}\pibar_{T,Z}^{ZZ}(q^2)\,\Bigr]\,,
\\[-2mm]
    \propagator{$W$}{$W$}           &\sim& \gwbar^2(q^2)     =\ghat^2
    \Bigl[\,1 -{\rm Re}\pibar_{T,W}^{WW}(q^2)\,\Bigr]\,,
  \end{eqnarray}
\esub
where 
$
     \pibar_{T,V}^{AB}(q^2)
      \equiv [\pibar_T^{AB}(q^2)\!-\!\pibar_T^{AB}(\mmv)]/(q^2\!-\!\mmv)
$
are the propagator correction factors that appear in the $S$-matrix elements 
after the weak boson mass renormalization is performed, and $\ehat \equiv 
\ghat\shat \equiv \gzhat^{}\shat\chat$ are the $\msbar$ couplings.
The `overlines' denote the inclusion of the pinch terms\cite{pinch,pinch2}, 
which make these effective charges useful\cite{kl89,hhkm,hms95,prw96} 
even at very 
high energies ($|q^2|\gg \mmz$). The amplitudes are then expressed in terms 
of these charge form factors plus appropriate vertex and box corrections. 
In our analysis\cite{hhkm} we assumed that all the vertex and box corrections 
are dominated by the SM contributions, except for the $\zbb$ vertex,
   \bea \label{zblbl}
   \Gamma_L^{Zbb}(q^2) = -\gzhat \Bigl\{ -\frac{1}{2}[1+\delb(q^2)]
                   +\frac{1}{3}\shat^2[1+\Gamma_1^{b_L}(q^2)] \Bigr\} ,
   \eea
for which the function  $\delb(\mmz )$ is allowed to take on an arbitrary 
value. Hence the charge form factors and $\delb$ can be directly extracted 
from the experimental data and their values be compared with the theoretical 
predictions. 
 
We {\it define}\cite{hhkm} the $S$, $T$, and $U$ variables of Ref.\cite{stu} 
in terms the effective charges,
  \begin{subequations}\label{stu_def}
  \begin{eqnarray}
    \frac{\sbar^2(\mmz)\cbar^2(\mmz)}{\bar{\alpha}(\mmz)} 
    -\frac{4\,\pi}{\gzbar^2(0)} &\equiv& \;\frac{S}{4} \,,
    \\
    \frac{\sbar^2(\mmz)}{\bar{\alpha}(\mmz)}\quad
    -\frac{4\,\pi}{\gwbar^2(0)} &\equiv& \frac{S+U}{4} \,,
    \\
    1\;-\,\frac{\gwbar^2(0)}{\mmw}\frac{\mmz}{\gzbar^2(0)}
    &\equiv& \;\alpha T \,,
  \end{eqnarray}
  \end{subequations}
where it is made manifest that these variables measure deviations from the 
tree-level universality of the electroweak gauge boson couplings. 
Here $\cbar^2=1-\sbar^2$ and $\bar{\alpha}(q^2)=\ebar^2(q^2)/4\pi$.  
They 
receive contributions from both the SM radiative effects and new physics 
contributions. The $S$, $T$, $U$ variables\cite{stu} as introduced by 
Peskin and 
Takeuchi are obtained\cite{hhkm} approximately by subtracting the SM 
contributions (at $\mh=1000$~GeV).
 
For a given electroweak model we can calculate the $S$, $T$, $U$ parameters 
($T$ is a free parameter in models without the custodial SU(2) symmetry), and 
the charge form factors are then fixed by the following identities\cite{hhkm}: 
  \begin{subequations}
    \label{gbarfromstu}
  \begin{eqnarray}
     \frac{1}{\gzbar^2(0)}
        &=& \frac{1+\delg -\alpha \,T}{4\,\sqrt{2}\,G_F\,\mmz} \,,
    \label{gzbarfromt}\\[2mm]
      \sbar^2(\mmz)
          &=& \frac{1}{2}
              -\sqrt{\frac{1}{4} -\bar{\alpha}(\mmz)
                    \biggl(\frac{4\,\pi}{\gzbar^2(0)} +\frac{S}{4}
                    \biggr)  }\,,
    \label{sbarfroms}\\[2mm]
       \frac{4\,\pi}{\gwbar^2(0)}
             &=& \frac{\sbar^2(\mmz)}{\bar{\alpha}^2(\mmz)}
                -\frac{1}{4}\,(S+U) \,.
    \label{gwbarfromu}
  \end{eqnarray}
  \end{subequations}
Here $\delg$ is the vertex and box correction to the muon 
lifetime\cite{del_gf} after subtracting the pinch term\cite{hhkm}: 
 \begin{eqnarray}
   G_F &=& \frac{\gwbar^2(0)+\ghat^2\delg}{4\,\sqrt{2}\mmw} \,.
   \label{gf}
 \end{eqnarray} 
In the SM, $\delg=0.0055$\cite{hhkm}. 
 
It is clear from the above identities that once we know $T$ and $\delg$ in a 
given model we can predict $\gzbar^2(0)$, and then with the knowledge of $S$ 
and $\bar{\alpha}(\mmz)$ we can calculate $\sbar^2(\mmz)$, and with the 
further knowledge of $U$ we can calculate $\gwbar^2(0)$. Since 
$\bar{\alpha}(0)=\alpha$ is known precisely, all four charge form factors are 
fixed at one $q^2$ point. The $q^2$-dependence of the form factors can 
also be calculated in a given model, but it is less dependent on physics at 
very high energies\cite{hhkm}. In the following analysis we assume that the 
SM contribution governs the running of the charge form factors between
$q^2=0$ and $q^2=\mmz$. We can then predict all the neutral-current amplitudes
in terms of $S$ and $T$, and the additional knowledge of $U$ gives the $W$ 
mass via Eq.~(\ref{gf}).
 
We should note here that our prediction for the effective mixing parameter 
$\sbar^2(\mmz)$ is not only sensitive to the $S$ and $T$ parameters but also 
on the precise value of $\bar{\alpha}(\mmz)$. This is the reason why our 
predictions for the asymmetries measured at LEP/SLC and, consequently, the 
experimental constraint on $S$ extracted from the asymmetry data are 
sensitive to  $\bar{\alpha}(\mmz)$. In order to keep track of the uncertainty 
associated with $\bar{\alpha}(\mmz)$ the parameter $\delta_\alpha$ was 
introduced in Ref.\cite{hhkm} as follows:
$1/\bar{\alpha}(\mmz)\!\equiv\! 4\pi/\ebar(\mmz)\!
=\!128.72\!+\!\delta_\alpha$.
We show in Table~4 the results of the four most 
recent updates\cite{mz94,swartz95,eidjeg95,bp95} 
on the hadronic contribution to the running of the effective 
QED coupling. Three definitions of the running QED coupling are compared.
The effective charge $\bar{\alpha}(\mmz)$ should be used in
Eqs.(\ref{stu_def}) and (\ref{gbarfromstu}), since the effective charges 
in (\ref{barcharges}) contain both fermionic and bosonic contributions
to the gauge boson propagator corrections. 
 
\begin{table}[t]
\caption{\protect\footnotesize\sl
The running QED coupling at the $\mz$ scale in the three schemes.
$1/\alpha(\mmz)_{\rm l.f.}$ contains only the light fermion
contributions to the running of the QED coupling constant
between $q^2=0$ and $q^2=\mmz$.
$1/\alpha(\mmz)_{\rm f}$ contains all fermion contributions
including the top-quark.  
The values 
$m_t=175$~GeV and $\alpha_s(\mz)=0.12$ in the perturbative
two-loop correction\cite{kniehl90} are assumed.
$1/\bar{\alpha}(\mmz)$ contains also the $W$-boson-loop
contribution\cite{hhkm} including the pinch term\cite{pinch,pinch2}.
}
\label{tab:alpha_mz}
\begin{center}
{\footnotesize
\begin{tabular}{|c|c|c|c|c|}
\hline
& $1/\alpha(\mmz)_{\rm l.f.}$
& $1/\alpha(\mmz)_{\rm f}$
& $1/\bar{\alpha}(\mmz)$
& $\delta_\alpha$
\\
\hline
Martin-Zeppenfeld '94\cite{mz94}
& $128.98\pm 0.06$ & $128.99\pm 0.06$ & $128.84\pm 0.06$ & $0.12\pm 0.06$
\\
Swartz '95\cite{swartz95}
& $128.96\pm 0.06$ & $128.97\pm 0.06$ & $128.82\pm 0.06$ & $0.10\pm 0.06$
\\
Eidelman-Jegerlehner '95\cite{eidjeg95}
& $128.89\pm 0.09$ & $128.90\pm 0.09$ & $128.75\pm 0.09$ & $0.03\pm 0.09$
\\
Burkhardt-Pietrzyk '95\cite{bp95}
& $128.89\pm 0.10$ & $128.90\pm 0.10$ & $128.76\pm 0.10$ & $0.04\pm 0.10$
\\
\hline
\end{tabular}
}
\end{center}
\end{table}
 
\begin{figure}[b]
 \begin{center}
 \leavevmode\psfig{file=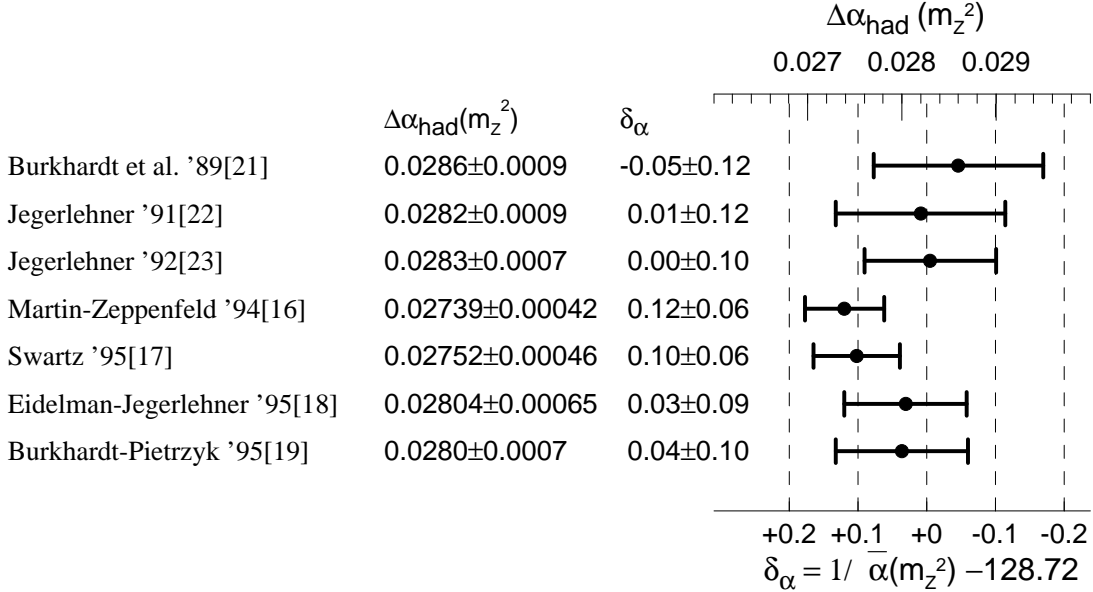,height=8cm,silent=0}
 \end{center}
\caption{\protect\footnotesize\sl
Various estimates of $\Delta\alpha_{\rm had}(\mmz)$
and the resulting $\bar{\alpha}(\mmz)$ in the 
minimal SM.
The parameter $\delta_\alpha$\cite{hhkm} is defined as
$\delta_\alpha \equiv 1/\bar{\alpha}(\mmz)-128.72$.
}
\label{fig:delta_had}
\end{figure}

The new and some earlier estimates\cite{burkhardt89,jeg91,jeg92} 
are also shown in Fig.~\ref{fig:delta_had}.
The analysis of Ref.\cite{hhkm} was based on the estimate \cite{jeg92}, 
$\delta_\alpha=0.00\pm 0.10$. The last four estimates made use of essentially 
the same total cross section data set for the process 
$\epem \to {\rm hadrons}$ 
between the two-pion threshold and the $Z$ mass scale. The estimates are 
slightly different reflecting different procedures adopted by each group 
to interpolate between the available data points. Eidelman and Jegerlehner
\cite{eidjeg95} and Burkhardt and Pietrzyk\cite{bp95} made no assumption on 
the shape ($s$-dependence) of the cross section, and hence their errors 
are conservative. Swartz\cite{swartz95} assumed smoothness of $s$-dependence 
of the cross section in order to profit from the smaller point-to-point 
errors within each experiment. Martin and Zeppenfeld\cite{mz94} also made use 
of the smaller experimental point-to-point errors by constraining the overall 
normalization on the basis of the perturbative QCD prediction with 
$\alpha_s(\mz )=0.118\pm 0.007$ down to $\sqrt{s}=3$~GeV. The smaller errors 
of these two estimates are obtained either because of the data point with the 
smallest normalization error\cite{swartz95} or because of replacing the large
normalization uncertainty by the small uncertainty of the perturbative QCD 
prediction\cite{mz94} in the region 3~GeV$<\sqrt{s}<7$~GeV. The mean values 
of the two estimates\cite{mz94,swartz95} 
are similar as a result of the fact that the measured 
cross section of the smallest normalization error in the above region agrees 
roughly with the perturbative QCD prediction. In the following analysis we 
adopt as a standard the conservative estimate of Ref.\cite{eidjeg95}, i.e. 
$\delta_\alpha=0.03\pm 0.09$ and investigate the sensitivity of our results 
to the deviation $\delta_\alpha-0.03$. We also show results of the analysis 
when the estimate\cite{mz94} $\delta_\alpha=0.12\pm 0.06$ is adopted instead.
 
Once $\bar{\alpha}(\mmz)$ is fixed the charge form factors 
in Eq.~(\ref{gbarfromstu}) can be calculated from $S$, $T$, $U$.
The following approximate formulae\cite{hhkm} are useful:
  \begin{subequations}
   \label{gbar_approx}
  \begin{eqnarray}
        \gzbar^2(0) 
         &\approx& 0.5456
        \hphantom{+0.0036\,S'}\;\, 
                  +0.0040\,T'\,,
   \label{gzbar_approx}\\
        \sbar^2(\mmz) 
         &\approx& 0.2334
                  +0.0036\,S'  
                  -0.0024\,T' \,\,,
         \qquad
   \label{sbar_approx}\\
        \gwbar^2(0) 
          &\approx& 0.4183
            -0.0030\,S'  
            +0.0044\,T'
            +0.0035\,U' \,\,, 
         \qquad
   \label{gwbar_approx}
  \end{eqnarray}
  \end{subequations}
where 
  \begin{subequations}
   \label{stu_prime}
  \begin{eqnarray}
      S' &=& S -0.72\,\delta_\alpha\,,\\
      T' &=& T +(0.0055-\delg)/\alpha\,,\\
      U' &=& U -0.22\,\delta_\alpha\,.
  \end{eqnarray}
  \end{subequations}
The values of $\gzbar^2(\mmz)$ and $\sbar^2(0)$ are then 
calculated from $\gzbar^2(0)$ and $\sbar^2(\mmz)$ above, 
respectively, by assuming the SM running of the form factors.
The $Z$ widths are sensitive to $\gzbar^2(\mmz)$, which can
be obtained from $\gzbar^2(0)$ in the SM approximately by
\begin{eqnarray}
\label{gzbarrunning}
\frac{4\pi}{\gzbar^2(\mmz )} \approx \frac{4\pi}{\gzbar^2(0)} 
-0.299 +0.031\log\biggl[1+\Bigl(\frac{26{\rm GeV}}{\mh}\Bigr)^2\biggr]. 
\end{eqnarray}
The approximation is valid to 0.001 provided $m_t > 160 \gev$ and 
$\mh > 40 \gev$.
On the other hand the low energy neutral current experiments 
are sensitive to $\sbar^2(0)$ which is obtained by assuming 
the SM running of the charge form factor 
$\sbar^2(q^2)/\bar{\alpha}(q^2)$: 
\begin{eqnarray}
\label{gbarrunning}
\frac{\sbar^2(0)}{\alpha} \approx 
\frac{\sbar^2(\mmz )}{\bar{\alpha}(\mmz )} 
+3.09 -\frac{\delta_\alpha}{2}.  
\end{eqnarray}

Finally, within the SM the $S$, $T$, $U$ parameters and 
the form factor $\delb=\bar{\delta}_b(\mmz )$ are functions of 
$\mt$ and $\mh$ which can be parametrized as
\begin{subequations}
\label{studb_approx}
\begin{eqnarray}
S_{\rm SM} &\approx& -0.233 -0.007\xt +0.091\xh -0.010x_H^2 \,,
\label{s_approx} \\
T_{\rm SM} &\approx& +0.879 +(0.130-0.003\xh)\xt +0.003\xt^2 
                     -0.079\xh -0.028x_H^2 +0.0026x_H^3 \,,
\nonumber \\   && 
\label{t_approx} \\
U_{\rm SM} &\approx& +0.362 +0.022\xt -0.002\xh \,,
\label{u_approx} \\
\delb_{\rm SM} &\approx&  -0.00995 -0.00087\xt -0.00002\xt^2  \,,
\label{db_approx}
\end{eqnarray}
\end{subequations}
where
$\xt=(\mt(\gev)-175)/10$ and $\xh=\log(\mh(\gev)/100)$.
The above approximate expressions are valid to $\pm 0.003$ 
for $S_{\rm SM}$, $T_{\rm SM}$ and $U_{\rm SM}$, and 
to $\pm 0.00007$ for $\delb_{\rm SM}$ 
in the domain $160~\gev <\mt<185~\gev$ and $40~\gev <\mh<1000~\gev$. 
They are evaluated after all the two-loop corrections 
included in Ref.\cite{hhkm} are taken into account, 
for $\alpha_s(\mz)=0.118$ in the two-loop ${\cal O}(\alpha_s)$ 
terms\cite{kniehl90}. 
The $\mh$-dependence of the $\delb(\mmz)_{\rm SM}$ function 
is found to be negligibly small for the above region of $m_t$. 

Note : Since the publication of our original paper\cite{hhkm} several
improvements have been achieved on the SM radiative corrections. Most notably, 
we now have the three-loop (order $\alpha_s^2$) QCD calculation of the $T$ 
parameter\cite{3loopt} as well as in the gauge boson propagator corrections
\cite{3loopvv}. These three-loop effects slightly modify the 
relationship between the electroweak $S$, $T$, $U$ parameters and the 
physical top quark mass $\mt$ in the above formulae 
(\ref{studb_approx}). 
After the completion of the present report
we took note of the new evaluation of non-factorizable QCD and electroweak 
corrections to the hadronic $Z$ boson decay rates\cite{qcdeltw}. 
A negative correction to the $Z$ hadronic width was found reducing the 
SM prediction for $\Gamma_h$ by 0.59~MeV
after summing over the four light quark flavors. 
The corresponding effect for the partial width 
$\Gamma(Z\rightarrow `b\bar{b}\mbox{'})$ has not been evaluated. 
This shift would in turn enhance the 
$\alpha_s$ value extracted from the electroweak data by 0.001. 
We refrain from modifying the numbers in the present report. 
If we assume that the corrections to the partial width
$\Gamma(Z\rightarrow `b\bar{b}\mbox{'})$ is small, 
the net effect 
for the numbers due to the above new calculations would be as follows :
\begin{itemize}
\item{
The three-loop corrections to the $T$ parameter\cite{3loopt}
modifies the relationship (\ref{t_approx}) between $T$ and the physical 
top quark mass.  By comparing\cite{3loopt} with \cite{hhkm}, we find 
\begin{eqnarray}	
  && \Bigl[m_t^{\rm (2-loop)}\Bigr]^2 
       \Bigl\{ 1 -\frac{2(3+\pi^2)}{9}\frac{\alpha_s(\mz)}{\pi} \Bigr\} 
   \nonumber \\
   & &\quad = \Bigl[m_t^{\rm (3-loop)}\Bigr]^2 
        \biggl\{ 1 -\frac{2(3+\pi^2)}{9} \frac{\alpha_s(\mt)}{\pi} 
           -14.594 \Bigl(\frac{\alpha_s(\mt)}{\pi}\Bigr)^2 \biggr\} \,.
   \label{mt3loop}
\end{eqnarray}
This can be approximated as 
  \begin{eqnarray}
   m_t^{\rm (2-loop)} = m_t^{\rm (3-loop)} 
   \biggl\{ 1 -(7.3-5.5\log\frac{\mt}{\mz})
           \Bigl(\frac{\alpha_s(\mz)}{\pi}\Bigr)^2 \biggr\}\,.
   \label{mt3loopapprox}
  \end{eqnarray}
For $\mt \sim 170\gev$, this corresponds to the replacement 
of $m_t^{\rm (2-loop)}$ by $m_t^{\rm (3-loop)}-1\gev$.  
Non-leading three-loop corrections 
calculated in \cite{3loopvv} modifies $S_{\rm SM}$, $U_{\rm SM}$ 
in (\ref{studb_approx}) and the running of the 
$\gzbar^2(q^2)$ charge (\ref{gzbarrunning}).  
Their effects are, however, much smaller 
than the leading effect as quoted above.  
Consequently, the three-loop ${\cal O}(\alpha_s^2)$ effects 
can be approximately 
taken into account by replacing all the $\mt$ symbols in this 
report by the r.h.s. combination of Eq.~(\ref{mt3loopapprox}), 
or roughly by $\mt-1\gev$.  In other words, the fitted 
$\mt$ value should be about $1\gev$ larger, while the results 
with an external $\mt$ constraint should correspond to those 
where the external $\mt$ is increased by about $1\gev$.  
}
\item{
The mixed QCD electroweak two-loop corrections of Ref.\cite{qcdeltw}
can be accounted for by replacing all $\alpha_s$ symbols 
in this report by $\alpha_s-0.001$.  
In other words, the fitted 
$\alpha_s$ value should be about $0.001$ larger, while the results 
with an external $\alpha_s$ 
constraint 
should correspond to those 
where the external $\alpha_s$ is increased by about $0.001$.  
This is because 
the $\alpha_s$ dependences in the corrections other than the 
hadronic width of the $Z$ are all negligibly small. 
}
\end{itemize}


 \section{Implications of the New Measurements}  
 
In this section the new results and their implications are discussed. Also a 
fit in terms of the $S$, $T$, $U$ parameters\cite{stu} of the electroweak
gauge boson propagator corrections as well as of the $Zb_Lb_L$ vertex form 
factor, $\delb(\mmz )$ is presented. The strengths of the QCD and QED 
couplings at the $\mz$ scale, $\alpha_s(\mz )$ and $\bar{\alpha}(\mmz )$, are
treated as external parameters in the fits, so that implications of their 
precise knowledge affecting the fit results are made explicit.
 
   \subsection{New LEP/SLC data} 
 
The updated $Z$ shape parameter measurements (see Tables~1--3) are used to 
extract the charge form factors. It is assumed that the vertex corrections 
except for the $Zb_Lb_L$ vertex function $\delb(\mmz )$ are dominated by the 
SM contributions.\footnote{%
We exclude from the fit the jet-charge asymmetry data in 
Table~1, since it allows an interpretation only within the
minimal SM.  It is included in our SM fit in section 5.
}
The free parameters are : $\gzbar^2(\mmz)$, $\sbar^2(\mmz)$, $\alpha_s'$ and 
$\delb(\mmz)$. The quantity $\alpha_s'$ is 
the combination%
\footnote{As will be explained in detail in the subsection~4.2, 
we modify the definition of $\alpha_s'$ in Refs.\cite{hhkm,sm95} by 
subtracting the SM contribution to $\delb(\mmz)$ at 
$m_t=175\gev$, $\delb(\mmz)=-0.00995$. See (\ref{db_approx}).
}
\begin{eqnarray}
\label{alpsprime}
\alpha_s' = \alpha_s(\mz )_{\msbar}\, +1.54\,[\delb(\mmz ) +0.00995]
\end{eqnarray}
that appears\cite{hhkm,sm95} in the theoretical 
prediction for $\Gamma_h$. The fit yields:
\fitofgzbsbalpspdb
The value of $\chi^2_{\rm min}$ is dominated by the 
contribution of the asymmetries 
which accounts for $14.1$ 
(cf. Eq.~(\ref{fitofsb96withleptonuniversality})). 
When allowing only $\sbar^2(\mmz)$ and $\gzbar^2(\mmz)$ 
to be fitted freely, and treating $\alpha_s'$ and $\delb(\mmz)$ 
as external parameters, we obtain an equivalent result:
\fitofgzbsb
Compared to the previous results in Ref.\cite{hhkm} the precision has 
increased by more than a factor of two. 
 
The fit can be qualitatively understood as follows. The asymmetries determine 
almost exclusively $\sbar^2(\mmz)$. The tiny difference between the above 
$\sbar^2(\mmz)$ and Eq.~(\ref{fitofsb96withleptonuniversality}) 
is due to the $\alpha_s'$-dependence
of $R_\ell$. The only quantity constraining $\gzbar^2(\mmz)$ is $\Gamma_Z$ 
which also depends on $\sbar^2(\mmz)$ and $\alpha_s'$, thus explaining the 
non-negligible error correlations above. The quantity $\alpha_s'$ is mainly 
determined by $R_\ell$ and also by $\sigma_h^0$. The observable $R_b$, i.e. 
the ratio of $\Gamma_b$ and $\Gamma_h$, is constraining $\delb(\mmz )$. It is 
interesting to note that the form factor $\delb(\mmz )$ is nearly 
uncorrelated from the other fit quantities because of our using the 
combination (\ref{alpsprime}). 
It is now straightforward to obtain the best value of $\alpha_s$ from 
$\alpha_s'$ and $\delb$ : 
\begin{eqnarray}
\label{alpswithdb}
\alpha_s = \alpha_s' -1.54\,[\delb+0.00995\,] = 0.1143 \pm 0.0057 \,. 
\end{eqnarray}
If on the other hand 
$R_b$ and $R_c$ are fixed to their SM predictions with $m_t=175$~GeV, i.e. 
$\delb= -0.00995$, one obtains $\alpha_s=0.1218\pm 0.0038$. This little
exercise demonstrates the crucial role of the $R_b$ and $R_c$ measurements
in obtaining information on $\alpha_s$ from the precision $Z$ experiments.
 
 \begin{figure}[b]
  \begin{center}
  \leavevmode\psfig{file=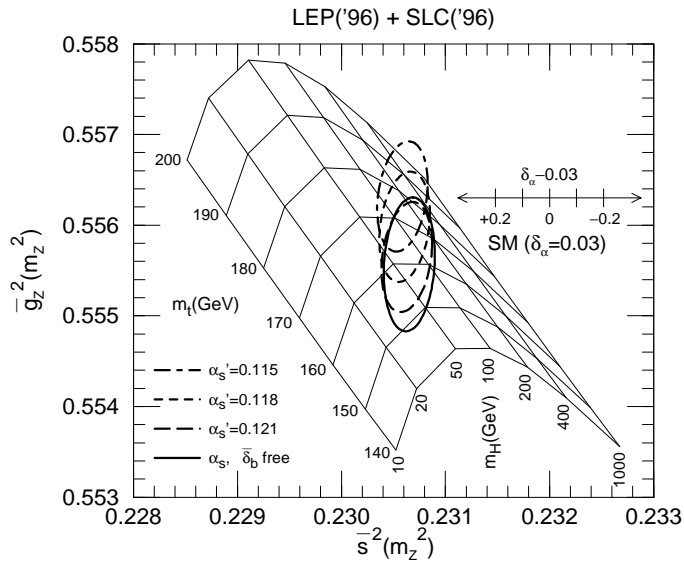,width=9cm,silent=0}
  \end{center}
 \caption{\protect\footnotesize\sl
 The 1-sigma allowed contours in 
($\protect\sbar^2(\mmz), \protect\gzbar^2(\protect\mmz)$) plane
 obtained from the fits to the $Z$ boson parameters. 
The solid contour is obtained by treating $\alpha_s'$ and $\delb$ as
free parameters in the fit.
 Also shown are the results by treating 
 $\alpha_s'$ as an external parameter. 
 Three values of $\alpha_s'$ ($=0.115,0.118,0.121$), 
 are chosen.
 The results are insensitive to the assumed $\delb$ value. 
 The grid illustrates 
 the SM predictions in the range 
 $140\protect\gev\!<\!m_t\!<\!200\protect\gev$ and 
 $10\protect\gev\!<\!\protect\protect\mh\!<\!1000\protect\gev$ at  
 $\delta_\alpha \!\equiv\! 1/\bar{\alpha}
 (\protect\mmz)\!-\!128.72 \!=\!0.03$, 
 where their dependences on $\delta_\alpha\!-\!0.03$
 are shown by a ``$\longleftrightarrow$'' symbol.
 }
 \label{fig:gzb2sb2}
 \end{figure}
 
Figure~\ref{fig:gzb2sb2} shows the fit result in the ($\sbar^2(\mmz),
\gzbar^2(\mmz)$) plane. 
The contours represent the 1-$\sigma$ (39\%CL) allowed region. 
The solid contour shows the result of the four-parameter fit 
(\ref{fitofgzbsbalpspdb}). 
Also shown are the results of the two-parameter fit in terms of 
$\gzbar^2(\mmz)$ and $\sbar^2(\mmz)$ treating $\alpha_s'$ as an external 
parameter. 
Three values of $\alpha_s'$ (0.115,0.118,0.121) are chosen in the 
figure, which correspond respectively to the $\alpha_s$ values in the SM 
at $\mt=175\gev$; see (\ref{alphas_prime_d}). 
The results are insensitive to the assumed $\delb$ value once the 
magnitude of the combination $\alpha_s'$ is fixed. 
The SM predictions for $\delta_\alpha=0.03$ and their dependence on 
$\delta_\alpha-0.03$ are also given. 
As expected from Eqs.~(\ref{gbar_approx}), 
(\ref{stu_prime}) and 
(\ref{studb_approx}), 
only the predictions for $\sbar^2(\mmz )$ is sensitive to
$\delta_\alpha$. 
 
\subsection{The heavy quark sector and $\alpha_s$}
 
\begin{figure}[b]
  \leavevmode\psfig{file=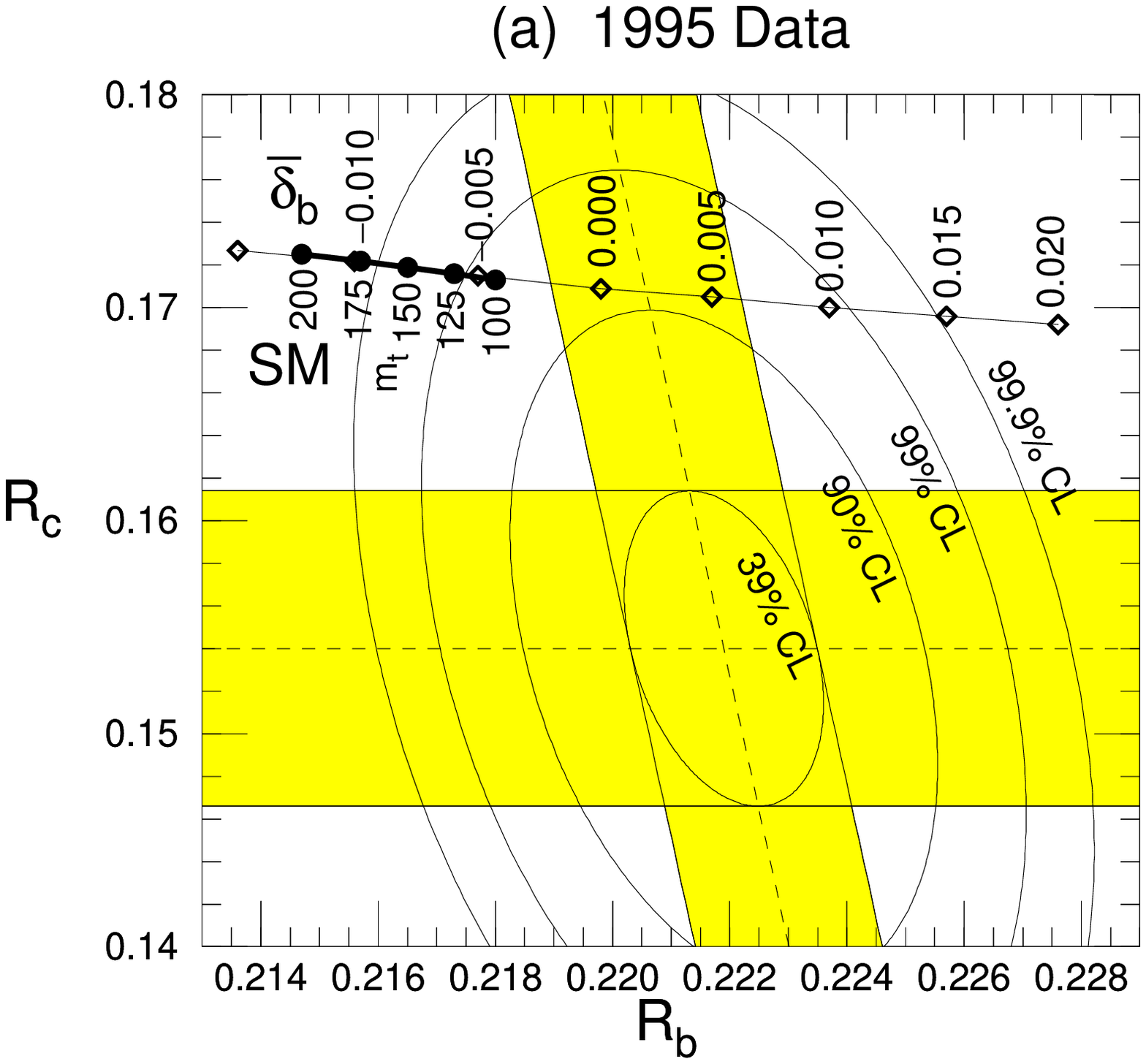,height=6.5cm,silent=0}
  \hfill
  \leavevmode\psfig{file=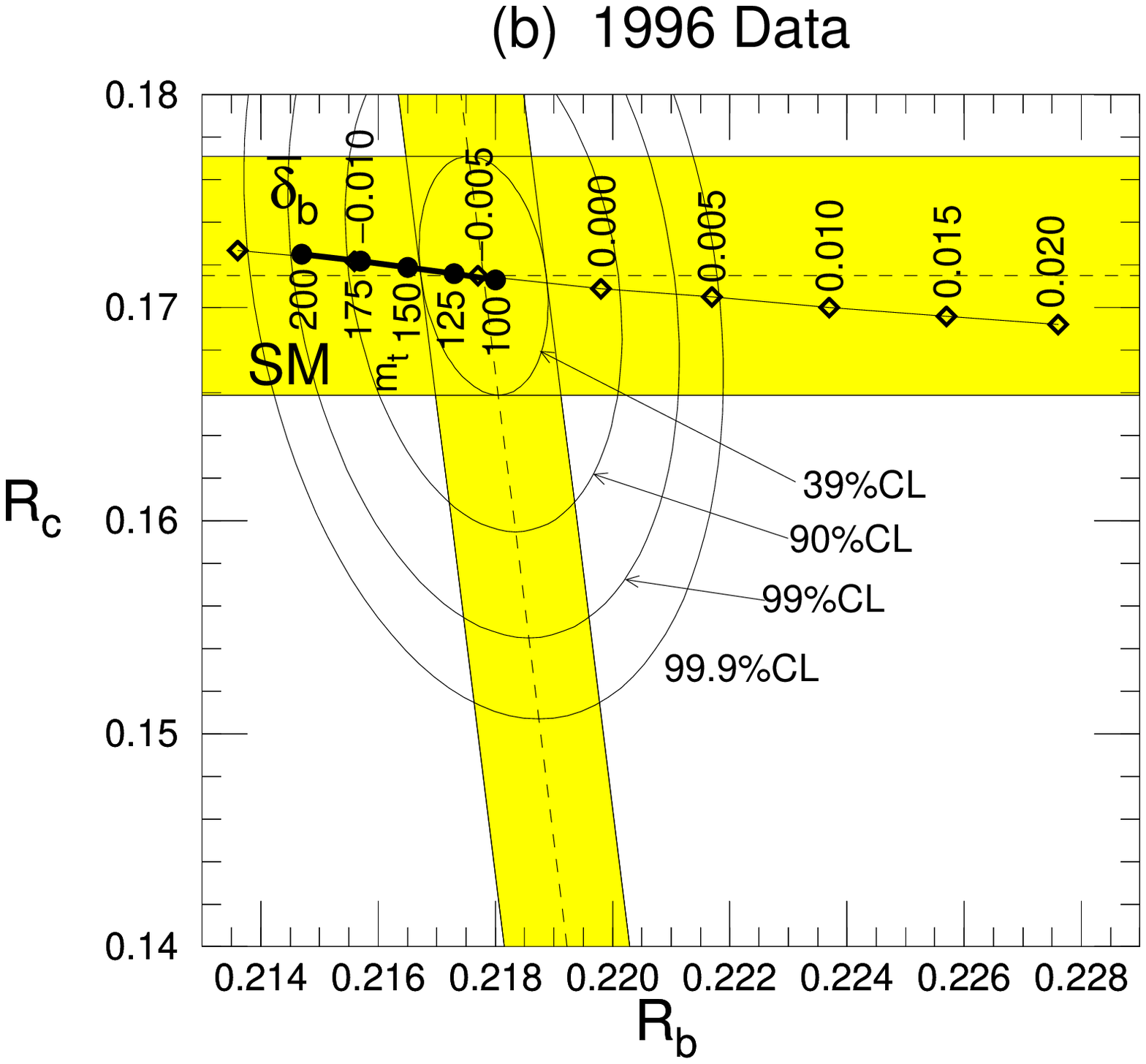,height=6.5cm,silent=0}
   \vspace{2mm}
        \caption{\protect\footnotesize\sl
        $R_b$ and $R_c$ data\cite{lepewwg95,lepewwg96}
        and the SM predictions\cite{hhkm}.
        }
        \label{fig:rbrc}
\end{figure}

The most striking results of the updated electroweak data are those of $R_b$ 
and $R_c$, which are shown in Fig.~\ref{fig:rbrc} juxtaposing the status
as of summer 1995 and 1996. The SM predictions to these ratios are shown by 
the thick solid line, where the value of the top-quark mass affecting the 
$Zb_Lb_L$ vertex correction is indicated by solid blobs. Although it was
tempting to conclude from the 1995 data on $R_b$ and $R_c$ that the SM is
excluded at 99.99\%CL, it was also clear\cite{khlp95,renton} 
that it would be  precocious to base 
such a far reaching conclusion on just these two 
measurements knowing how complex the analyses are
and how critical the role of systematic effects is.
 
\begin{figure}[b]
  \leavevmode\psfig{file=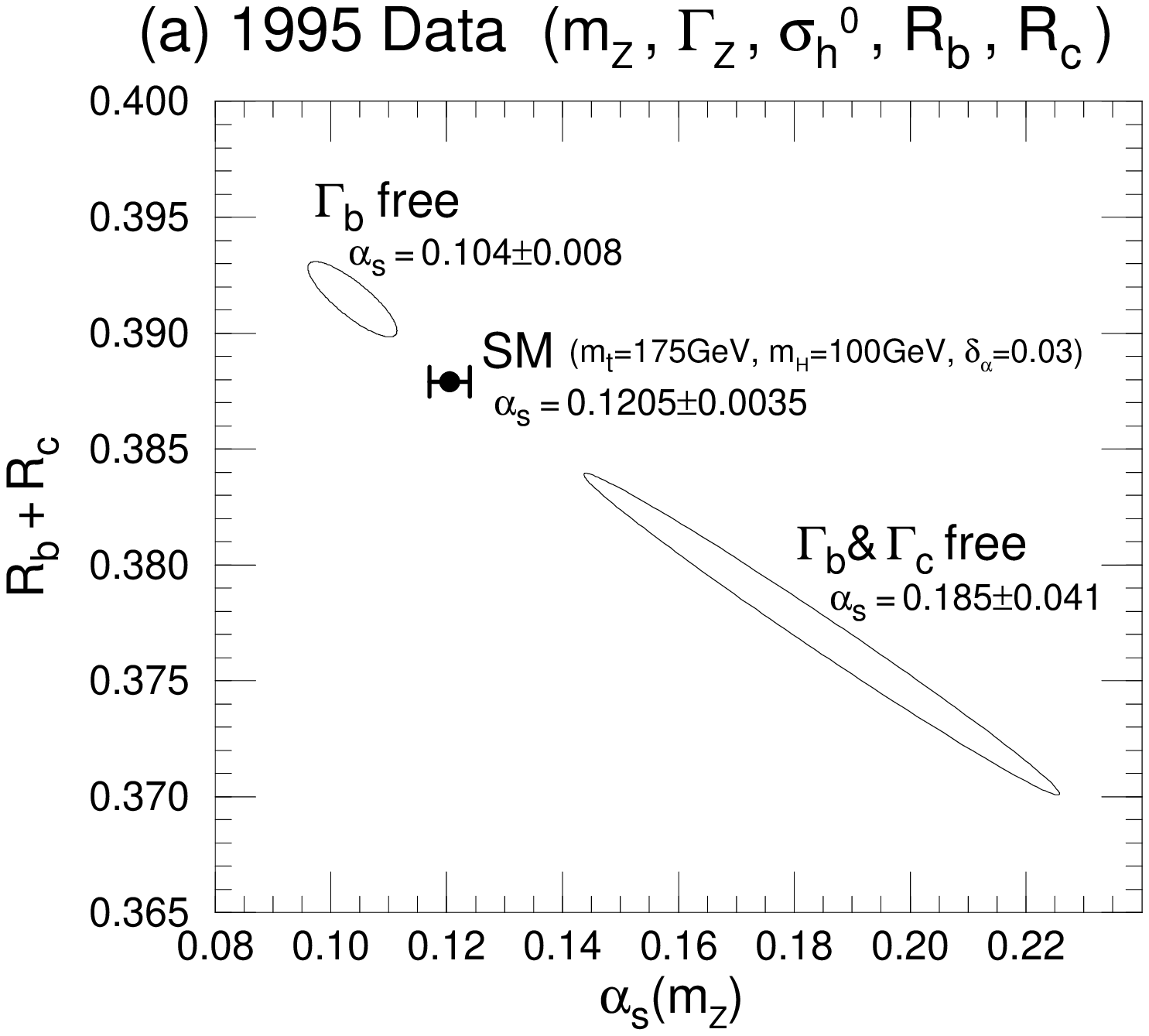,height=6.5cm,silent=0}
  \hfill
  \leavevmode\psfig{file=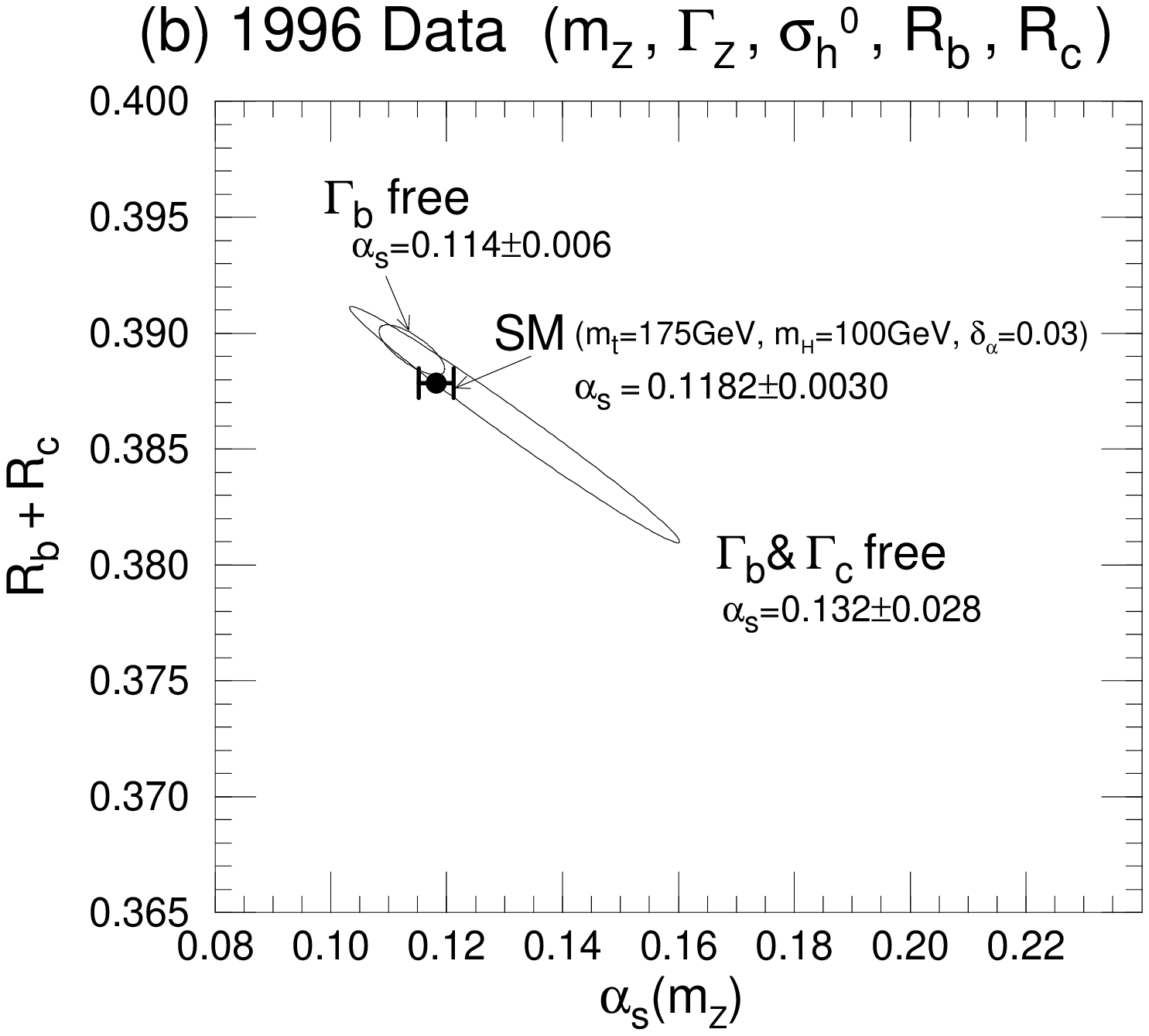,height=6.5cm,silent=0}
   \vspace{2mm}
        \caption{\protect\footnotesize\sl
        $R_b+R_c$ vs $\alpha_s$.
        {}From the 1995 data\cite{lepewwg95}(a) 
           and the 1996 date\cite{lepewwg96}(b).
        }
        \label{fig:rbrc_vs_alphas}
\end{figure}

It is useful to note the fact that the three most accurately measured 
line-shape parameters, $\Gamma_Z$, $\sigma_h^0$ and $R_\ell$ in Table~1, 
determine accurately the $Z$ partial widths $\Gamma_l$, $\Gamma_h$ and 
$\Gamma_{\rm inv}$, 
because they are 
three independent combinations of the three partial widths, i.e.
$\Gamma_Z =\Gamma_h+3\Gamma_l+\Gamma_{\rm inv}$, $R_l=\Gamma_h/\Gamma_l$, and
$\sigma_h^0 = (12\pi/\mmz )\Gamma_h\Gamma_l/\Gamma_Z^2$. 
We find 
\bea
\label{del_zwidths} 
  \begin{array}{c@{\;}c@{\;}l}
     \Delta \Gamma_h/(\Gamma_h)_{\rm SM} &=& \phantom{-}0.0011 \pm 0.0014 \\
     \Delta \Gamma_l/(\Gamma_l)_{\rm SM} &=& -0.0013 \pm 0.0013 \\
     \Delta \Gamma_{\rm inv}/(\Gamma_{\rm inv})_{\rm SM} 
      &=& -0.0050 \pm 0.0040 
  \end{array}
 \quad
  \rho_{\rm corr} =\left(
     \begin{array}{rrr}
       1.00 & 0.49 & -0.41\\
            & 1.00 &  0.23\\
            &      &  1.00
     \end{array}\,\right)\,,
\eea
where
$(\Gamma_h)_{\rm SM}=1743.4\mev$, 
$(\Gamma_\ell)_{\rm SM}=84.03\mev$ and 
$(\Gamma_{\rm inv})_{\rm SM}=501.9\mev$ 
are the SM predictions\cite{hhkm} for $m_t=175\gev$, $\mh=100\gev$, 
$\alpha_s=0.118$ and $\da=0.03$. 
The high precision of 0.14\% of the hadronic $Z$ partial width, 
$\Gamma_h$, strongly restricts any attempt to modify theoretical predictions 
for the ratios $R_b$ and $R_c$\cite{khlp95}.
To see this, $\Gamma_h$ is approximately expressed as
\bea 
        \Gamma_h &=& \Gamma_u +\Gamma_d +\Gamma_s +\Gamma_c +\Gamma_b 
                +\Gamma_{\rm others}
\nonumber\\
                 &\sim& \{ \Gamma_u^0 +\Gamma_d^0 +\Gamma_s^0 
        +\Gamma_c^0 +\Gamma_b^0 \} 
        \times [1+\frac{\alpha_s}{\pi}+ {\cal O}(\frac{\alpha_s}{\pi})^2 ] ,
\label{gamma_h_appr}
\eea
where $\Gamma_q^0$'s are the partial widths in the absence of the final state 
QCD corrections. Hence, to a good approximation, the ratios $R_q$ can be 
expressed as ratios of $\Gamma_q^0$ and their sum. A decrease in $R_b$ and an 
increase in $R_c$ should then imply a decrease and an increase of $\Gamma_b^0$
and $\Gamma_c^0$, respectively, from their SM predicted values. The strong 
interaction coupling $\alpha_s$ acts like a flavor independent adjustment 
parameter. This is clearly borne out in Fig.~\ref{fig:rbrc_vs_alphas}, where, 
once {\em both} $\,\Gamma_b^0$ {\em and} $\Gamma_c^0$ are 
left free in the fit,
the above $\Gamma_h$ drives $\alpha_s$ for the 1995 data\cite{lepewwg95}  
to an unacceptably large value, while for the 1996 update\cite{lepewwg96}  
a consistent picture emerges. On the 
other hand, if we allow only $\Gamma_b^0$ to vary by assuming the SM value of 
$\Gamma_c^0$ (the straight line of the extended SM in Fig.~\ref{fig:rbrc}),
then the $\Gamma_h$ constraint gives a slightly smaller value of 
$\alpha_s$, see Eq.~(\ref{alpswithdb}), 
though still compatible with the global average\cite{pdg96}, 
$\alpha_s=0.118\pm 0.003$.

In general, if we introduce a fractional change 
in the bare hadronic width 
  \begin{eqnarray}
       \frac{\delta \Gamma_h^0}{(\Gamma_h^0)_{\rm SM}} 
       \approx
       \frac{\sum_q \delta \Gamma_q^0}{\sum_q (\Gamma_q^0)_{\rm SM}}\,,
  \end{eqnarray}
one measures to a good approximation from the $Z$-line 
shape parameters the combination 
  \begin{eqnarray}
    \alpha_s + \pi \frac{\delta\Gamma_h^0}{(\Gamma_h^0)_{\rm SM}}\,.
  \end{eqnarray}
In other words, the effective parameter $\alpha_s'$ 
  \begin{eqnarray}
     \alpha_s' \equiv \alpha_s(\mz )_{\msbar}
     + 3.186 \frac{\delta\Gamma_h^0}{(\Gamma_h^0)_{\rm SM}}\,.
  \end{eqnarray}
is constrained by the $Z$ parameters.
The coefficient in front of the fractional width ratio is slightly 
larger than $\pi$ because of the higher-order QCD corrections.  
For definiteness, we use the SM prediction 
$(\Gamma_h^0)_{\rm SM}=1678.7\mev$ evaluated at 
$(\mt,\mh)=(175,100)\gev$.  
If only the $\zbb$ vertex is 
allowed to deviate from the SM prediction, 
  \begin{subequations}
     \label{alphas_prime}
  \begin{eqnarray}
     \alpha_s' &=& \alpha_s + 3.186 \frac{\delta R_b}{1-R_b} \\
          &\approx& 
          \alpha_s 
          +1.54 \Bigl(\delb -[(\delb)_{\rm SM}]_{\mt=175\gev}\Bigr) \\
          &\approx& 
          \alpha_s +1.54 (\delb +0.00995) \\
           &\approx&
          \alpha_s + 0.00134\xt +1.54\,[\delb]_{\rm New Physics}
          \label{alphas_prime_d}
  \end{eqnarray}
  \end{subequations}
in agreement with the expression (\ref{alpsprime}).
The last equality is obtained by inserting the SM expression 
(\ref{db_approx}) for $\delb$ where we neglect the small 
quadratic term. 
If both $R_b$ and $R_c$ are modified, 
it is the combination 
 \begin{eqnarray}
   \alpha_s' = \alpha_s 
     + 3.186 \frac{\delta R_b +\delta R_c}{1 - R_b - R_c}
 \end{eqnarray}
which is constrained by the $\Gamma_h$ data. 

\begin{figure}[b]
\begin{center}
  \leavevmode\psfig{file=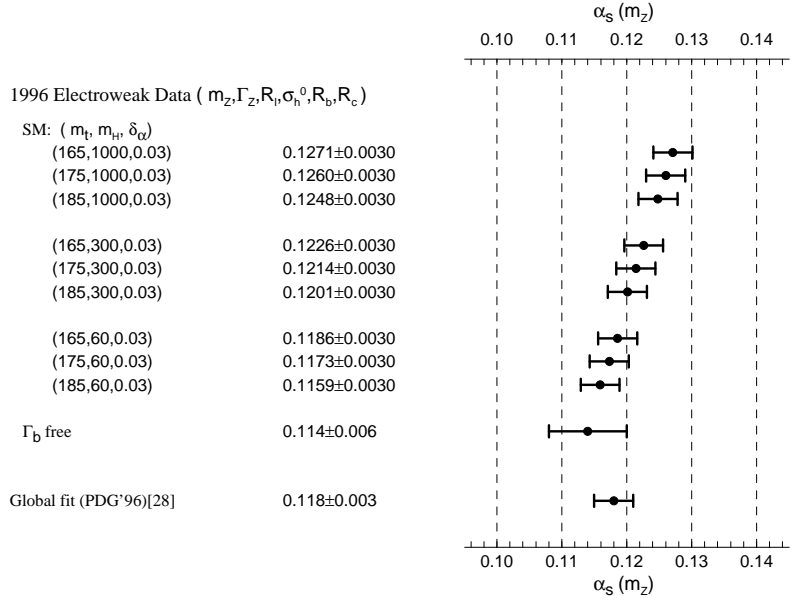,height=8cm,silent=0}
\vspace{2mm}
\caption{\protect\footnotesize\sl
Constraint on $\alpha_s$ from the electroweak $Z$ boson data 
by assuming the SM for various $\mt$ and $\mh$ at $\da=0.03$.  
Also shown is the result of a more general
fit, where $\Gamma_b$ is a free parameter. 
For comparison, the global average as obtained by 
the Particle Data Group\protect\cite{pdg96} is shown. 
}
\label{fig:alphas}
\end{center}
\end{figure}
 
At present, the LEP Collaborations have not yet completed their analyses of 
$R_b$ and $R_c$ by including the latest runs. 
However, there are new precise analyses of OPAL on $R_c$\cite{rbrc_opal96} 
and $R_b$\cite{rb_opal96} and one by ALEPH on $R_b$\cite{rb_aleph96}. 
The new analyses aim at reducing as much as
possible the use of information not directly obtainable from
experiment itself. The increased number of tags in the ALEPH analysis
implies also a smaller correlation between $R_b$ and $R_c$. The
preliminary values quoted at the 1996 summer conferences\cite{lepewwg96} 
roughly 
agree with the SM expectation and 
it may now be meaningful to compare the constraints 
on the strong coupling constant $\alpha_s$ from the $Z$-pole data 
with those
from other sectors
\cite{pdg96} (see Fig.~\ref{fig:alphas}).  
We find the following parametrization for the $\mt$, $\mh$ and $\da$ 
dependences of the SM fit to $\alpha_s$:
\begin{eqnarray}
 \label{alpsfit}
 \alpha_s &=& 0.1182 \pm 0.0030 
                        -0.00075\xt 
                        +0.0023\xh +0.00046x_H^2 
                        -0.00074\xa 
\end{eqnarray}
where $\xt=(\mt(\gev)-175)/10$, $\xh=\log(\mh(\gev)/100)$, 
and $\xa=(\da-0.03)/0.09$. 
The parametrization is valid in the range 
$150<m_t(\gev )<200$, $60<\mh(\gev )<1000$ and 
$|\delta_\alpha|<0.2$. 
It is remarkable that the electroweak data alone imply an intrinsic 
precision of $\pm 0.003$ (disregarding new physics contribution to 
the $Z$ partial widths) which is deteriorated by the imprecise 
knowledge of the external parameters, i.e. the masses of the top and 
Higgs and also by the 
running ``QED'' coupling $\alpha(\mmz)$ (see also Section 5.1). 
It can be seen from Fig.~\ref{fig:alphas} and the above 
parametrization that the agreement between the SM fit 
to the $Z$ parameters and the present world average 
of direct measurements, $\alpha_s=0.118\pm 0.003$, is good 
only for a relatively light Higgs boson 
($\mh\simlt 300\gev$).  

\subsection{New Neutrino Data}  
 
\begin{figure}[b]
 \begin{center}
 \leavevmode\psfig{file=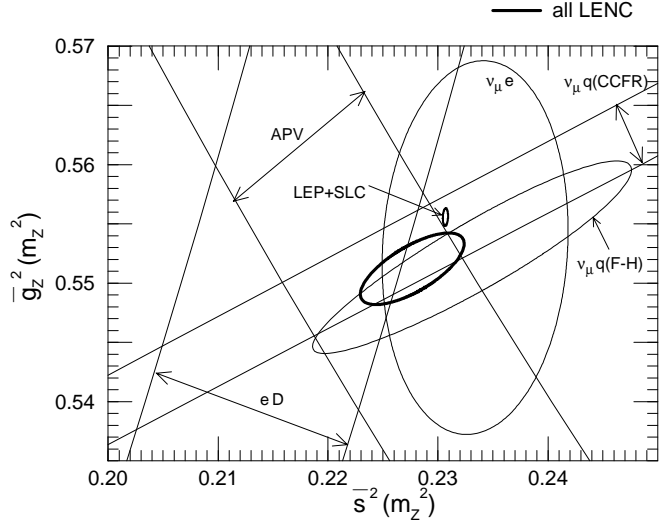,height=7cm,silent=0}
 \end{center}
\caption{\protect\footnotesize\sl
Fit to the low-energy neutral-current data in terms of the 
two universal charge form factors $\protect\sbar^2(\mmz )$ and 
$\protect\gzbar^2(\mmz )$ which are calculated from $\protect\sbar^2(0)$ and 
$\protect\gzbar^2(0)$ by assuming the SM running of the charge form factors. 
1-$\sigma$ (39\%CL) contours are shown separately for
the old\cite{fh88} and the new\cite{ccfr95_kevin} $\nu_\mu$--$q$ data,
the $\nu_\mu$--$e$ data, the atomic parity violation (APV) data,
and the SLAC $e$--${\rm D}$ polarization asymmetry data: 
see Ref.\cite{hhkm}
The 1-$\sigma$ contour of the combined fit, 
Eq.~(\protect\ref{fitoflencatmz}), 
is shown by the thick ellipse. 
The little ellipse
represents the 1-sigma constraint from LEP/SLC data corresponding to the
solid ellipse of Fig.~\protect\ref{fig:gzb2sb2}.
}
\label{fig:lenc}
\end{figure}

A new piece of information in the low-energy neutral current sector comes 
from the CCFR collaboration\cite{ccfr95_kevin} which measured the 
neutral- to charged-current cross section ratio in $\nu_\mu$ scattering off 
nuclei. Using the model-independent parameters of Ref.\cite{fh88},
they constrain the following linear combination,
\bea \label{k_ccfr}
K &=& 1.732 g_L^2
    + 1.119 g_R^2
    - 0.100 \delta_L^2
    - 0.086 \delta_R^2 \,,
\eea
and obtain 
%
\dataofnqccfrorig
%
with $m_c=(1.31\pm 0.24)\gev$. 
Because of the larger $\langle Q^2 \rangle_{\rm CCFR} = 36{\gev}^2$ 
in the CCFR experiments compared to the old data\cite{fh88}
($\langle Q^2 \rangle_{\rm HF} = 20{\gev}^2$), the measurement is first
expressed in terms of $\sbar^2(0)$ and $\gzbar^2(0)$ and then combined
with the old data. Figure~\ref{fig:lenc} shows the CCFR-band
together with the ellipse of all previous $\nu q$-data. 
 
The CCFR data (\ref{dataofnqccfrorig}) 
being obtained after correcting for the external photonic 
corrections lead to the constraint :
\fitofnqccfr
It should be noted that the old data\cite{fh88} were also corrected for
external photonic corrections.\footnote{%
The $\delta_{c.c.}$ correction in Ref.\cite{hhkm} was hence erroneously 
counted twice. The fit Eq.~(4.17) of Ref.\cite{hhkm} has therefore been 
revised here.}
We find 

\fitofnqfh
%
The combination of the new CCFR\,data\cite{ccfr95_kevin}
with the previous neutrino data\cite{fh88} yields:
%
\fitofnq
%
The combined fit to all the low-energy neutral current data 
including those studied in Ref.\cite{hhkm} gives :
\fitoflenc
For later convenience  these results are also expressed at the shifted scale
$q^2 = \mmz$. 
Here we assume no significant new physics contributions to the running 
of the charge form factors from 0 to $m_Z^2$. 
Uncertainty from the $\mh$-dependence of the running of 
$\gzbar^2(\mmz)$, Eq.~(\ref{gzbarrunning}), is negligibly small 
for $\mh>70\gev$. 
The result is then :
\fitoflencatmz
%
Fig.~\ref{fig:lenc} shows the individual contributions to the fit. The data
agree well with each other. Also shown is the combined LEP/SLC fit (the solid 
ellipse 
of Fig.~\ref{fig:gzb2sb2}). Although the low energy data 
are far less precise than 
those from the $Z$ resonance, they nevertheless
constrain possible new interactions beyond the ${\rm SU(2)_L\times U(1)_Y}$ 
gauge interactions, such as those from an additional $Z$ 
boson\cite{langacker94}.
 
We may now combine the constraints from the $Z$ parameters, 
Eqs.~(\ref{fitofgzbsbalpspdb}) and (\ref{fitofgzbsb}), and 
those from the low energy neutral current 
experiments, Eq.~(\ref{fitoflencatmz}): 
\fitofncgzbsbalpspdb
for the four-parameter fit, and 
\fitofncgzbsb
for the two-parameter fit with external $\alpha_s'$ and $\delb$. 
The net effect of the low energy data is to move the mean value of 
$\gzbar^2(\mmz)$ down by 0.00032, i.e. nearly half a standard deviation. 
As can be seen from Fig.~\ref{fig:lenc}, this downward shift is mainly a 
consequence of the old $\nu_q$--$q$ scattering data\cite{fh88}. 

Future results from the NUTEV Collaboration, 
succeeding to the CCFR Collaboration, are 
expected to improve considerably 
the constraints on 
the low energy form factors.

\subsection{The (S,T,U)-Fit}

\begin{figure}[b]
 \begin{center}
 \leavevmode\psfig{file=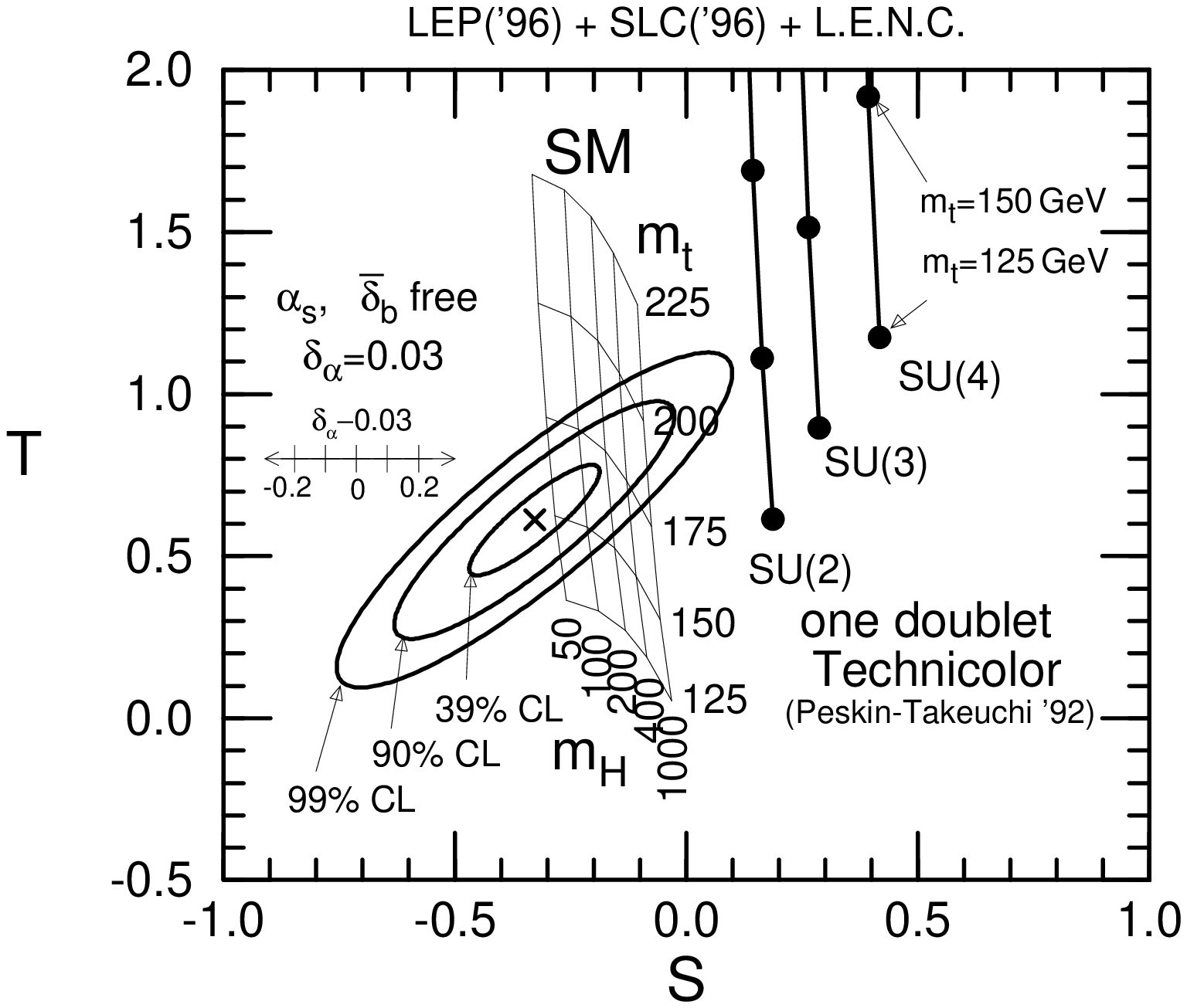,height=7cm,silent=0}
 \end{center} 
\caption{\protect\footnotesize\sl
Constraints on ($S$, $T$) from the five-parameter fit to
all the electroweak data for $\delta_\alpha=0.03$ and 
$\protect\delg=0.0055$. 
Together with $S$ and $T$, the $U$ parameter, 
the $\protect\zbb$ vertex form factor, 
$\protect\delb(\protect\mmz)$, and the QCD coupling, 
$\alpha_s(\mz)$, are allowed to vary in the fit.
Also shown are the SM predictions in the range 
$125\protect\gev\!<\!m_t\!<\!225\protect\gev$ 
and $50\protect\gev\!<\!\protect\mh\!<\!1000\protect\gev$. 
The predictions\protect\cite{stu} of one-doublet 
${\protect\rm SU}(N_c)$--TC models are shown for $N_c=2,3,4$. 
}
\label{fig:st}
\end{figure}
 
All neutral current data are summarized in Eq.~(\ref{fitoflenc}) for low 
energy ($q^2\approx 0$) and in Eq.~(\ref{fitofgzbsbalpspdb}) for the 
$Z$-shape parameters. 
In addition, the slightly improved $W$ mass\cite{mw96} in Table~1,
 \begin{equation}
    \mw = 80.356 \pm 0.125 \gev
 \end{equation}
gives
%
\fitofmw
for $\delg=0.0055$ in Eq.~(\ref{gf}).
 
Using Eq.~(\ref{stu_def}) or (\ref{gbarfromstu}) a three-parameter fit to all 
the electroweak data, i.e. the $Z$ parameters, the $W$ mass and the low-energy 
neutral-current data, is performed in terms of $S$, $T$, $U$, 
while $\alpha_s'$ 
and $\delb$ are treated as external parameters. To be specific the top and
Higgs masses required in the mild running of the charge form factors (see
Eq.~(\ref{gzbarrunning})) are set to $175$~GeV and $100$~GeV. The fit yields :
%
\fitofstu
%
The dependence of the $S$ and $U$ parameters upon $\delta_\alpha$ may be 
understood from Eq.~(\ref{gbar_approx}) and (\ref{stu_prime}). 
For an arbitrary value of $\delg$ the 
parameter $T$ should be replaced by 
$T'\!\equiv\! T\!+\!(0.0055\!-\!\delg)/\alpha$\cite{hhkm}. 
Note that the uncertainty in $S$ coming from 
$\delta_\alpha=0.03\pm 0.09$\cite{eidjeg95} 
is of the same order as that from the uncertainty in 
from $\alpha_s=0.118\pm 0.003$\cite{pdg96}; 
they are not at all negligible when compared to the overall error.
The $T$ parameter has little $\delta_\alpha$ dependence, but is sensitive 
to $\alpha_s$. 
 
The above results, together with the SM predictions, are shown in 
Fig.~\ref{fig:st} as the projection onto the ($S,\,T$) plane. Accurate 
parametrizations of the SM contributions to the $S$, $T$, $U$ parameters are 
found in Ref.\cite{hhkm}, while their compact parametrizations valid in the 
domain $160\gev<\mt<185\gev$ and $40\gev<\mh<1000\gev$ are given 
in Eq.~(\ref{studb_approx}).
Also shown are the predictions\cite{stu} of the minimal (one-doublet) 
SU($N_c$) Technicolor (TC) models with $N_c\!=\!2,3,4$, where QCD-like 
spectra of Techni-bosons with the large $N_c$ scaling and a specific 
top-quark mass generation mechanism is assumed. Obviously the 
current experiments provide a fairly stringent constraint on 
the simple TC 
models. Any TC model to be realistic must provide an additional negative 
contribution to $S$\cite{appelquist93} and at the same time a rather 
small contribution to $T$.
Our results confirm the observations\cite{khlp95,ellis95} 
based on the previous data, and are consistent with those of 
other recent updates\cite{altarelli96,langacker97,rosner97}.   
%
 
\begin{table}[b]
\caption{\protect\footnotesize\sl
Constraints on the parameters 
$S_{\rm new}$, $T_{\rm new}$, $U_{\rm new}$ 
which are obtained by subtracting the SM contribution 
$S_{\rm SM}$, $T_{\rm SM}$, $U_{\rm SM}$ from 
$S$, $T$, $U$ for $\alpha_s=0.118$ and 
$\delta_\alpha=0.03$. 
Correlations among errors are the same as in 
Eq.~(\protect\ref{fitofstu_mean}). 
}
\label{tableofstunew}
\begin{center}
{\footnotesize
\begin{center}
\begin{tabular}{|c@{}c|c@{\,}c|c@{\,}c|c|}
   \hline
   \begin{tabular}{cc} $m_t$\\(GeV) \end{tabular}
   & 
   \begin{tabular}{cc} $m_H^{}$\\(GeV) \end{tabular}
   & 
   $\displaystyle{\left(
   \begin{array}{@{}c@{}} S \\ T \\ U \end{array}
   \right)}$
   & 
   $\chi^2_{\rm min}/({\rm d.o.f.})$ 
   &
   $\displaystyle{\left(
   \begin{array}{@{}c@{}}S_{\rm SM}\\T_{\rm SM}\\U_{\rm SM}\end{array}
   \right)}$
   & 
   $\chi^2/({\rm d.o.f.})$
   &
   $\displaystyle{\left(
   \begin{array}{@{}c@{}}S_{\rm new}\\T_{\rm new}\\U_{\rm new}\end{array}
   \right)}$
   \\
    \hline
      169 &  100 &
    $\displaystyle{\begin{array}{r}
     -0.27\pm  0.13\\  0.71\pm  0.14\\  0.41\pm  0.38
    \end{array}}$
    &
      \begin{tabular}{c}
      23.9/21 \\ (30\% CL)
      \end{tabular}
    &
    $\displaystyle{\begin{array}{r}
     -0.23\\  0.80\\  0.35
    \end{array}}$
    &
      \begin{tabular}{c}
      24.5/24 \\ (43\% CL)
      \end{tabular}
    &
    $\displaystyle{\begin{array}{r}
     -0.05\pm  0.13\\ -0.09\pm  0.14\\  0.06\pm  0.38
    \end{array}}$
    \\ 
    \hline
      169 & 1000 &
    $\displaystyle{\begin{array}{r}
     -0.28\pm  0.13\\  0.70\pm  0.14\\  0.41\pm  0.38
    \end{array}}$
    &
      \begin{tabular}{c}
      23.9/21 \\ (30\% CL)
      \end{tabular}
    &
    $\displaystyle{\begin{array}{r}
     -0.07\\  0.51\\  0.34
    \end{array}}$
    &
      \begin{tabular}{c}
      57.5/24 \\ (0.01\% CL)
      \end{tabular}
    &
    $\displaystyle{\begin{array}{r}
     -0.21\pm  0.13\\  0.19\pm  0.14\\  0.07\pm  0.38
    \end{array}}$
    \\ 
    \hline
      175 &  100 &
    $\displaystyle{\begin{array}{r}
     -0.26\pm  0.13\\  0.73\pm  0.14\\  0.39\pm  0.38
    \end{array}}$
    &
      \begin{tabular}{c}
      25.1/21 \\ (24\% CL)
      \end{tabular}
    &
    $\displaystyle{\begin{array}{r}
     -0.23\\  0.88\\  0.36
    \end{array}}$
    &
      \begin{tabular}{c}
      28.1/24 \\ (26\% CL)
      \end{tabular}
    &
    $\displaystyle{\begin{array}{r}
     -0.03\pm  0.13\\ -0.15\pm  0.14\\  0.03\pm  0.38
    \end{array}}$
    \\ 
    \hline
      175 & 1000 &
    $\displaystyle{\begin{array}{r}
     -0.27\pm  0.13\\  0.72\pm  0.14\\  0.40\pm  0.38
    \end{array}}$
    &
      \begin{tabular}{c}
      25.1/21 \\ (24\% CL)
      \end{tabular}
    &
    $\displaystyle{\begin{array}{r}
     -0.08\\  0.58\\  0.36
    \end{array}}$
    &
      \begin{tabular}{c}
      48.4/24 \\ (0.2\% CL)
      \end{tabular}
    &
    $\displaystyle{\begin{array}{r}
     -0.20\pm  0.13\\  0.14\pm  0.14\\  0.04\pm  0.38
    \end{array}}$
    \\ 
    \hline
      181 &  100 &
    $\displaystyle{\begin{array}{r}
     -0.25\pm  0.13\\  0.75\pm  0.14\\  0.38\pm  0.38
    \end{array}}$
    &
      \begin{tabular}{c}
      26.4/21 \\ (19\% CL)
      \end{tabular}
    &
    $\displaystyle{\begin{array}{r}
     -0.24\\  0.96\\  0.38
    \end{array}}$
    &
      \begin{tabular}{c}
      34.2/24 \\ (8\% CL)
      \end{tabular}
    &
    $\displaystyle{\begin{array}{r}
     -0.02\pm  0.13\\ -0.21\pm  0.14\\  0.00\pm  0.38
    \end{array}}$
    \\ 
    \hline
      181 & 1000 &
    $\displaystyle{\begin{array}{r}
     -0.26\pm  0.13\\  0.74\pm  0.14\\  0.38\pm  0.38
    \end{array}}$
    &
      \begin{tabular}{c}
      26.5/21 \\ (19\% CL)
      \end{tabular}
    &
    $\displaystyle{\begin{array}{r}
     -0.08\\  0.66\\  0.37
    \end{array}}$
    &
      \begin{tabular}{c}
      41.3/24 \\ (2\% CL)
      \end{tabular}
    &
    $\displaystyle{\begin{array}{r}
     -0.18\pm  0.13\\  0.08\pm  0.14\\  0.01\pm  0.38
    \end{array}}$
    \\ 
    \hline
\end{tabular}
\end{center}
}
\end{center}
\end{table}

To be 
more quantitative 
Table~\ref{tableofstunew} provides the values of $S$, $T$ 
and $U$ after subtracting the SM contributions ($S_{\rm new}\equiv 
S-S_{\rm SM}$, etc.). The $m_t$- and $\mh$-dependences of the extracted $S$, 
$T$ and $U$ values result from the fact that the SM prediction for $\delb$ 
being strongly $\mt$ dependent has been assumed in $\alpha_s'$ for a fixed 
$\alpha_s=0.118$; see (\ref{alphas_prime_d}) 
with $[\delb]_{\rm New Physics}=0 $.  
All values in the table are obtained by setting $\alpha_s=0.118$ and 
$\delta_\alpha=0.03$. The values for different choices of $\alpha_s$ and 
$\delta_\alpha$ together with the error correlation matrix can be
read-off 
from Eq.~(\ref{fitofstu}). It is worth pointing out that the SM fit provides
only a poor fit (less than 1\%CL) when $\mh=1000\gev$ and 
$\mt < 170\gev$.
New physics contributions of {\em both} $S_{\rm new}\approx -0.2$ {\em and} 
$T_{\rm new}\approx 0.2$ may then be needed because of the large correlation 
of 0.86 between the two quantities. 
In fact, once $S_{\rm new}$ is given by a model of dynamical
symmetry breaking, the $T_{\rm new}$ should be severely constrained
by the data ; 
$T_{\rm new} -1.1\, S_{\rm new} =0.37 \pm 0.073$
for $m_t=169\gev$ and $\mh=1000\gev$. 
The necessity of an additional positive 
$T_{\rm new}$ contribution cannot easily be read off from Fig.~\ref{fig:st}, 
where the projection of the fit (\ref{fitofstu}) onto the $(S,T)$ plane is 
shown when the combination $\alpha_s'$ (\ref{alpsprime}) of the $Zb_Lb_L$ 
vertex form factor $\delb$ and $\alpha_s$ are allowed to vary. 
The most stringent constraint on the $S$, $T$, $U$ parameters is 
obtained as an eigenvector of the correlation matrix of 
(\ref{fitofstu}):
\begin{eqnarray}
  T'-1.10 S'+0.04 U'= 0.99 \pm 0.073.
\label{stu_constraint}
\end{eqnarray}
Fit results for 
$S_{\rm new}$, $T_{\rm new}$ and $U_{\rm new}$ for other choices of $\mt$, 
$\mh$, $\alpha_s$ and $\delta_\alpha$ can easily be obtained from the result 
(\ref{fitofstu}) and the parametrization (\ref{studb_approx}).
 
Finally, regarding the point $(S,T,U)=(0,0,0)$ 
as the one with no-electroweak
corrections (a more precise treatment will be given in section 
5.2) $\chi^2_{\rm min}/({\rm d.o.f.})=141/(22)$ is found. 
On the other hand, if also the remaining electroweak corrections to 
$G_F$ are switched off by setting $\delg=0$, 
then $T'=0.0055/\alpha=0.75$ is found and the point
$(S,T',U)=(0, 0.75, 0)$ gives $\chi^2_{\rm min}/({\rm d.o.f.})=34.2/(22)$ 
being barely (5\%CL) consistent with the data.
As emphasized in Ref.\cite{novikov93}, the genuine electroweak correction 
is not trivially established
in this analysis because of the cancellation 
between the large $T$ parameter from $m_t \sim 175$~GeV and the non-universal 
correction $\delg$ to the muon decay constant in the observable 
combination\cite{hhkm} $T'=T+(0.0055-\delg)/\alpha$. 


\section{ The Minimal Standard Model Confronting the Electroweak Data }
Throughout this section all radiative corrections are assumed to be
dominated by the SM contributions. Within the minimal SM all electroweak
quantities are uniquely predicted as functions of $m_t$ and $\mh$. 
A careful investigation is done to elucidate the role of the  
input parameters $\alpha_s$ and $\bar{\alpha}(\mmz)$
required for the interpretation. 

A brief discussion on the significance of bosonic radiative corrections 
containing the weak-boson self-couplings is also given.

\subsection{ 4-parameter fit }
Within the Minimal Standard Model the electroweak precision data are
expressed in terms of the two mass parameters $m_t$ and $\mh$. 
In a first, and most general, attempt also the parameters $\alpha_s$ 
and $\delta_\alpha$ are left free. 
The result of the 4-parameter fit yields :
\fitofmtxhalpsda
Instead of fitting $\mh$ itself it is more appropriate to fit 
$\xh=\log(\mh/100\gev)$; otherwise the uncertainties are too asymmetric. 
It is remarkable that the fitted $\alpha_s$ value agrees well with the global 
fit result\cite{pdg96} and that its uncertainty is as low as 0.003. 
Also the fitted $\bar{\alpha}(\mmz)$ agrees within the large errors 
with other recent measurements \cite{mz94,swartz95,eidjeg95,bp95}. 
The fitted $m_t$ value is about 2-$\sigma$ below the present Tevatron
measurement, $m_t=175\pm 6\gev$\cite{mt96}. 
The relatively low $\mh$ value, $\mh=60^{+210}_{-50}\gev$, 
is a consequence of this. 
$\mh$ and $\delta_\alpha$ appear to be strongly anti-correlated as 
a consequence of the strong asymmetry constraint which is 
sensitive to $\delta_\alpha$.  
Large $\delta_\alpha$ (large $1/\bar{\alpha} (\mmz)$) implies 
small $\mh$. 

Next we present results of the 4-parameter fit on the 
electroweak data when external constraints on $\alpha_s$, 
$\alpha_s=0.118 \pm 0.003$\cite{pdg96}, 
and those on $\da$ are imposed. 
For $\delta_\alpha=0.03\pm 0.09$\cite{eidjeg95}, we obtain 
\fitofmtxhalpsdawithalpsdaej
while for $\delta_\alpha=0.12\pm 0.06$\cite{mz94}, we obtain 
\fitofmtxhalpsdawithalpsdamz
Because of the strong correlation between $\xh$ and  $\da$ in 
(\ref{fitofmtxhalpsda}), 
the error of $\xh$ is reduced by about a factor of two. 
At the same time, a strong positive correlation between the errors of
$m_t$ and $\xh$ appears. 
Larger $\da$ (smaller $\bar{\alpha}(\mmz)$) implies larger 
$\xh$ and larger $m_t$. 
The fitted $m_t$ value is still somewhat smaller than the observed
Tevatron value\cite{mt96}. 
This is partly due to the average $R_b$ value, which is presently
about 2-$\sigma$ larger than the SM prediction assuming $m_t=175\gev$.
The fit (\ref{fitofmtxhalpsdawithalpsdaej}) without $R_b$ and 
$R_c$ data yields
%
\fitofmtxhalpsdawithalpsdaejwithoutrbrc
%
The discrepancy is now reduced to the 1-$\sigma$ level. 
Although the above elliptic parametri\-zations reproduce the $\chi^2$ function 
only approximately, we find that the preferred ranges of $\mt$ and $\mh$ in 
Eq.~(\ref{fitofmtxhalpsdawithalpsdaejwithoutrbrc}) agree well with the 
corresponding results of Ref.\cite{ellis96}. 

\begin{figure}[b]
 \begin{center}
  \leavevmode\psfig{file=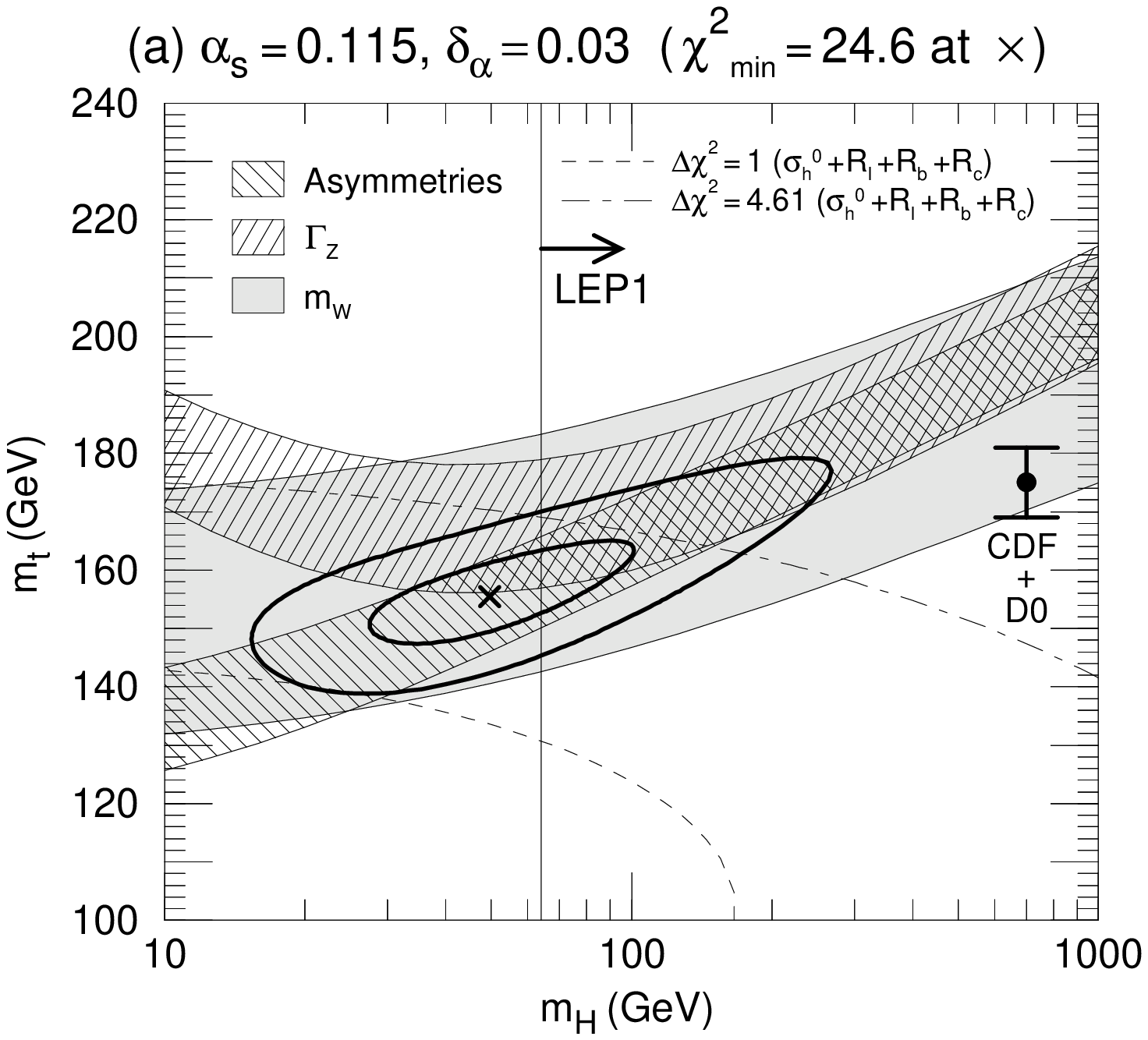,height=5.3cm,silent=0}\hfil
  \leavevmode\psfig{file=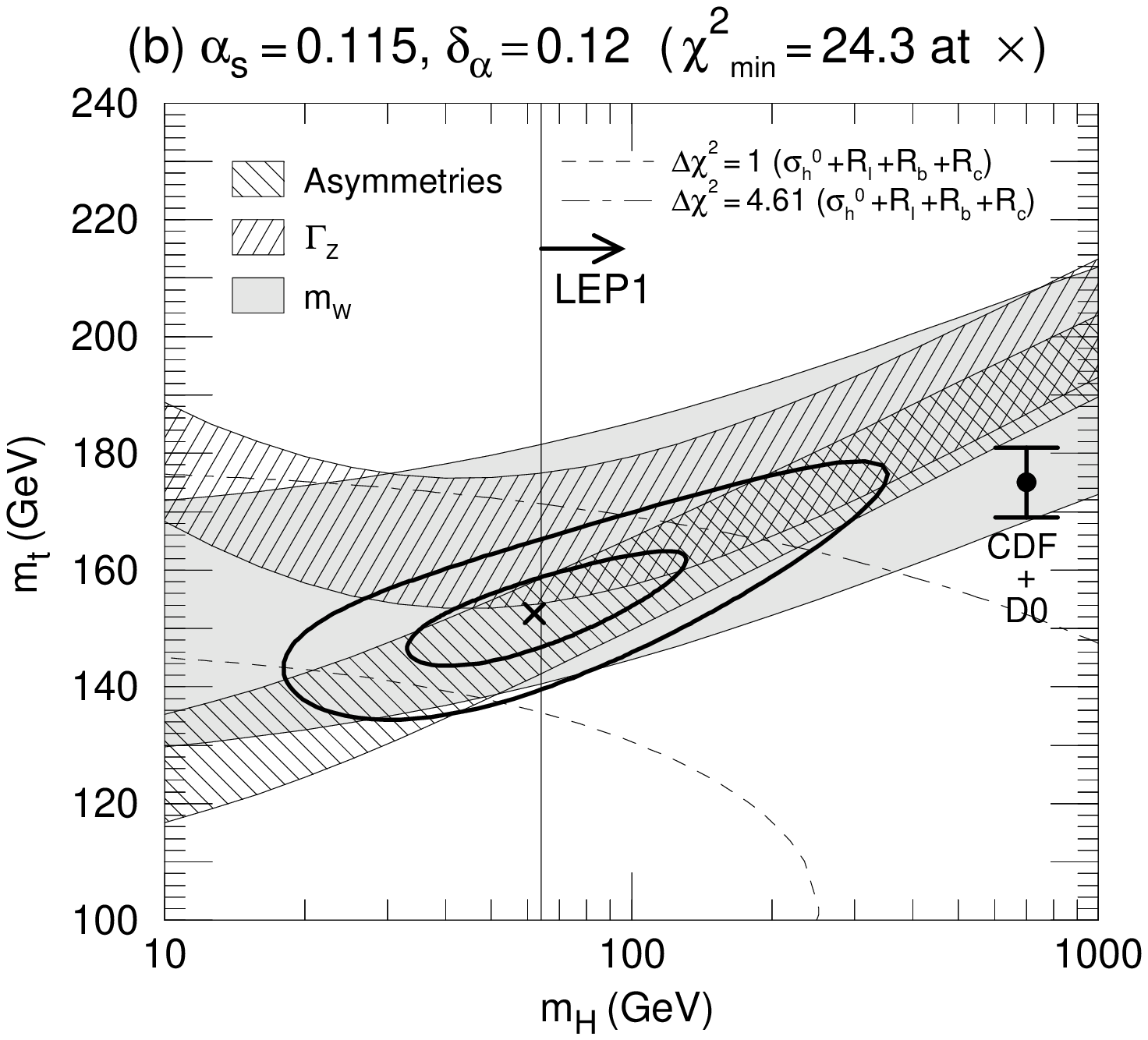,height=5.3cm,silent=0}\\
  \leavevmode\psfig{file=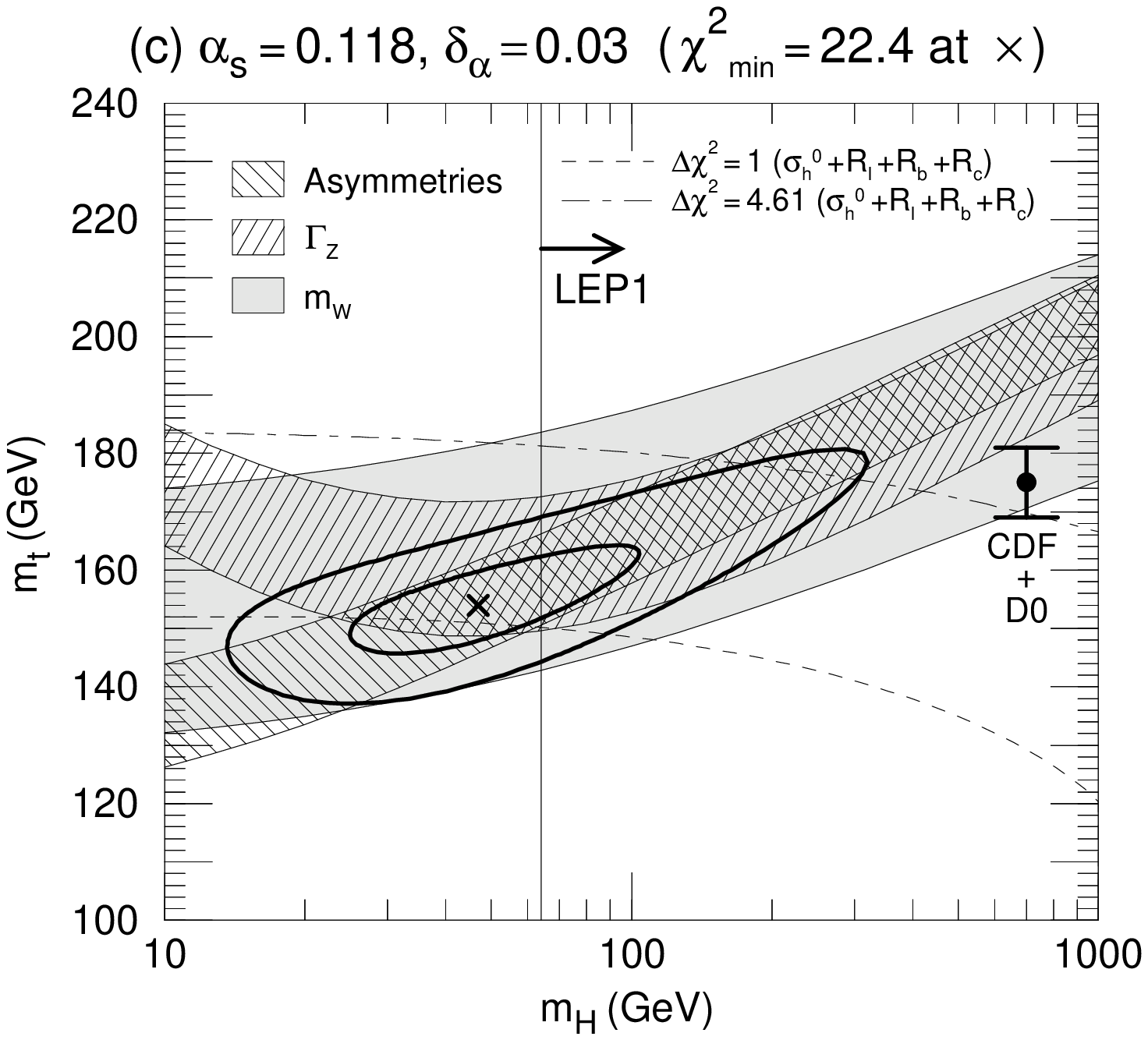,height=5.3cm,silent=0}\hfil
  \leavevmode\psfig{file=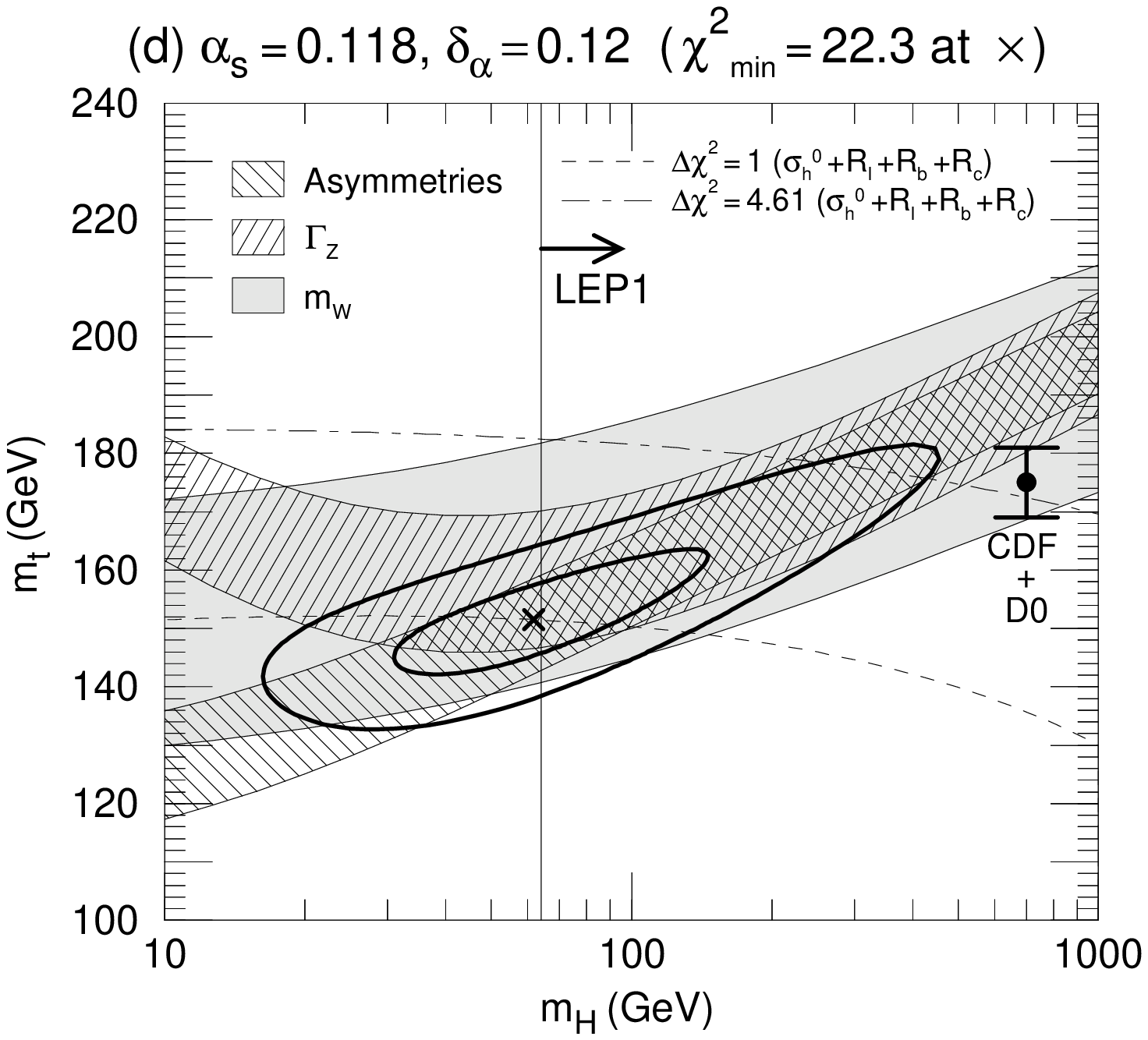,height=5.3cm,silent=0}\\
  \leavevmode\psfig{file=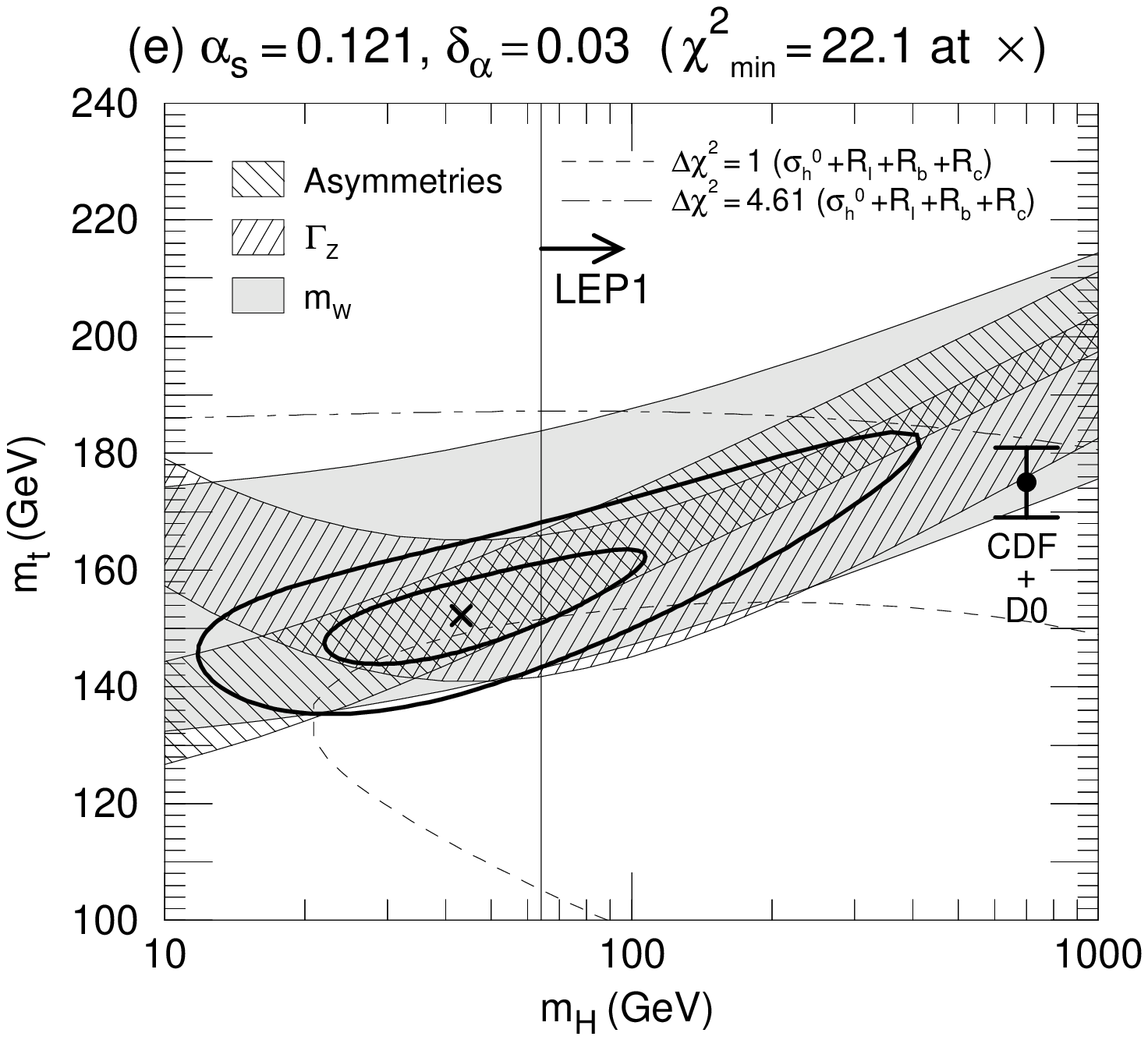,height=5.3cm,silent=0}\hfil
  \leavevmode\psfig{file=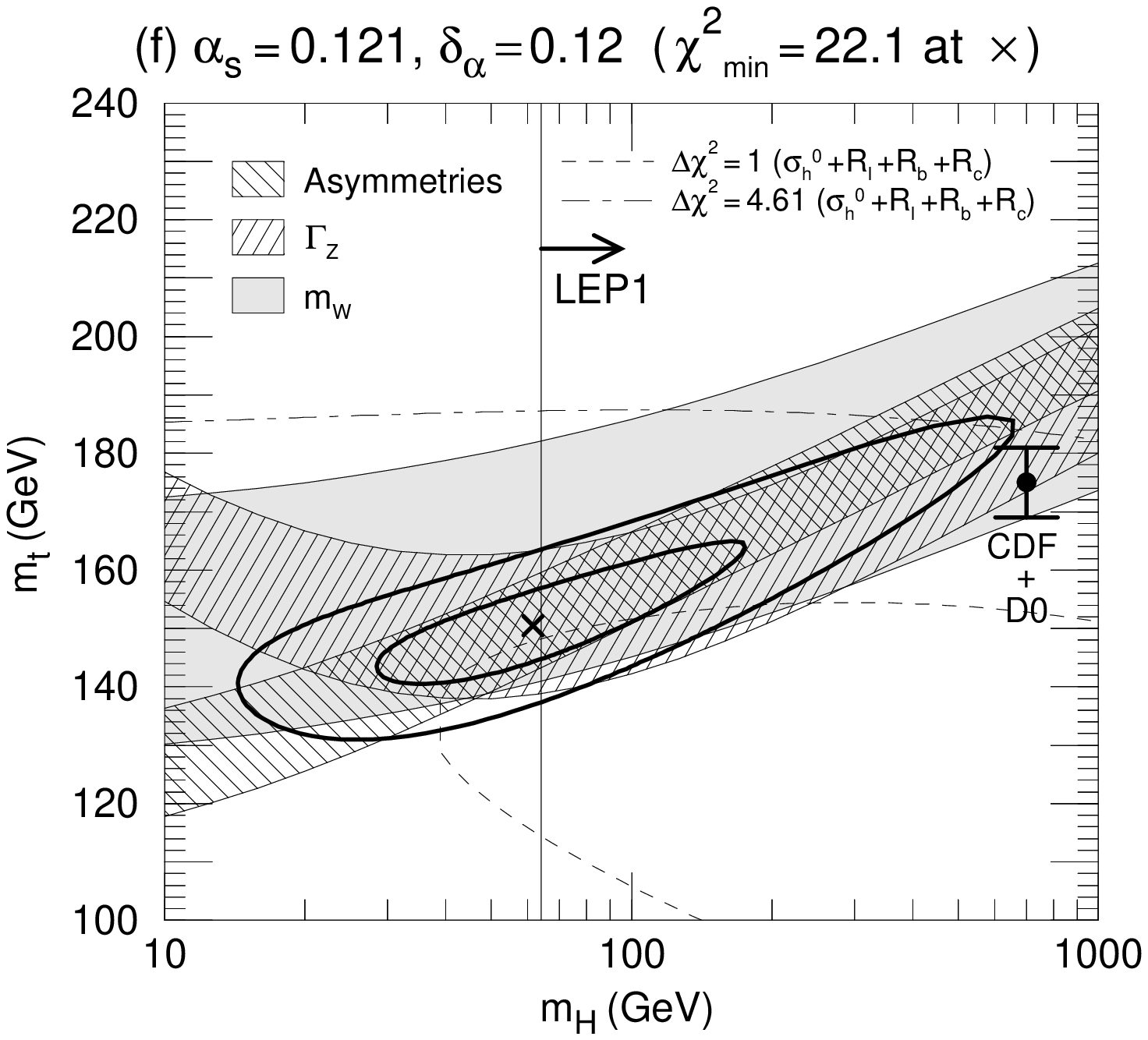,height=5.3cm,silent=0}
 \end{center}
\vspace{-3mm}
\caption{\protect\footnotesize\sl
The SM fit to all electroweak data in the ($\protect\mh,\,m_t$) 
plane for various choices of $(\alpha_s,\delta_\alpha)$ :
(a) (0.115,0.03) , (b) (0.115,0.012) ,  
(c) (0.118,0.03) , (d) (0.118,0.012) ,  
(e) (0.121,0.03) , (f) (0.121,0.012) ,  
where $\delta_\alpha=1/\bar{\alpha}(\mmz )-128.72$\cite{hhkm}. 
The thick inner and outer contours correspond to 
$\Delta\chi^2=1$ ($\sim$ 39\% CL), 
and $\Delta\chi^2=4.61$ ($\sim$ 90\%~CL), respectively. 
The minimum of $\chi^2$ is marked by the sign ``$\times$''. 
Also shown are the 1-$\sigma$ bands from the $Z$-pole asymmetries, 
$\Gamma_Z$ and $\mw$. 
The dashed lines show the constraint from the sum of $\sigma_h^0$, 
$R_\ell$, $R_b$ and $R_c$.
}
\label{fig:mtmh}
\end{figure}

Throughout 
(\ref{fitofmtxhalpsda})--(\ref{fitofmtxhalpsdawithalpsdaejwithoutrbrc}), 
the fitted $\alpha_s$ value agree 
well 
with the global average,
$\alpha_s=0.118\pm0.003$\cite{pdg96}. 
A slightly smaller value of $\mh\,(\sim 50\gev)$ is favored with the error 
of order 1 for $\log \mh$, and slightly smaller value of $m_t$ is favored 
as compared to the Tevatron 
measurement. 
The best-fit value of $\mh$ is sensitive to $\da$, 
whereas that of $m_t$ is sensitive to the $R_b$ data. 
The 4-parameter fit results given in 
Eqs.~(\ref{fitofmtxhalpsda})--(\ref{fitofmtxhalpsdawithalpsdaejwithoutrbrc}) 
are intended 
to illustrate qualitatively our understanding of the SM fit to the
electroweak. The errors are not fully elliptic. 
More accurate constraints on these parameters can be obtained from 
the parametrization of the $\chi^2$-function given 
below in Eq.~(\ref{total_chisqsm}). 

In conclusion,
the fits are stable and agree
with the a priori knowledge on $\alpha_s$ and $\delta_\alpha$.  
It is justified to proceed with an in-depth study based on 
the two parameters $m_t$ and $\mh$, 
where now $\alpha_s$ and $\delta_\alpha$ play the role of external parameters.

\subsection{ Constraints on $m_t$ and $\mh$ as functions of 
                $\alpha_s$ and $\bar{\alpha}(\mmz)$ }

\begin{figure}[b]
\begin{center}
 \leavevmode\psfig{file=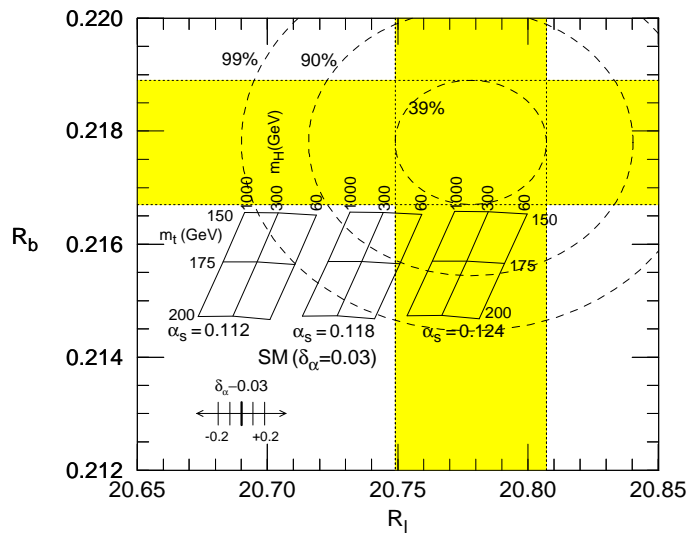,width=9cm,silent=0}
\end{center}
\caption{\protect\footnotesize\sl
The $R_b$ vs $R_\ell$ plane. The SM predictions are shown in the range 
$120\protect\gev\!<\!m_t\!<\!240\gev$, and $60\protect\gev\!<\!\protect
\mh\!<\!1\tev$, for three cases of $\alpha_s$ ($\alpha_s$=$0.11$, $0.12$ and 
$0.13$). These predictions are for $\delta_\alpha=0.03$, and their 
dependences on $\delta_\alpha$ are also indicated. Also shown are the 39\%,
90\% and 99\%CL contours obtained by combining only the $R_\ell$ and 
$R_b$ data. 
}
\label{fig:rlrb}
\end{figure}

\noindent 
In the minimal SM all relevant form factor values, 
i.e. $\gzbar^2(\mmz)$, $\sbar^2(\mmz)$, 
$\gzbar^2(0)$, $\sbar^2(0)$, $\gwbar^2(0)$ and $\delb(\mmz)$, 
are predicted uniquely in terms of 
on the two mass parameters $m_t$ and $\mh$. 
A convenient parametrization of the SM contributions to these form factors 
is given in Eqs.~(\ref{gbar_approx})--(\ref{studb_approx}), 
as functions of 
$\xt=(\mt(\gev )-175)/10$, $\xh=\log(\mh(\gev)/100)$
together with $\alpha_s$ and $\delta_\alpha$. 
Figure~\ref{fig:mtmh} shows the result of the fit to all
electroweak data in the ($\mh,\,m_t$)-plane for choices of $\alpha_s$ and  
$\delta_\alpha$ representative of their present knowledge. 
The figure exhibits to what extent the best-fit values 
as well as the size and 
orientation of the corresponding error ellipses 
($\Delta\chi^2\equiv\chi^2-\chi^2_{\rm min}=$ 1 and 4.61) depend on 
the knowledge of the external parameters $\alpha_s$ and $\delta_\alpha$. 

In order to understand how the fit comes about the 1-$\sigma$ constraints 
from the individual observables are shown separately. 
The narrow ``asymmetry'' band is sensitive to $\delta_\alpha$, whereas 
the ``$\Gamma_Z$'' band is sensitive to $\alpha_s$.  
The asymmetries constrain $m_t$ and $\mh$ through $\sbar^2(\mmz)$, 
while $\Gamma_Z$ does so through all the three form  factors 
$\gzbar^2(\mmz)$, $\sbar^2(\mmz)$ and $\delb(\mmz)$. 
It is most remarkable\cite{sm95} that in the SM $\Gamma_Z$ depends upon almost 
the same combination of $m_t$ and $\mh$ as the one measured through 
$\sbar^2(\mmz)$ provided $\mh$ is larger than about 60~GeV, which is 
indeed the range not excluded by the LEP1 experiments. 
The reason can be traced back to the approximate 
cancellation of the quadratic 
$m_t$-dependence of $\gzbar^2(\mmz)$ and $\delb$. 
Thus, the asymmetries and $\Gamma_Z$ alone, despite their quite small 
experimental errors, are constraining only a narrow band in the 
($m_t$,$\mh$)-plane. 
The present constraint due to the $\mw$ measurement overlaps this band.

Additional information is required to disentangle the above $m_t$-$\mh$ 
correlation. 
This is provided by $R_\ell$, $\sigma_h$, $R_b$ and is
shown in Fig.~\ref{fig:mtmh} by dashed lines corresponding to a 
$\Delta\chi^2$ of 1 ($\sim$39\%~CL) and 4.61 ($\sim$90\%~CL). 
The constraints due to $R_\ell$ and $R_b$ can also be seen in 
Fig.~\ref{fig:rlrb}. 
$R_\ell$ is sensitive to the assumed value of $\alpha_s$, and, 
for $\alpha_s=0.118$, the data favors small $\mh$.  
$R_b$ is neither sensitive to $\alpha_s$ nor $\mh$ and the 
present average 
disfavors large $m_t$. 
 
\begin{figure}[t]
\begin{center}
 \leavevmode\psfig{file=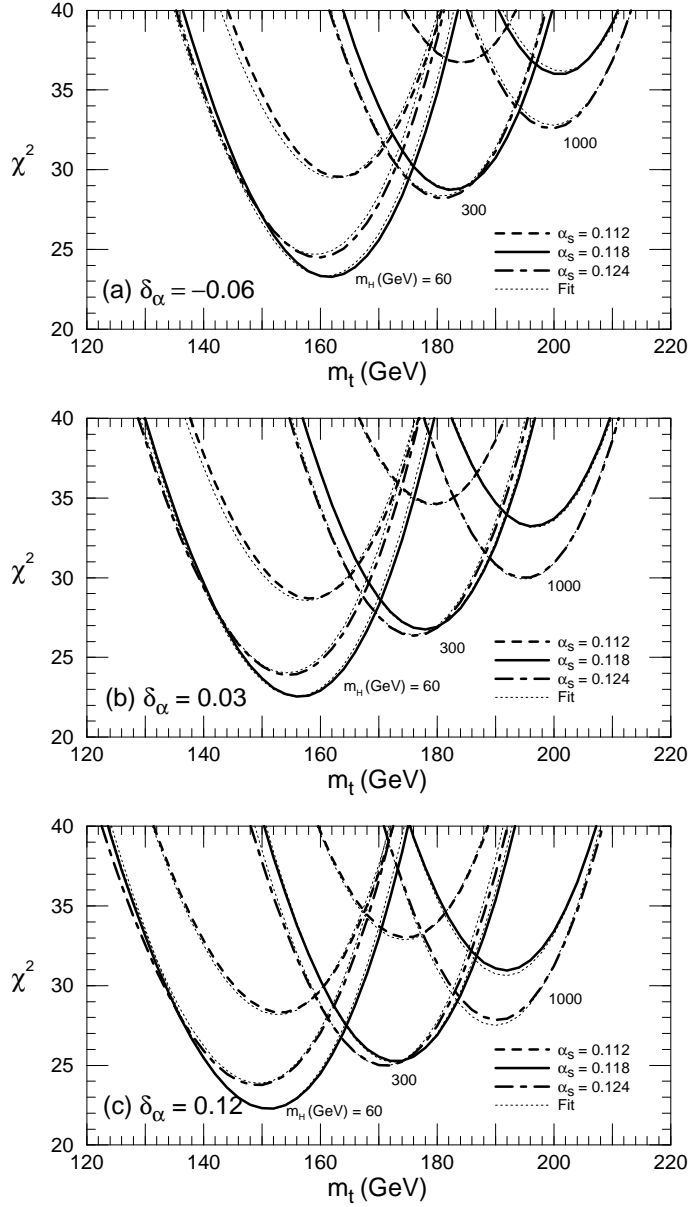,width=9cm,silent=0}
\end{center}
\caption{\protect\footnotesize\sl
Total $\chi^2$ of the SM fit to all the electroweak data
as functions of $m_t$ for $\mh=60$, 300, 1000$\gev$ and
$\alpha_s(\mz)=0.112,\,0.118,\,0.124$.
The uncertainty $\delta_\alpha$ in the hadronic vacuum polarization
contribution to the effective charge $1/\bar{\alpha}(\mmz)$ is shown
for three cases, $\delta_\alpha=-0.06$ (a), $0.03$ (b), $+0.12$ (c).
The dotted lines are obtained by using the approximate formula
(\protect\ref{total_chisqsm}).
The number of degrees of freedom is 25. 
}
\label{fig:chisqvsmt}
\end{figure}

Without the data on $R_\ell$, $\sigma_h$  and $R_b$ the region of large 
$\mh$-values in the ($m_t,m_H$)-band of Fig.~\ref{fig:mtmh} 
$(\mh\sim 1\tev)$ would not be excluded at all, as far as the 
electroweak data are concerned.
It is worth noting that in comparing Fig.~9 (a) with (e) 
(or (b) with (f)) the $\Gamma_Z$-band is shifted downwards 
by more than 10~GeV in the top quark mass when one increases 
$\alpha_s$ from 0.115 to 0.121, but despite of this shift
the best-fit point moves only marginally downwards by about 
1.7~GeV (see also the parametrization (\ref{fitofmt}) below).  
This is mainly because the constraint from $\sigma_h^0$, $R_\ell$ 
and $R_b$ allows larger $\mt$ for larger $\alpha_s$, as can be seen 
from dashed contours in Fig.~\ref{fig:mtmh}, or from Fig.~\ref{fig:rlrb}.  
The fit improves slightly at larger $\alpha_s$, because the $\Gamma_Z$ 
constraint then favors lower $m_t$ which in turn is favored by the 
$R_b$ data. 
On the other hand the change in $\delta_\alpha$ from the mean value 
of the estimate of Ref.~\cite{eidjeg95}, 0.03, to that of Ref.~\cite{mz94}, 
0.12, lowers the best-fit $\mt$ value by about 5~GeV and enhances that 
of $\mh$ slightly (by about 15~GeV), whereas the overall fit quality 
remains  unchanged.  
The $\chi^2$ function of the fit to all 
electroweak data can be parametrized in terms of the four parameters 
$m_t$, $\mh$, $\alpha_s$ and $\delta_\alpha$ :
%
\chisqsm
%
with
%
\fitofmt
%
and 
%
\chisqhsm
%
Here $m_t$ and $\mh$ are measured in GeV, 
and ${\rm d.o.f}=25$. 
This parametrization reproduces the exact $\chi^2$ function within 
a few percent accuracy in the range $100\gev<m_t<250\gev$, 
$60\gev<\mh<1000\gev$ and $0.10<\alpha_s(\mz)<0.13$. 
The best-fit value of $m_t$ for a given set of $\mh$, $\alpha_s$ and 
$\delta_\alpha$ is readily obtained from Eq.~(\ref{fitofmtbest}) with its 
approximate error of (\ref{fitofmterror}). 
See dotted curves in Fig.~\ref{fig:chisqvsmt}. 
 
\begin{figure}[b]
\begin{center}
  \leavevmode\psfig{file=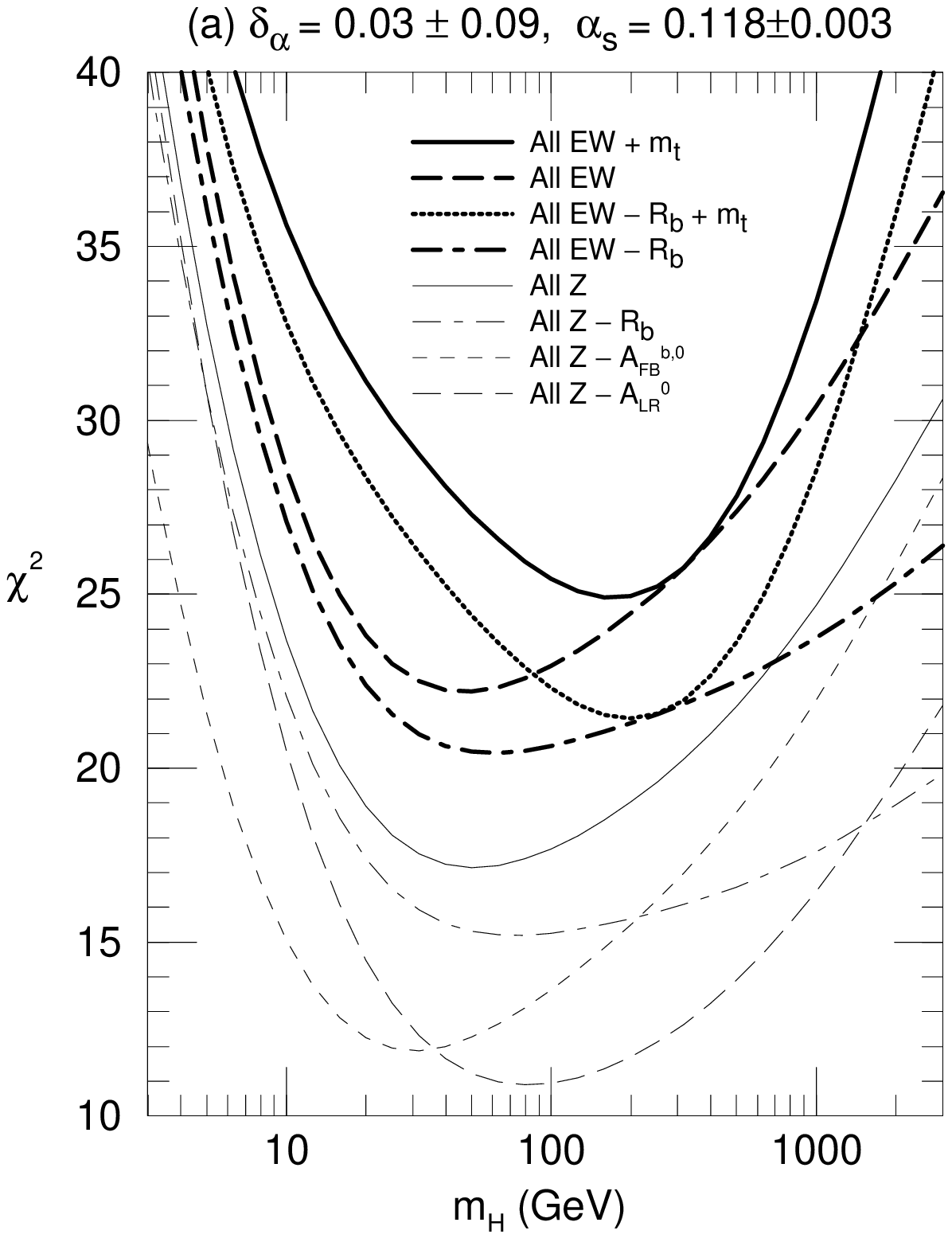,width=7cm,silent=0}
  \hfill
  \leavevmode\psfig{file=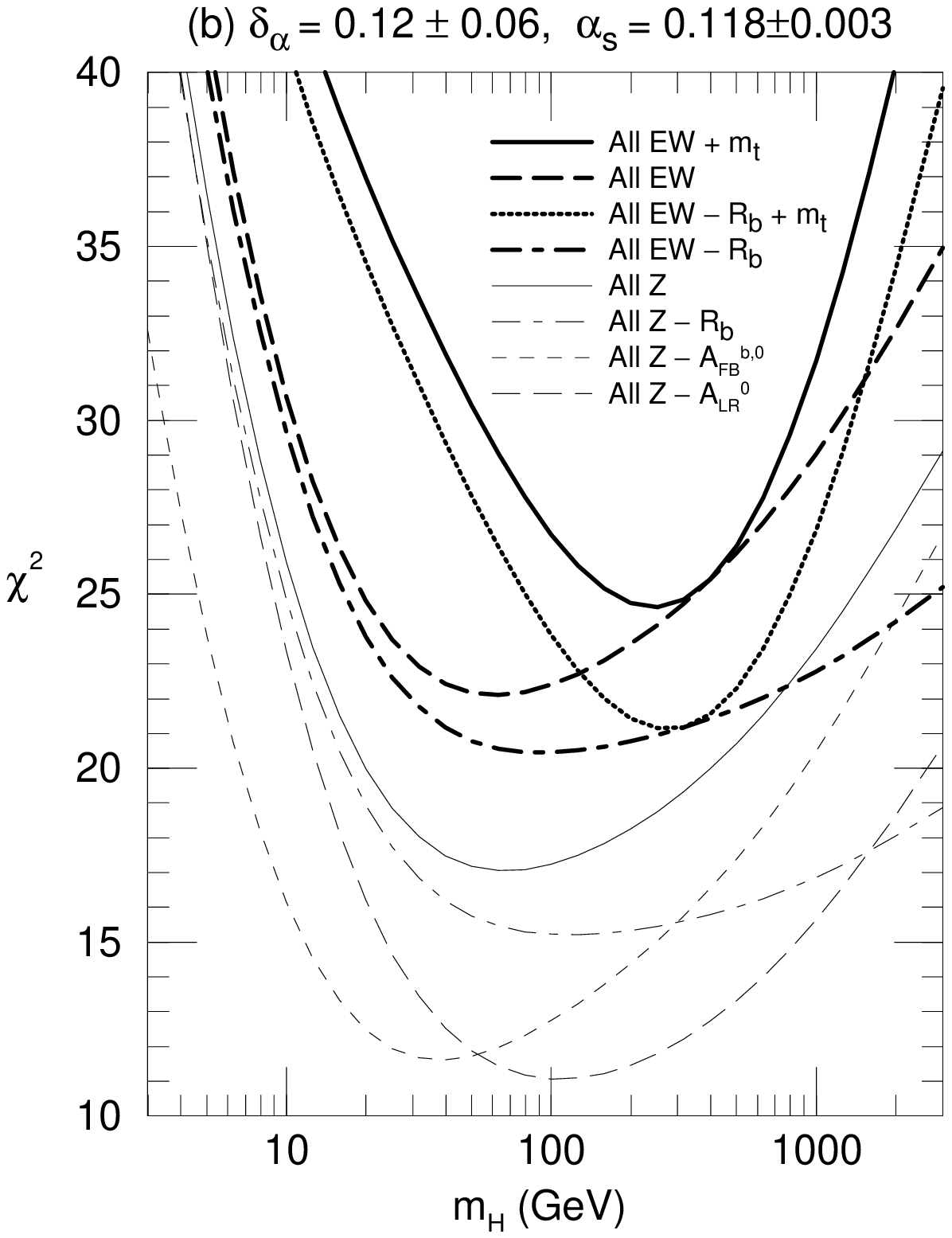,width=7cm,silent=0}
\end{center}
\caption{\protect\footnotesize\sl
Total $\chi^2$ of the SM fit to all the electroweak data
as functions of $\mh$ when $m_t$ is allowed to vary, 
with $\alpha_s(\mz)=0.118\pm 0.003$\cite{pdg96} 
for $\delta_\alpha=0.03\pm 0.09$\cite{eidjeg95} (a) 
and $\delta_\alpha=0.12\pm 0.06$\cite{mz94} (b).
Results for various sets of the electroweak data with 
or without the Tevatron $m_t$ data, $m_t=175\pm 6\gev$\cite{mt96} 
are given.
The degrees of freedom is 25 for `All EW $+m_t$' case.
}
\label{fig:chisqvsmh}
\end{figure}

For $\mh=60,300,1000\gev$, $\alpha_s=0.118\pm 0.03$ and $\delta_\alpha=0.03
\pm 0.09$, one obtains 
%
\fitofmtstandard
%
where the mean value is for $\mh=300$~GeV. 
The fit (\ref{mtfit_standard}) agrees with the best value from CDF and D0
\bea \label{mt_tevatron}
        m_t = 175 \pm 6~\gev.
\eea
This agreement strongly suggests that the electroweak theory respects  
the gauge invariance, since otherwise the quantum corrections could 
not be calculated.  
An elaboration on this point follows in the next subsection.  
Furthermore, the successful prediction of $\mt$ based on the 
simple SM radiative corrections strongly 
supports  the presence of the custodial SU(2) symmetry 
as part of physics responsible for spontaneous symmetry breaking.
Without custodial SU(2) 
there would have been no prediction for $\mt$.  
Furthermore, 
the mechanism that leads to the large mass splitting of the third 
generation quarks should give rise to 
a $T$ value which is similar 
to its standard model value.
Therefore, 
the success of the SM prediction not only 
suggests
the presence 
of the custodial SU(2) symmetry, but also constrains the mechanism 
of the fermion mass generation.  

Due to the quadratic form of Eq.~(\ref{total_chisqsm}) 
one can readily integrate out the unwanted terms, for instance
those containing $\alpha_s$ and/or $\delta_\alpha$, and render 
the result independent of them.
Also, additional constraints on the external parameters $\alpha_s$ and 
$\delta_\alpha$, such as those from future improved measurements or the 
constraint from the grand unification of these couplings may be added 
without difficulty. 
 
 As discussed above, the value for $\mh$ resulting from the Standard Model
 fit is correlated with the value of mt. The present value for $R_b$ which
 disfavors large masses of the top quark induces therefore a small value
 of the Higgs mass. It should also be noted that the choice of the value
 of $\alpha_s$ as an external parameter implies via 
 Eq.~(\ref{fitofgzbsbalpspdb}) a constraint on the  vertex form factor 
 $\delb$ and influences in turn the fit value for $\mh$. 
Shown in Fig.~\ref{fig:chisqvsmh} are the $\mh$-dependence 
of $\chi^2$ under various assumptions. 
We present in Table~6 the corresponding 95\%CL upper and lower bounds 
on $\mh$(GeV) from the electroweak data.
A low mass Higgs boson is clearly favored. 
However, this trend disappears for $\alpha_s=0.118\pm 0.003$\cite{pdg96}, 
once we ignore the $R_b$ data. 
If we adopt the estimate $\delta_\alpha=0.03\pm 0.09$\cite{eidjeg95}
for $\bar{\alpha}(\mmz)$, the 95\%CL upper bound on $\mh$ is $270\gev$
from all the $Z$ boson data, while it weakens to $1200\gev$,
if 
the $R_b$ data
are ignored. 
The corresponding upper bounds with the estimate
$\delta_\alpha=0.12\pm0.06$\cite{mz94} 
are $370\gev$ and $1900\gev$, respectively. 
An addition of the low energy neutral current data slightly 
lowers the upper $\mh$ bound, mainly because the combined fit, 
(\ref{fitoflencatmz}), 
gives slightly 
smaller $\gzbar^2(\mmz)$, i.e. smaller $T$, than the $Z$ parameters
would give alone. 
Just like smaller $R_b$ favors smaller $\mt$, smaller $T$ favors 
smaller $\mt$ and through the strong $\mt$ and $\mh$ correlation 
from the $\Gamma_Z$ and the asymmetry data smaller $\mh$ is favored. 
It is hence the direct measurement 
from 
the 
Tevatron, 
$m_t=175\pm 6\gev$, 
that essentially constrains the allowed $\mh$,
$\mh<480\gev$ for $\delta_\alpha=0.03\pm 0.12$ or 
$\mh<590\gev$ for $\delta_\alpha=0.09\pm 0.06$ at 95\%CL,  
given the $\Gamma_Z$ and the asymmetry constraint. 

\begin{table}[t]
\begin{center}
\caption{\protect\footnotesize\sl
 95\%CL upper and lower bounds of $\mh$(GeV) 
for $\alpha_s=0.118\pm 0.003$\cite{pdg96}, 
$\delta_\alpha=0.03\pm0.09$\cite{eidjeg95} or
$\delta_\alpha=0.12\pm0.06$\cite{mz94}. 
Results for various sets of the electroweak data with 
or without the Tevatron $m_t$ data, $m_t=175\pm 6\gev$\cite{mt96} 
are given.
} 
\vspace*{1mm}
{\footnotesize
\begin{tabular}{|l|cccc|cccc|}
\hline
& \multicolumn{8}{|c|}{ $\alpha_s=0.118 \pm 0.003 $ }
\\
\cline{2-9}
& \multicolumn{4}{|c|}{ $\delta_\alpha = 0.03 \pm 0.12 $ }
& \multicolumn{4}{|c|}{ $\delta_\alpha = 0.12 \pm 0.06 $ }
\\
\cline{2-9}
& best-fit 
& \begin{tabular}{c}lower\\ bound \end{tabular} 
& \begin{tabular}{c}upper\\ bound \end{tabular} 
& $\chi^2_{\rm min}$
& best-fit 
& \begin{tabular}{c}lower\\ bound \end{tabular} 
& \begin{tabular}{c}upper\\ bound \end{tabular} 
& $\chi^2_{\rm min}$\\
\hline
 EW$\,+\,m_t$             & 170& 46&  480& 24.9& 240&  87&  590& 24.6\\
 EW$\,+\,m_t\,-\,R_b$     & 200& 54&  550& 21.4& 280& 100&  670& 21.1\\
 EW                       &  51& 17&  270& 17.1&  67&  21&  370& 17.1\\
 EW$\,-\,R_b$             &  60& 17&  730& 20.4&  90&  22& 1200& 20.4\\
 $Z$                      &  51& 17&  270& 17.1&  67&  21&  370& 17.1\\
 $Z\,-\,R_b$              &  72& 18& 1200& 15.2& 120&  24& 1900& 15.2\\
 $Z\,-\,A_{\rm FB}^{b,0}$ &  30& 11&  140& 11.9&  38&  13&  200& 11.6\\
 $Z\,-\,A_{\rm LR}^0$     &  82& 23&  450& 10.9& 110&  29&  590& 11.1\\
\hline
\end{tabular}
}
\end{center}
\end{table}

\begin{table}[b]
\begin{center}
\caption{\protect\footnotesize\sl
 95\%CL upper and lower bounds of $\mh$(GeV) 
when $m_t$ is fixed externally. 
Two estimates of $\bar{\alpha}(\mmz)$ are examined, 
$\delta_\alpha=0.03\pm 0.09$\cite{eidjeg95} and 
$\delta_\alpha=0.12\pm 0.06$\cite{mz94}, 
for $\alpha_s=0.118\pm 0.003$\cite{pdg96}. 
} 
\vspace*{1mm}
{\footnotesize
\begin{tabular}{|c|cccc|cccc|}
\hline
& \multicolumn{8}{|c|}{ $\alpha_s=0.118 \pm 0.003 $ }
\\
\cline{2-9}
& \multicolumn{4}{|c|}{ $\delta_\alpha = 0.03 \pm 0.09 $ }
& \multicolumn{4}{|c|}{ $\delta_\alpha = 0.12 \pm 0.06 $ }
\\
\hline
$m_t$
& best-fit 
& \begin{tabular}{c}lower\\ bound \end{tabular} 
& \begin{tabular}{c}upper\\ bound \end{tabular} 
& $\chi^2_{\rm min}$
& best-fit 
& \begin{tabular}{c}lower\\ bound \end{tabular} 
& \begin{tabular}{c}upper\\ bound \end{tabular} 
& $\chi^2_{\rm min}$\\
\hline
 160 &   72 & 22  &  190 & 22.7 &  110 &  45 &  240 & 22.6\\
 165 &  110 & 33  &  260 & 23.4 &  150 &  67 &  320 & 23.3\\
 170 &  150 & 54  &  360 & 24.3 &  220 &  99 &  430 & 24.0\\
 175 &  220 & 87  &  490 & 25.3 &  300 & 140 &  580 & 24.9\\
 180 &  310 & 130 &  660 & 26.4 &  410 & 210 &  780 & 26.0\\
 185 &  430 & 200 &  900 & 27.6 &  550 & 290 & 1000 & 27.1\\
 190 &  590 & 280 & 1200 & 28.9 &  740 & 390 & 1400 & 28.5\\
\hline
\end{tabular}
}
\end{center}
\end{table}

\begin{figure}[t]
\begin{center}
  \leavevmode\psfig{file=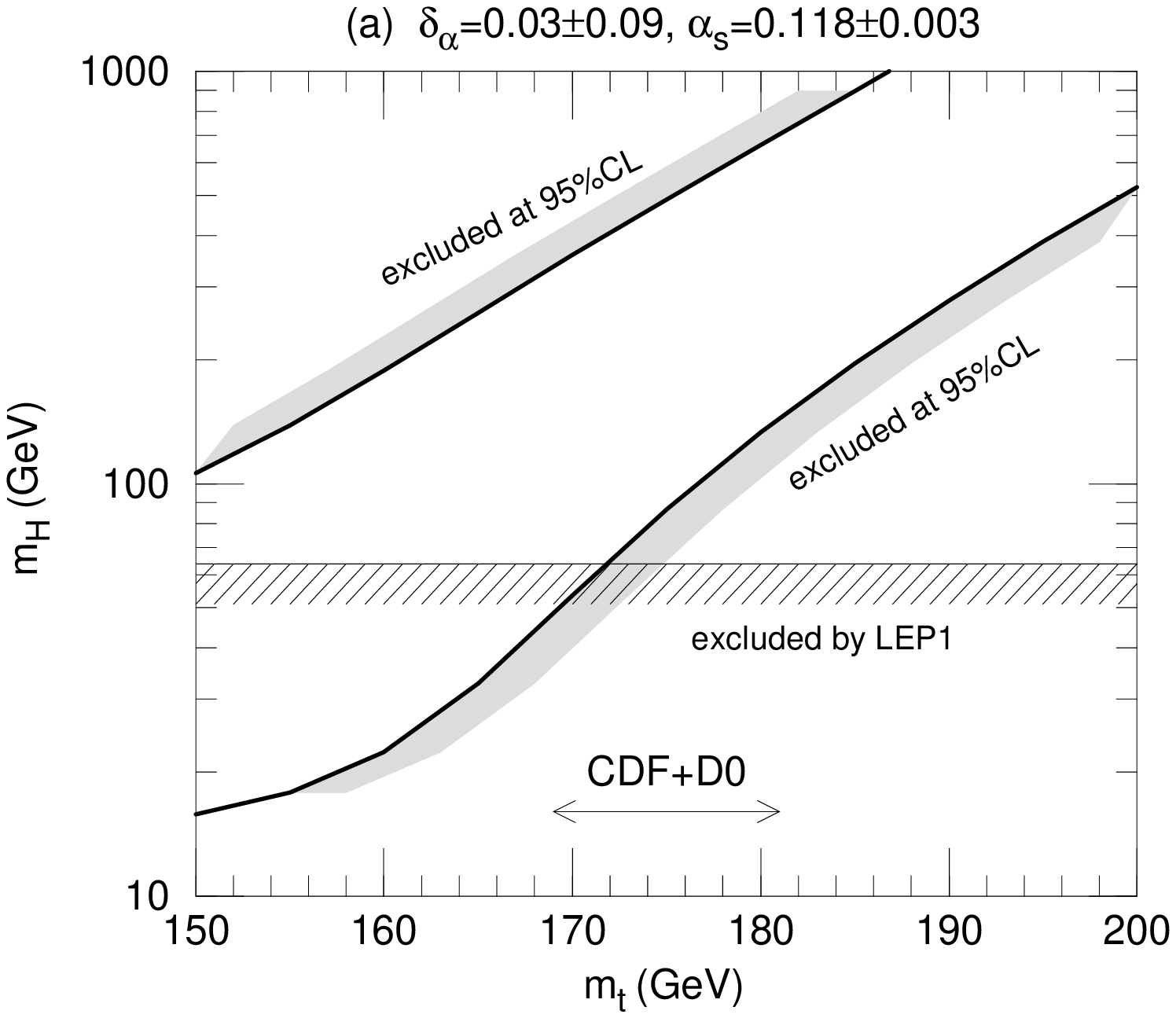,width=7cm,silent=0}
  \hfill
  \leavevmode\psfig{file=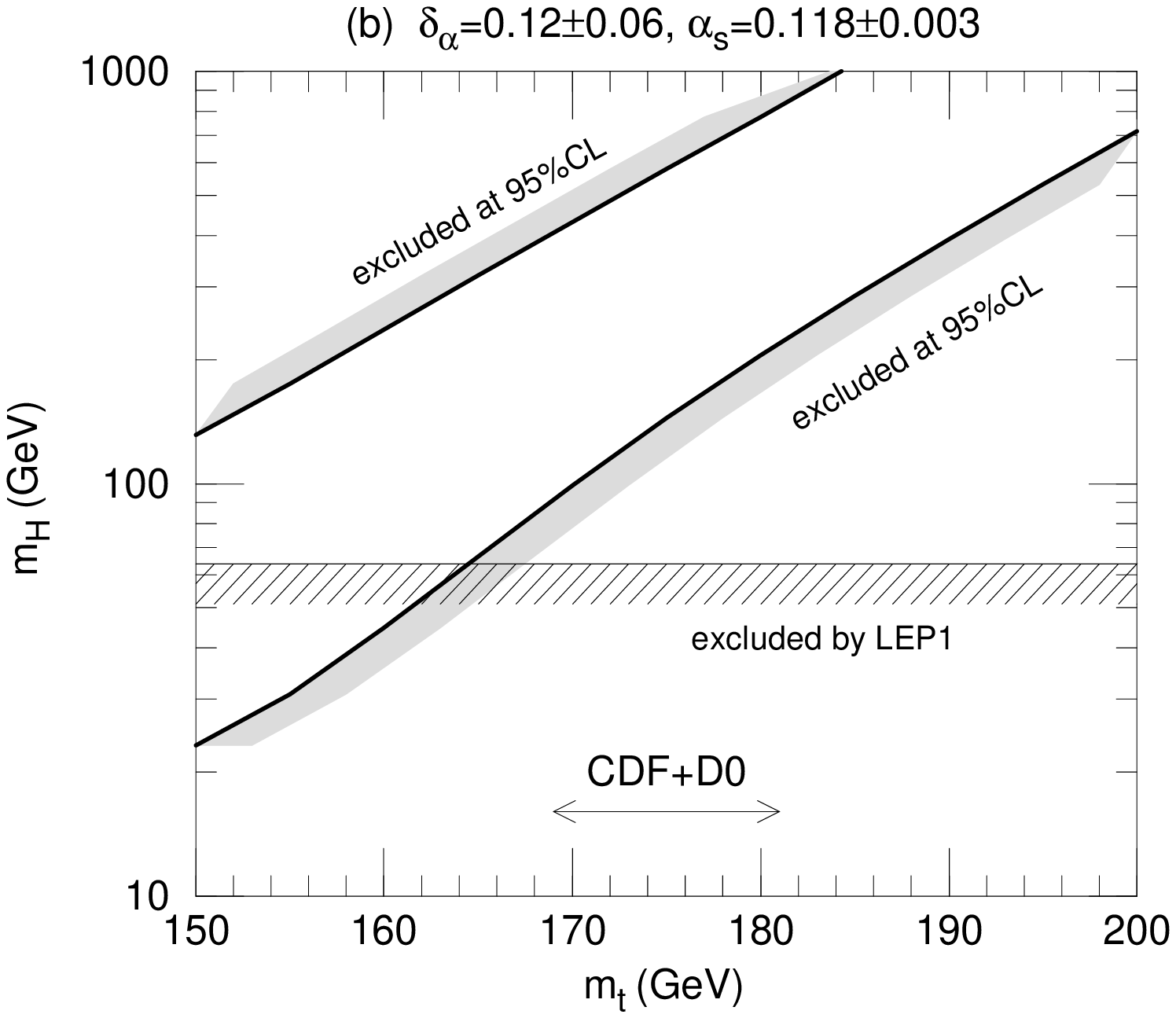,width=7cm,silent=0}
\end{center}
\caption{\protect\footnotesize\sl
Constraints on the Higgs mass in the SM from all the electroweak data. 
Upper and lower bounds of the Higgs mass at 95\%~CL are shown as functions of 
the top mass $m_t$, where $m_t$ is treated as an external parameter with 
negligible uncertainty. 
The results are shown for $\alpha_s=0.118\pm 0.003$\cite{pdg96} and
for $\delta_\alpha=0.03\pm 0.09$\cite{eidjeg95} (a) 
and $\delta_\alpha=0.12\pm 0.06$\cite{mz94} (b).
Also shown are the direct lower bound on $\mh$ from LEP1
and the Tevatron data $m_t=175\pm 6\gev$\cite{mt96}.
}
\label{fig:mhlimit}
\end{figure}

The constraints on $\mh$ become much stronger once the top 
quark mass is known precisely, either due to more precise 
measurements or due to deeper theoretical insight.
Lower and upper bounds on $\mh$ are 
shown in Fig.~\ref{fig:mhlimit} and in Table~7 as functions 
of $m_t$ for the two estimates of $\bar{\alpha}(\mmz)$. 
With the estimate of E-J\cite{eidjeg95}, $\delta_\alpha=0.03\pm0.12$, 
a lower $\mh$ is favored ($\mh<360\gev$ at 95\%CL), 
if $m_t < 170\gev$, while an intermediate $\mh$ is favored
($\mh>140\gev$ at 95\%CL) for $m_t>180\gev$. 
With the estimate of M-Z\cite{mz94}, $\delta_\alpha=0.12\pm 0.06$, 
similar constraints are found at about $5\gev$ smaller $m_t$. 
It is hence rather crucial for models where the Higgs boson is light 
(e.g. $\mh<130\gev$ in the MSSM\cite{mhmssm}) 
to have $m_t$ smaller than the actual 
present mean value, $m_t\sim 175\gev$.  

Finally, we repeat the 4-parameter fits, 
(\ref{fitofmtxhalpsda})--(\ref{fitofmtxhalpsdawithalpsdamz})
on the 
electroweak data when 
the direct 
$m_t$ measurement, $m_t=175\pm 6\gev$ (Tevatron), 
is taken into account.  
Without external constraints on $\alpha_s$ and $\da$, we find
\fitofmtxhalpsdawithmt
The top quark mass appears now basically determined by 
the direct measurement, while the mean $\delta_\alpha$ 
grows considerably and, consequently, a larger $\mh$, 
$\mh=530^{+1600}_{-170}\gev$ is favored. 
The shifted best value of $m_t$ slightly affects 
the sensitivity to $\alpha_s$ (see Eq.~(\ref{fitofmtxhalpsda})).
The value of $\alpha(\mmz)$ obtained from the electroweak measurements 
agrees roughly with that of Ref.\cite{langacker97}, which may be expressed as 
$\delta_\alpha = 0.21^{+0.25}_{-0.32}$.   

Adding the external constraint $\alpha_s=0.118 \pm 0.003$~\cite{pdg96} 
does not significantly alter the situation, 
because the fit (\ref{fitofmtxhalpsdawithmt}) results in 
the $\alpha_s$ value consistent with the world average.

\begin{figure}[t]
\begin{center}
  \leavevmode\psfig{file=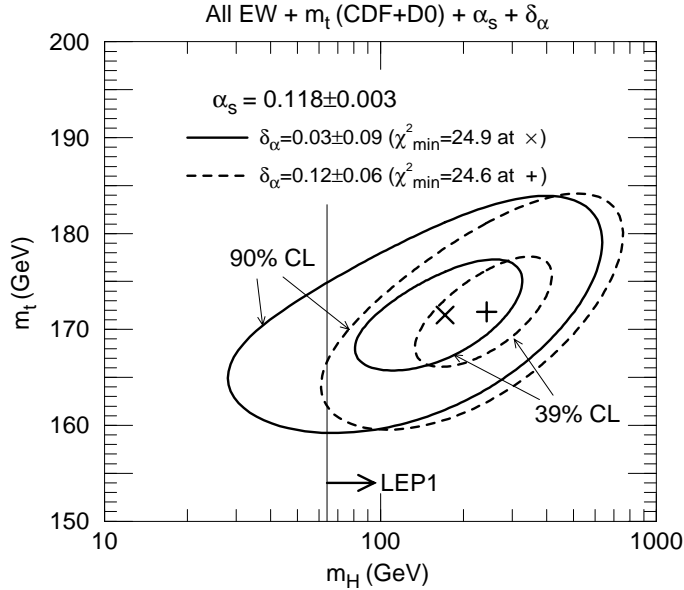,width=9cm,silent=0}
\end{center}
\caption{\protect\footnotesize\sl
The SM fit to all electroweak data in the ($\protect\mh,\,m_t$) 
plane with external constraints on $m_t$ from Tevatron, 
$m_t=175\pm 6 \gev$\cite{mt96}, $\alpha_s=0.118\pm 0.003$\protect\cite{pdg96}, 
and two estimates\cite{eidjeg95,mz94} for 
$\delta_\alpha=1/\bar{\alpha}(\mmz )-128.72$. 
The inner and outer contours correspond to 
$\Delta\chi^2=1$ ($\sim$ 39\%~CL), 
and $\Delta\chi^2=4.61$ ($\sim$ 90\%~CL), respectively. 
The minimum of $\chi^2$ is marked by the sign ``$\times$'' 
for the $\bar{\alpha}(\mmz)$ estimate of \cite{eidjeg95} 
and by the sign ``$+$'' 
for the estimate \cite{mz94}.
Also shown is the direct lower bound on $\mh$ from LEP1.
}
\label{fig:mtmh_withmt}
\end{figure}

Because the best-fit value of $\delta_\alpha$ in 
(\ref{fitofmtxhalpsdawithmt}) is slightly larger than the 
estimate $\delta_\alpha=0.03\pm 0.09$ by E-J\cite{eidjeg95}, 
the strong negative correlation between 
$\Delta\delta_\alpha$ and $\Delta\xh$ in 
(\ref{fitofmtxhalpsdawithmt}) 
entails a sizeably lower $\mh$:
\fitofmtxhalpsdawithmtalpsdaej
Since the four parameter fits are not fully elliptic, we show in 
Fig.~\ref{fig:mtmh_withmt} both the 1-$\sigma$ and 90\%CL allowed 
regions in the $(\mh,\, \mt)$ plane by solid contours.  
The 1-$\sigma$ preferred range $\mh=170^{+150}_{-90}\gev$ 
agrees roughly with the estimates of Refs.\cite{lepewwg96,ellis96,boer96}. 
Similar results with slightly larger $\mh$ are found, 
if we adopt the M-Z estimation\cite{mz94} 
$\delta_\alpha=0.12\pm 0.06$:
\fitofmtxhalpsdawithmtalpsdamz
The corresponding allowed ranges in the $(\mh,\, \mt)$ plane are 
given by dashed contours in Fig.~\ref{fig:mtmh_withmt}.  
The preferred $\mh$ range is now $\mh=240^{+180}_{-110}\gev$. 
We note here that once the external $m_t$ data is included in the
fit, the relative importance of the $R_b$ 
on the SM fit 
decreases. The above fit (\ref{fitofmtxhalpsdawithmtalpsdaej}) and 
(\ref{fitofmtxhalpsdawithmtalpsdamz})
are barely affected by excluding the $R_b$ 
data. 

The above exercises demonstrate well the overall consistency of 
the electroweak radiative corrections in the SM and emphasize at the
same time the importance of an improved $\bar{\alpha}(\mmz)$ estimate for 
constraining $\mh$ in fits based on electroweak precision experiments. 

\subsection{Is there already indirect evidence for the 
standard $W$ self-coupling?}
 
The success of the SM in describing all precision electroweak experiments at 
the quantum level may be taken as indirect evidence of the non-Abelian nature 
of the electroweak theory, or respectively of the standard universal 
gauge-boson self-couplings, because it is the non-Abelian gauge symmetry of 
the SM which ensures its renormalizability. 
  
Any alternative\cite{altmodels} to gauge models should necessarily have the 
new physics (cut-off) scale of order $\mw$, whereas the universality of the 
weak interactions may be associated with the underlying symmetry of the 
fundamental theory and the vector boson dominance which require relatively 
high ($\gg \mw$) scales for new physics. 
The fact that the SM works well at the quantum level indicates that 
the weak boson interactions do not deviate significantly from their 
gauge theory form at least up to the scale of $2\mt$. 
Therefore, it is instructive to study in more detail which part
of the standard radiative corrections is supported by experiment and 
whether indeed there is evidence for the gauge boson self-couplings.  
 
It is not straightforward to answer this question, since we have to 
identify which finite portion of the quantum corrections is sensitive 
to the weak-boson self-interactions. 
Usually one splits the complete SM radiative corrections into just 
two separately gauge invariant pieces, namely the fermionic loop 
contributions to the gauge-boson self-energies and the rest. 
It can then be stated unambiguously that neither of the corrections 
alone is sufficient to describe the data, and that only the inclusion 
of both contributions ensures the success of the SM radiative 
corrections\cite{sirlin94}. 
As a matter of fact, the bosonic part of the correction contains the 
weak boson self-interactions as an essential part and in this sense 
it is indirect evidence for universal couplings.
 
In a more detailed attempt\cite{kh_ucla95} at understanding the importance 
of bosonic contributions due to the $WWZ$ and $WW\gamma$ couplings, 
it should be elucidated to what extent these finite bosonic correction 
terms depend on the splitting of the gauge bosons into themselves. 
For instance, the box diagrams do not contain gauge-boson self-couplings. 
It is useful to split the bosonic corrections into three separately 
gauge-invariant pieces, namely `box-like', `vertex-like' and 
`propagator-like' pieces by appealing to the S-matrix pinch 
technique\cite{pinch}. 
It is then only the `vertex-like' and `propagator-like' pieces which 
contain the gauge boson self-couplings. 
Schematically we separate the SM radiative corrections into the 
following five pieces:
  \SetScale{0.2}
  \begin{eqnarray}
     \begin{array}{lll@{\quad\hspace{1mm}\quad}l}
      {\cal M}\quad=& 
       \quad \mbox{QED/QCD} & & (A)
       \\[0.3mm]
      &\mbox{+ fermion-loop} &
       \begin{picture}(40,20)(2,8)
         \CArc(120,50)(40,0,180)
         \CArc(120,50)(40,180,360)
         \Photon(20,50)(80,50){7}{3}   
         \Photon(160,50)(220,50){7}{3} 
       \end{picture}
       & (B)
       \\[0.3mm]
      &\mbox{+ box} &
       \begin{picture}(40,20)(-12,8)
         \Line(0,0)(0,100)
         \Line(100,0)(100,100)
         \Photon(0,25)(100,25){7}{5}
         \Photon(0,75)(100,75){7}{5}
       \end{picture}
       & (C)
       \\[0.3mm]
       &\mbox{+ vertex} & 
       \begin{picture}(40,20)(-12,8)
         \Line(30,0)(30,100)
         \PhotonArc(30,50)(30,90,270){7}{4}
         \Photon(30,50)(100,50){7}{4}
       \end{picture}
       +
       \begin{picture}(40,20)(-12,8)
         \Line(0,0)(0,100)
         \Photon(50,50)(100,50){7}{3}
         \Photon(0,20)(50,50){7}{3}
         \Photon(0,80)(50,50){7}{3}
       \end{picture}
       & (D)
       \\[0.3mm]
       &\mbox{+ bosonic-loop} &
       \begin{picture}(45,20)(2,8)
         \PhotonArc(120,50)(40,0,180){7}{6}
         \PhotonArc(120,50)(40,180,360){7}{6}
         \Photon(20,50)(80,50){7}{3}  \Vertex(80,50){1} 
         \Photon(160,50)(220,50){7}{3} \Vertex(160,50){1}
       \end{picture}
       +
       \begin{picture}(40,20)(5,8)
         \PhotonArc(120,50)(40,0,360){7}{11}
         \Photon(36,0)(211,0){7}{9}
       \end{picture}
       +
       \begin{picture}(45,15)(2,8)
         \DashCArc(120,50)(40,0,180){10}
         \PhotonArc(120,50)(40,180,360){7}{6}
         \Photon(20,50)(80,50){7}{3}  \Vertex(80,50){1} 
         \Photon(160,50)(220,50){7}{3} \Vertex(160,50){1}
       \end{picture}
     & (E)
     \end{array}\label{abcde}
  \end{eqnarray}
Details of this separation for each radiative correction term 
may be obtained straightforwardly from the analytic expressions 
presented in Ref.\cite{hhkm}.
By confronting these `predictions' with the electroweak data the results of 
Table~8 are obtained. 
 
The `no-EW' entry confronts the tree-level predictions of the SM where only 
QCD and external QED corrections $(A)$ are applied. 
In this column $\bar{\alpha}(m_Z^2)$ is calculated by including only 
contributions from light quarks and leptons with 
$\delta_h=0.03$\cite{hhkm,eidjeg95} for the hadronic uncertainty. 
It is quite striking to re-confirm the observation\cite{novikov93} 
that these `no-EW' predictions agree well with experiments at LEP1/SLC. 
In fact, it reduces the $\chi^2$ over the SM, partly because of the 
$R_b$ data, which prefer no electroweak corrections $\delb(\mmz )=0$ 
compared to the SM prediction $\delb(\mmz )=-0.00995$ for $\mt=175\gev$. 
It is only the $\mw$ value\cite{novikov94} and the $Z$ boson width 
which give significantly higher $\chi^2$ compared to the SM.

This can be understood as follows\cite{kh_ucla95}. 
The three most accurately constrained electroweak parameters are 
$\sbar^2(\mmz )$ from the asymmetries, $\gzbar^2(\mmz )$ from $\Gamma_Z$ 
at LEP1/SLC experiments, and $\mw$ from Tevatron experiments. 
In terms of the `observable' combinations $(S', T', U')$ of 
Eq.~(\ref{stu_prime}), 
they can be expressed as
\begin{subequations}
\label{approx}
\bea
\gzbar^2(0) &\approx& 0.5456    + 0.0040T'
\,\\
\sbar^2(\mmz ) &\approx& 0.2324 +0.0036S' -0.0024T'
\,\\
\label{approx_mw}
\mw (\gev ) &\approx& 79.84 -0.28S' +0.42T' +0.33U' -0.29(T'-T)\,.
\eea
\end{subequations}
In the absence of electroweak corrections, the predictions are obtained by 
setting $S=T=U=\delg=0$ and also by setting $\gzbar^2(\mmz )-\gzbar^2(0)=0$.  
The purely light flavor value of 
$1/\bar\alpha(\mmz )=1/\alpha(\mmz )_{\rm l.f.}=128.89$ (see Table 4)
corresponds to $\da=0.17$. 
These input values give rise to $(S',T',U')=(-0.12, 0.75, -0.04)$ 
which is not far from their SM values $(-0.23, 0.88, 0.36)$ for 
$\mt=175\gev$, $\mh=100\gev$ and $\delg=0.0055$, or from the $(S,T,U)$ fit 
result of Eq.~(\ref{fitofstu}).  
The `no-EW' case thus gives almost the same predictions for the three charge
form factors, $\gzbar^2(0)$, $\sbar^2(\mmz )$, and $\gwbar^2(0)$ with those 
of the SM. 
All the asymmetry data at LEP1/SLC are hence reproduced well. 
The low energy neutral current experiments are also reproduced well, 
since the running of the $\sbar^2(q^2)$ charge below the $\mz$ scale 
is essentially governed by the `QED' effects. 
The `no-EW' model predicts significantly smaller $\Gamma_Z$ by about 
3 to 4 $\sigma$ for $\alpha_s=0.118\pm 0.003$, because the running of 
the $\gzbar^2$ charge, $\gzbar^2(\mmz )-\gzbar^2(0)\sim 0.05$ 
has been ignored. 
Furthermore, it fails to predict the measured $\mw$-value
by about 3 $\sigma$, because its prediction is sensitive directly 
to the $\mu$ decay correction factor $\delg$ in Eq.~(\ref{gf}).  
This results in the last term in Eq.~(\ref{approx_mw}), 
$-0.29(T'-T)$, which lowers the $\mw$ prediction by more than 300~MeV.  

The next `${\rm + fermion}$' column\footnote{%
$\mh=100$~GeV is chosen to fix the negligible two-loop 
contributions in the `${\rm + fermion}$' and 
`${\rm + vertex}$' columns.}~gives the result of $(A)+(B)$ 
in Eq.~(\ref{abcde}). 
If we include only the fermionic corrections the $T$ parameter grows from 
zero to 1.14, while the factor $\delg$ remains zero. 
The combination $T'$ then becomes $T'=1.14+0.75=1.89$ which gives 
a too large $\gzbar^2(0)$ and a too small $\sbar^2(\mmz )$ as can be 
read off from Eq.~(\ref{approx}). 
The fermionic loop gives a dominant contribution to the running of 
$\gzbar^2$ below $\mz$, and the resulting 
$\gzbar^2(\mmz )\approx\gzbar^2(0)+0.005$ makes the $Z$ boson width 
unacceptably large. 
{}From Table~8, we find that about half of $\chi^2\sim 500$ in the 
`$+$fermion' entry comes from $\Gamma_Z$ and the rest from the 
LEP1/SLC asymmetries. 
In contrast, we find excellent agreement for $\mw$ in the same column. 
This is mainly because $\mw$ is more sensitive to $T$ rather than to 
$\delg$ when $\alpha$, $G_F$ and $\mz$ are fixed:  
$0.42T'-0.29(T'-T) = 0.42T+0.13(T'-T)$. 
Even though there are fortuitous cancellations among the remaining terms, 
we find no further improvement in the $\mw$ fit by adding extra 
radiative effects.    
 
It turned out that the `box-like' corrections to the $\mu$-decay matrix 
elements amount to almost 80\% of the total $\delg$ value: 
\bsub \label{dgsm}
  \bea
    [\delg]_{\rm SM} &=& \delgbox + \delgvertex \,,
\\ \label{dgsm_box}
    \delgbox &=& \frac{\gzhat^2}{16\pi^2}
    \Bigl( \frac{5}{2} -5\shat^2+\shat^4 \Bigr)
    \frac{\mmw}{\mmz-\mmw}\log\frac{\mmz}{\mmw} \,, 
\\ \label{dgsm_vertex}
    \delgvertex &=& \frac{\ghat^2\chat^2}{16\pi^2}
    \Bigl( 2-\frac{\mmz+\mmw}{\mmz-\mmw}\log\frac{\mmz}{\mmw} \Bigr)
    + \frac{\ehat^2}{8\pi^2} \,.
  \eea
\esub
Hence by adding the `box-like' corrections, $\delgbox =0.00429$, 
we have $(S',T',U')=(-0.20,1.30,-0.02)$, and the fit improves significantly.  
The $T'$ value is still slightly too large, and the `$+$box' entry 
still gives too large $\Gamma_Z$ and too small $\sbar^2(\mmz )$.  
This can be seen from the column of `${\rm + box}$',
where we give results of $(A)+(B)+(C)$ corrections in Eq.~(\ref{abcde}).

These electroweak effects do not affect much the fit of the low energy 
neutral current experiments because of their larger experimental errors. 
It is worth noting here that among the electroweak radiative 
corrections, the 'box-like' ones, especially the $WW$-box contribution,  
are most significant in the atomic parity violation experiments. 
Indeed the fit for $Q_W(C_s)$ improves significantly by adding 
the 'box-like' corrections.  

Up to this stage no contribution from quantum fluctuations with the 
weak-boson self-couplings are counted. 
Next the column `${\rm + vertex}$' is considered, where the results of 
A+B+C+D corrections are listed and where we may hope to see their effects. 
It turns out that the effects of the remaining 20\% correction to 
$\delg$ and the effects in part from the vertex corrections in the 
$Z$-decay matrix elements considerably reduce the $\chi^2$ in the LEP1/SLC
sector of the experiments from about 200 down to 30. 
The effect of the full $\delg$ is to change the charge form factor inputs to 
$(S',T',U')=(-0.20,1.14,-0.02)$, which reduces $\gzbar^2(\mmz )$ by only 
0.1\%, 
increases $\sbar^2(\mmz)$ by 0.2\%. 
The predicted $\Gamma_Z$ is reduced by 1\% and excellent agreement with 
the data is found (compare the relevant entries in the `$+$box' and 
`$+$vertex' entries).  
The major effect of the vertex corrections to $\Gamma_Z$ is actually coming 
from the corrections to the $Zff$ vertices in which the corrections from 
the diagrams with the $WWZ$ vertex, $\overline{\Gamma}^f_2$ in Table~3 of 
Ref.~\cite{hhkm}, are 
most significant. 
The prediction $\sbar^2(\mmz )=0.22995$ 
is still by about 3-$\sigma$ away from the fit 
(\ref{fitofsb96withleptonuniversality}).  
 
Inclusion of the `propagator-like' corrections either improves or worsens the 
fit depending on the Higgs boson mass. 
The improvement is sizeable only when the Higgs boson mass is not too large, 
as can be seen from the last column in Table~8.  

It is therefore tempting to conclude that the effect of the `vertex-like' 
corrections, and hence that of the standard $WWV$ self-interactions is 
essential for the success of the SM at the quantum correction level.
Once the gauge invariance of the weak boson interactions is assumed, 
quantum fluctuations at very short distances become universal and hence 
they can be renormalized by precisely measured quantities.  
Remaining finite parts of the quantum corrections hence measure the 
effects of the intermediate scale physics which can be sensitive to 
the symmetry breaking physics.   
With further improvement of the electroweak data, we will therefore learn 
more about physics of 100~GeV to 1~TeV that could affect these finite 
correction terms.  
The precision electroweak physics may still give us hints of new 
physics at the energy region which is not yet explored directly by 
high energy experiments. 


\begin{table}[t]
\begin{center}
\caption{\protect\footnotesize\sl
The electroweak data and the SM predictions.
The three predictions for $\Gamma_Z$, $\sigma_h^0$ and $R_\ell$
are for $\alpha_s=0.115$, 0.118 and 0.121.
}
 \def\equnit{$\times 10^{-5}$}       
 \def\afb{A_{\rm FB}}                
 \def\non{$\;\;$------}              
{ \scriptsize
 \begin{tabular}{|l|r|l|l|l|l|l@{\,\,}l@{\,\,}l@{\,}|}
 \hline                              
  & data \hspace{7mm} & no-EW &      
  +fermion  & +box& +vertex &        
  \multicolumn{3}{|c|}{+propagator}\\
 \hline
         $m_t$ (GeV)   &               
& \non  &    175 &    175 &    175 &    175 &    175 &    175\\
         $m_H$ (GeV)   &               
& \non  &    100 &   \non &    100 &     60 &    300 &   1000\\
 \hline
$S$                    &               
 &  \non  & -0.067 & -0.067 & -0.067 & -0.283 & -0.146 & -0.075\\
$T$                    &               
 &  \non  &  1.136 &  1.136 &  1.136 &  0.910 &  0.762 &  0.583\\
$U$                    &               
 &  \non  &  0.017 &  0.017 &  0.017 &  0.364 &  0.359 &  0.358\\
$\bar{\delta}_G$       &               
 &  \non  &  \non  &0.00429 &0.00549 &0.00549 &0.00549 &0.00549\\
$1/\bar{\alpha}(m_Z^2)$&               
 & 128.89 & 128.90 & 128.90 & 128.90 & 128.75 & 128.75 & 128.75\\
$\bar{s}^2(m_Z^2)$     &               
 &0.23114 &0.22815 &0.22955 &0.22995 &0.23009 &0.23094 &0.23163\\
$\bar{g}_Z^2(m_Z^2)$   &               
 &0.54863 &0.55812 &0.55569 &0.55502 &0.55639 &0.55592 &0.55518\\
$\bar{\delta}_b(m_Z^2)$&               
& \non & \non  & \non  &-0.00996 &-0.00997 &-0.00994 &-0.01000\\
$\bar{s}^2(0)$         &               
 &0.23866 &0.23584 &0.23716 &0.23753 &0.23850 &0.23930 &0.23995\\
$\bar{g}_Z^2(0)$       &               
 &0.54863 &0.55321 &0.55083 &0.55017 &0.54926 &0.54867 &0.54795\\
$\bar{g}_W^2(0)$       &               
 &0.42182 &0.42713 &0.42452 &0.42379 &0.42449 &0.42339 &0.42238\\
 \hline
$\Gamma_Z$(GeV) &  2.4946 $\pm$ 0.0027 
 & 2.4836 & 2.5346 & 2.5198 & 2.4905 & 2.4963 & 2.4920 & 2.4868\\
                &                      
 & 2.4853 & 2.5364 & 2.5215 & 2.4922 & 2.4980 & 2.4937 & 2.4885\\
                &                      
 & 2.4870 & 2.5381 & 2.5233 & 2.4939 & 2.4997 & 2.4953 & 2.4902\\
$\sigma_h^0$(nb)&  41.508 $\pm$  0.056 
 & 41.507 & 41.500 & 41.502 & 41.489 & 41.490 & 41.493 & 41.496\\
                &                      
 & 41.491 & 41.484 & 41.486 & 41.473 & 41.474 & 41.477 & 41.481\\
                &                      
 & 41.475 & 41.468 & 41.470 & 41.457 & 41.458 & 41.461 & 41.465\\
$R_\ell$        &  20.778 $\pm$  0.029 
 & 20.768 & 20.817 & 20.795 & 20.733 & 20.731 & 20.716 & 20.703\\
                &                      
 & 20.788 & 20.837 & 20.815 & 20.753 & 20.751 & 20.736 & 20.723\\
                &                      
 & 20.808 & 20.857 & 20.835 & 20.773 & 20.771 & 20.756 & 20.743\\
$\afb^{0,\ell}$ &  0.0174 $\pm$ 0.0010 
 & 0.0169 & 0.0224 & 0.0198 & 0.0175 & 0.0172 & 0.0157 & 0.0145\\
$A_\tau$        &  0.1401 $\pm$ 0.0067 
 & 0.1500 & 0.1732 & 0.1624 & 0.1516 & 0.1505 & 0.1439 & 0.1384\\
$A_e$           &  0.1382 $\pm$ 0.0076 
 & 0.1500 & 0.1732 & 0.1624 & 0.1516 & 0.1505 & 0.1439 & 0.1384\\
$R_b$           &  0.2178 $\pm$ 0.0011 
 & 0.2182 & 0.2181 & 0.2182 & 0.2156 & 0.2156 & 0.2157 & 0.2157\\
$R_c$           &  0.1715 $\pm$ 0.0056 
 & 0.1717 & 0.1719 & 0.1718 & 0.1722 & 0.1722 & 0.1721 & 0.1721\\
$\afb^{0,b}$    &  0.0979 $\pm$ 0.0023 
 & 0.1054 & 0.1219 & 0.1142 & 0.1064 & 0.1056 & 0.1008 & 0.0969\\
$\afb^{0,c}$    &  0.0735 $\pm$ 0.0048 
 & 0.0753 & 0.0883 & 0.0822 & 0.0764 & 0.0758 & 0.0721 & 0.0691\\
  $\sin^2\theta^{lept}_{eff}(\langle Q_{\rm FB}\rangle)$
&  0.2320 $\pm$ 0.0010 
 & 0.2311 & 0.2282 & 0.2296 & 0.2309 & 0.2311 & 0.2319 & 0.2326\\
$A_{\rm LR}$    &  0.1542 $\pm$ 0.0037 
 & 0.1500 & 0.1732 & 0.1624 & 0.1516 & 0.1505 & 0.1438 & 0.1384\\
$A_b({\rm LR})$ &   0.863 $\pm$  0.049 
 &  0.936 &  0.938 &  0.937 &  0.935 &  0.935 &  0.934 &  0.934\\
$A_c({\rm LR})$ &   0.625 $\pm$  0.084 
 &  0.669 &  0.679 &  0.674 &  0.670 &  0.669 &  0.666 &  0.664\\
    $\chi^2$        &$(\alpha_s=0.115)\quad $
 &   37.1 &  455.3 &  181.6 &   32.3 &   27.3 &   24.9 &   49.5\\
    (d.o.f.=14)     &$(\alpha_s=0.118)\quad $
 &   32.4 &  475.3 &  194.0 &   29.0 &   26.5 &   21.6 &   43.4\\
                    &$(\alpha_s=0.121)\quad $
 &   29.8 &  497.2 &  208.4 &   27.7 &   27.6 &   20.2 &   39.2\\
 \hline
$g_L^2$         &  0.2980 $\pm$ 0.0044 
 & 0.2955 & 0.3027 & 0.3049 & 0.3067 & 0.3049 & 0.3036 & 0.3024\\
$g_R^2$         &  0.0307 $\pm$ 0.0047 
 & 0.0309 & 0.0307 & 0.0307 & 0.0298 & 0.0300 & 0.0301 & 0.0302\\
$\delta_L^2$    & -0.0589 $\pm$ 0.0237 
 &-0.0601 &-0.0606 &-0.0652 &-0.0645 &-0.0645 &-0.0645 &-0.0644\\
$\delta_R^2$    &  0.0206 $\pm$ 0.0160 
 & 0.0186 & 0.0184 & 0.0184 & 0.0179 & 0.0180 & 0.0180 & 0.0181\\
$\chi^2 $       &                      
 &    0.4 &    1.8 &    3.9 &    5.5 &    3.5 &    2.4 &    1.5\\
 \hline
$K$ (CCFR)      &  0.5626 $\pm$ 0.0060 
 & 0.5519 & 0.5641 & 0.5685 & 0.5702 & 0.5673 & 0.5653 & 0.5632\\
$\chi^2 $       &                      
 &    3.2 &    0.1 &    1.0 &    1.6 &    0.6 &    0.2 &    0.0\\
 \hline
$s^2_{eff}$     &   0.233 $\pm$  0.008 
 &  0.239 &  0.236 &  0.235 &  0.229 &  0.230 &  0.231 &  0.231\\
$\rho_{eff}$    &   1.007 $\pm$  0.028 
 &  1.000 &  1.008 &  1.016 &  1.015 &  1.013 &  1.012 &  1.011\\
$\chi^2$        &                      
 &    0.6 &    0.1 &    0.1 &    0.4 &    0.2 &    0.1 &    0.1\\
 \hline
$Q_W$           &  -71.04 $\pm$   1.81 
 & -74.73 & -74.74 & -72.96 & -72.92 & -73.01 & -73.10 & -73.14\\
$\chi^2$        &                      
 &    4.2 &    4.2 &    1.1 &    1.1 &    1.2 &    1.3 &    1.3\\
 \hline
$2C_{1u}-C_{1d}$&   0.938 $\pm$  0.264 
 &  0.709 &  0.725 &  0.730 &  0.729 &  0.724 &  0.721 &  0.718\\
$2C_{2u}-C_{2d}$&  -0.659 $\pm$  1.228 
 &  0.081 &  0.099 &  0.103 &  0.112 &  0.106 &  0.101 &  0.097\\
$\chi^2$        &                      
 &    1.9 &    1.2 &    1.1 &    1.1 &    1.2 &    1.4 &    1.5\\
 \hline
$m_W$ (GeV)     &  80.356 $\pm$  0.125 
 & 79.957 & 80.459 & 80.384 & 80.363 & 80.429 & 80.325 & 80.229\\
$\chi^2$        &                      
 &   10.2 &    0.7 &    0.0 &    0.0 &    0.3 &    0.1 &    1.0\\
 \hline
  $\chi^2_{\rm tot}$&$(\alpha_s=0.115)\quad $
 &   57.6 &  463.3 &  188.8 &   41.9 &   34.4 &   30.3 &   54.9\\
  (d.o.f.=25)       &$(\alpha_s=0.118)\quad $
 &   52.9 &  483.3 &  201.3 &   38.6 &   33.6 &   27.0 &   48.8\\
                    &$(\alpha_s=0.121)\quad $
 &   50.3 &  505.2 &  215.7 &   37.3 &   34.7 &   25.7 &   44.7\\
 \hline
 \end{tabular}
}
\end{center}
\end{table}
%


\section{Impact of future improved measurements}
 
Constraints on various electroweak quantities are expected to be 
improved in the near future.  
Their impact on the knowledge of the top and the Higgs masses 
is discussed in the following subsections. 
The last subsection deals with 
future constraints on the 
$S$, $T$, $U$ parameters. 

According to the discussions in the previous section
the constraints on the top and the Higgs masses are basically 
obtained from three quantities,
$\Gamma_Z$, $\sbar^2(\mmz )$ from various $Z$-pole asymmetries, 
and $\mw$.  
After the completion of the LEP1 experiments no further
improvement on $\Gamma_Z$ is expected.
Significant improvements in $\sbar^2(\mmz )$ may be expected from 
SLC, Tevatron, LHC, and future Linear $\epem$ Colliders (LC).  
Improved measurements on $\mw$ are also expected from LEP2, 
Tevatron, LHC, and LC.  
The top quark mass will be measured accurately at Tevatron, 
LHC, and LC.  
The Higgs boson mass can be measured at LEP2, LHC and LC, provided
it exists and its mass lies in the accessible energy range of
these machines.
Finally, a more precise value of $\bar{\alpha}(\mmz )$ will be 
obtained from experiments at Novosibirsk, DA$\Phi$NE, B factories 
at KEK, SLAC and DESY, 
and possibly at the Beijing $\tau$-charm factory (BTCF).  

In order to assess the impact of
such future improvements in the 
electroweak 
sector, 
we found the following approximate 
formulae for the SM predictions useful:
%
%
%
\begin{subequations}\label{gzsb2mw_para}
\begin{eqnarray}
  \Gamma_Z(\mev ) &\approx &
      2497.1 +(2.51-0.01\,\xh)\, x_t
             -2.29\,\xh -0.65\,x_H^2
   \nonumber
   \\&& \hphantom{2497.1}
		+0.6\,x_\alpha +1.6\,x_s \,,
   \label{gammaz_approx}
   \\
   \sbar^2(\mmz) &\approx&
      0.23034 -(0.000335-0.000001\,\xh)\, x_t
      +0.000518\,\xh+0.000017\,x_H^2 \quad
   \nonumber
   \\&& \hphantom{0.23034}
		-0.00023\,x_\alpha +0.00001\,x_s \,,
	\label{sb2_approx}
   \\
    \mw (\gev ) &\approx&
       80.400 +(0.0635-0.0001\,\xh)\,x_t 
      -0.0603\,\xh -0.0062\,x_H^2 
   \nonumber
   \\&& \hphantom{80.400}
		+0.012\,x_\alpha -0.002\,x_s \,.
	\label{mw_approx}
\end{eqnarray}
\end{subequations}
%
Here $x_t = (\mt -175\gev )/10\gev$, $\xh = \log(\mh/100\gev )$, 
$x_\alpha = (\da -0.03)/0.09$, and $x_s = (\alpha_s-0.118)/0.003$.  
The approximations are valid to 0.2~MeV (\ref{gammaz_approx}), 
0.00002 (\ref{sb2_approx}) and 0.003~GeV(\ref{mw_approx}), respectively, 
in the region 
$|\xt|<1$, $|\xa|<1$, $|\xs|<1$ and 70~GeV$<\mh<$700~GeV.  
It is instructive to 
recall
that $\Gamma_Z$ and $\mw$ measure 
approximately the same combination of $\mt$, $\mh$ and $\da$: 
$x_t -0.9\,\xh +0.2\,x_\alpha$.  
The asymmetry parameter $\sbar^2(\mmz )$ constrains a 
different combination, which may be approximated as 
$x_t -1.5\,\xh +0.7\,x_\alpha$.
Therefore, we need improvements in both $\mw$ and $\sbar^2(\mmz )$ 
to reduce the electroweak constraints on $\mt$ and $\mh$.  
With the improved direct determination of $\mt$, each of the above 
experiments will lead to a significantly better constraint on $\mh$.  
The $\mh$ constraint can be 
strengthened by more precise estimates of
$\bar{\alpha}(\mmz )$.

\subsection{Asymmetries}
%
\begin{figure}[b]
\begin{center}
 \leavevmode\psfig{file=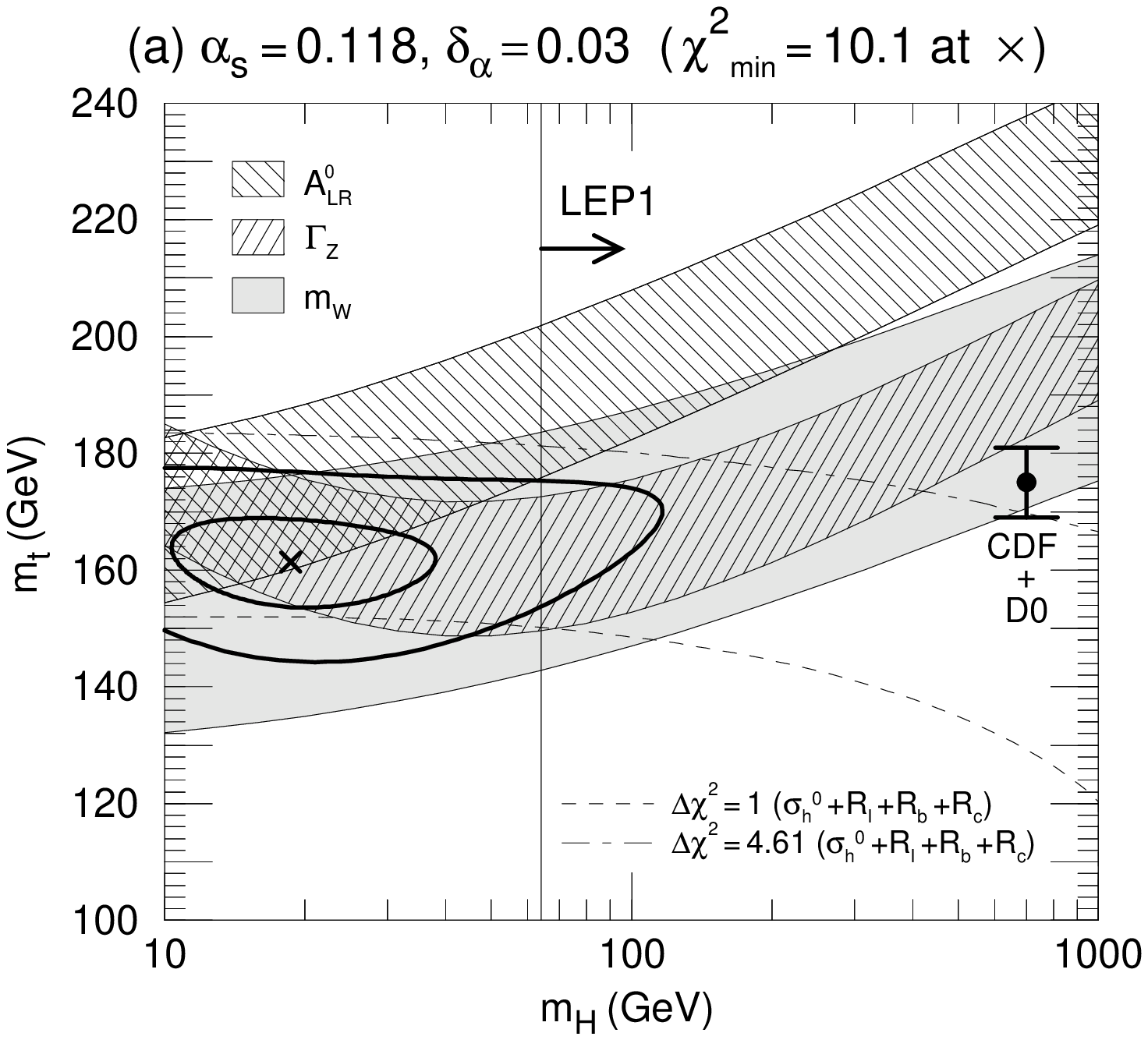,width=7cm,silent=0}
 \leavevmode\psfig{file=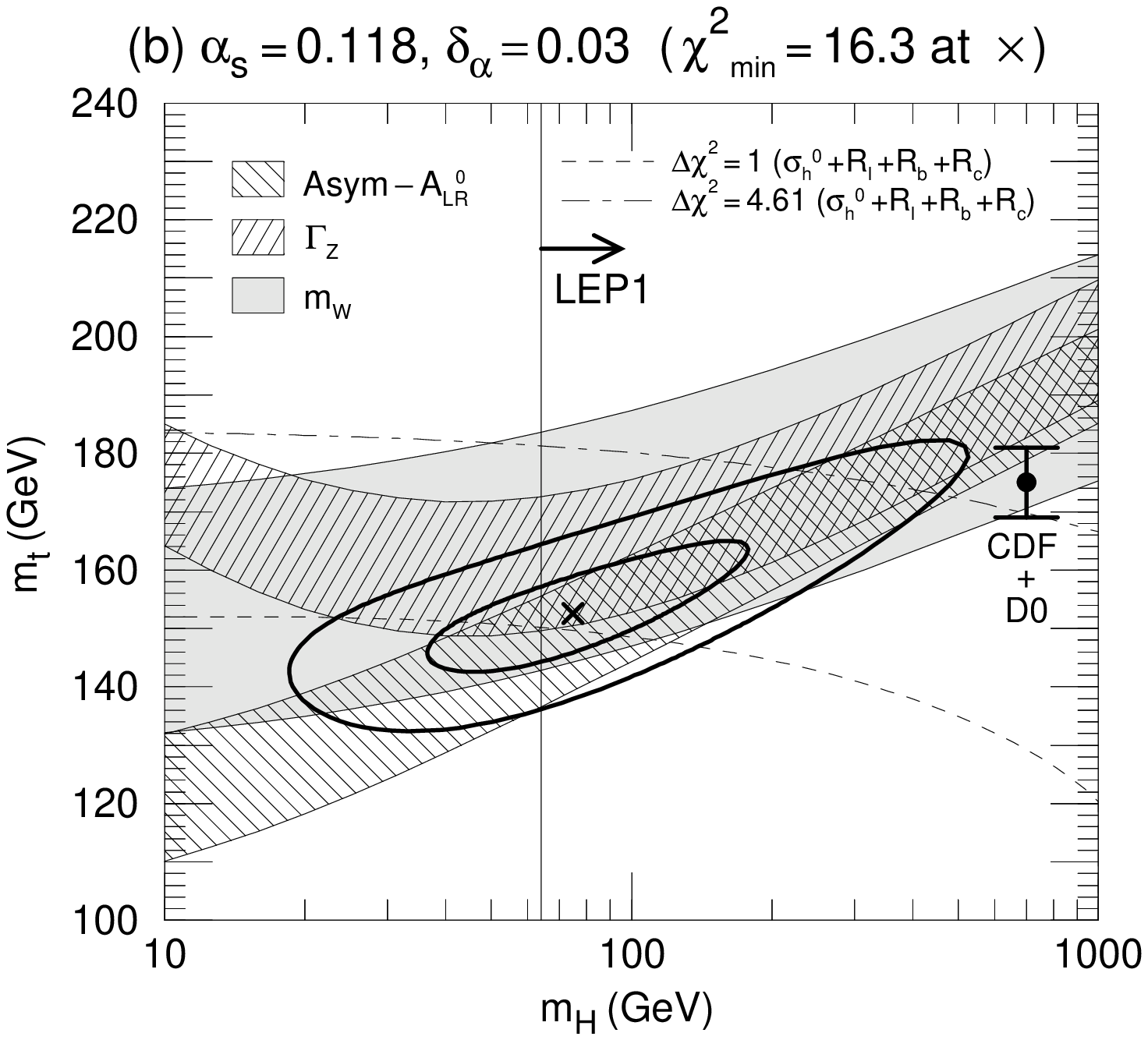,width=7cm,silent=0}
\end{center}
\caption{\protect\footnotesize\sl
SM-fit using (a) only the left-right asymmetry by the SLD Collaboration and
(b) using all other asymmetry data.
}
\label{fig:aslr}
\end{figure}
%
The different asymmetry measurements from LEP and SLC are in 
agreement with each other, 
although showing a large dispersion. 
The SLD collaboration has contributed the most precise individual 
determination, namely 
$\sbar^2(\mmz) = 0.2294 \pm 0.0005$. 
The result is dominated by statistics and thus allows for 
substantial improvement. 
The
average of all other measurements yields 
$\sbar^2(\mmz)= 0.2317 \pm 0.0003$. 
It is 
instructive to repeat the above SM fit once with $A_{LR}^0$ alone and then 
with all other asymmetry data. Figure~\ref{fig:aslr} shows the result in the
$(m_H,m_t)$-plane. 
Due to the somewhat high value of $A_{LR}^0$, i.e. low value of  
$\sbar^2(\mmz)$, the best-fit value for the Higgs mass turns out to 
be rather low and most of the allowed region is already excluded by 
the 
result of the Higgs searches at LEP1. 
The complementary fit leads to a best-fit Higgs mass of about 75~GeV, 
but with a low value for the top quark mass of 152~GeV. 
The 90\% CL allowed region 
overlaps significantly 
with the direct information on $\mh$ and 
$m_t$. 
The change in size and orientation of the error ellipses can be 
understood by considering the SM grid in Fig.~\ref{fig:gzb2sb2}. 
%

\begin{figure}[t]
\begin{center}
 \leavevmode\psfig{file=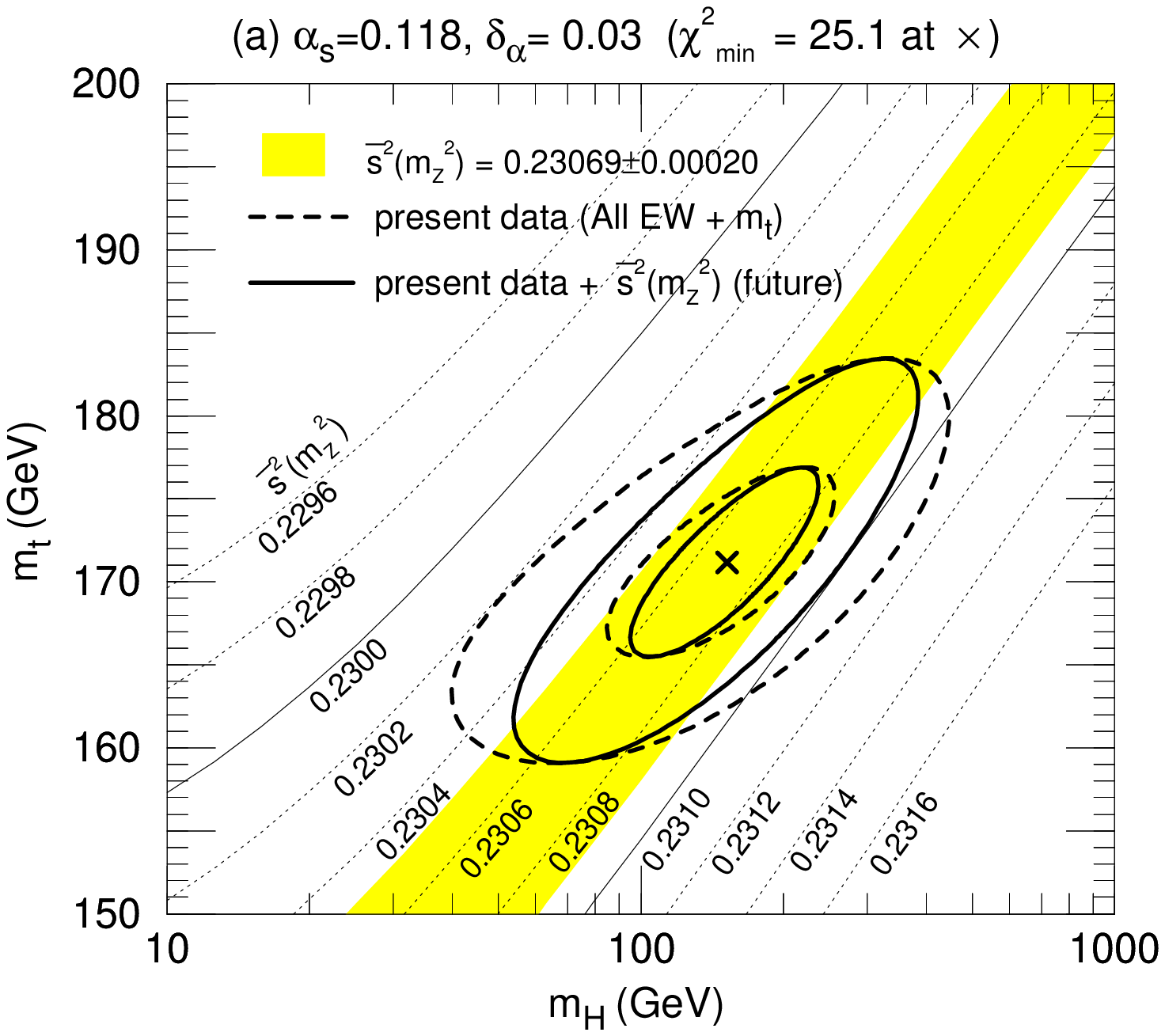,width=7cm,silent=0}
 \leavevmode\psfig{file=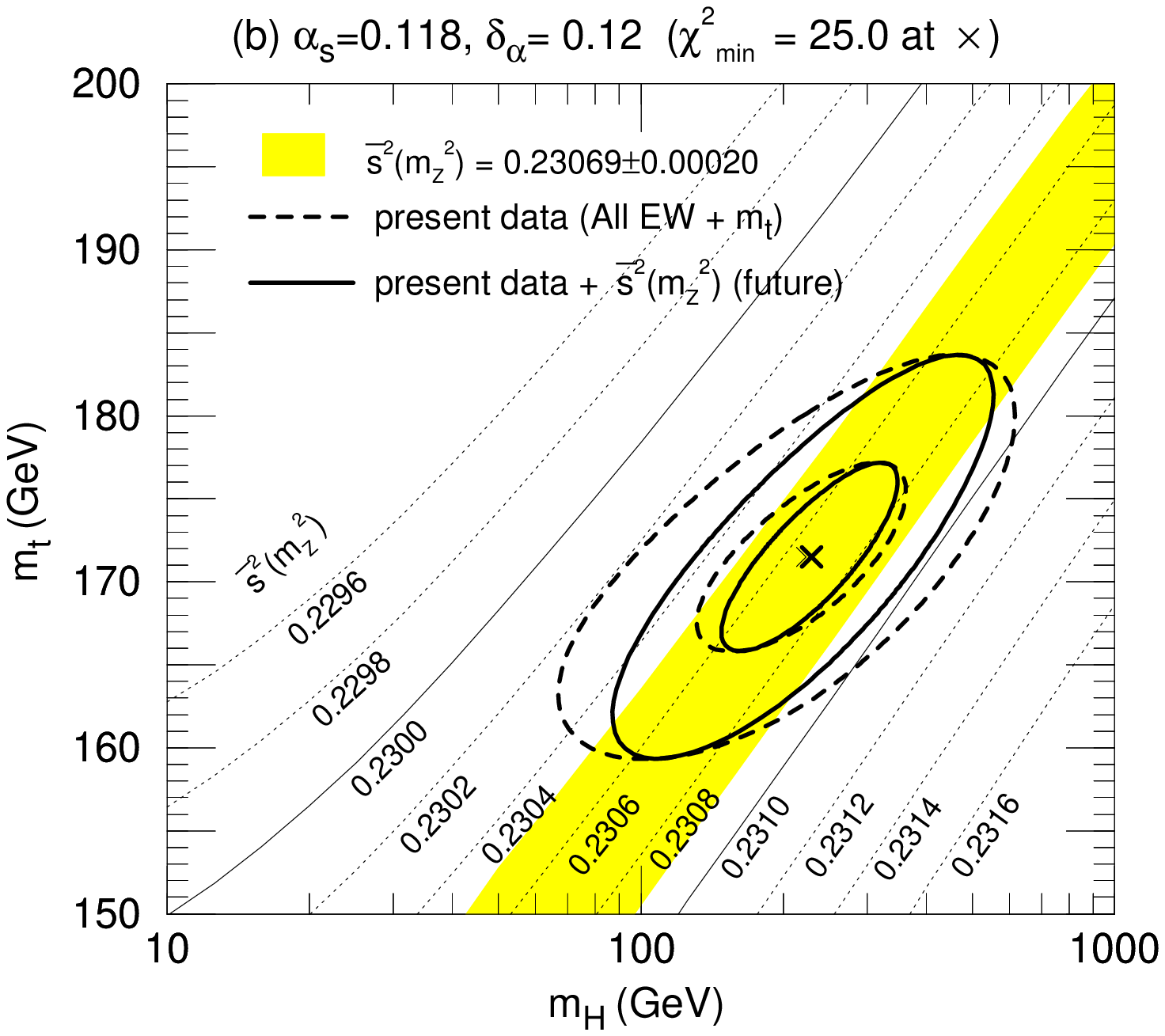,width=7cm,silent=0}
\end{center}
\caption{\protect\footnotesize\sl
Impact of future improvement in $\sbar^2(\mmz)$ 
in the ($\protect\mh,\,m_t$) plane, for $\alpha_s=0.118$ 
and $\delta_\alpha=0.03$(a), $\delta_\alpha=0.12$(b). 
An assumed future data $\sbar^2(\mmz)=0.23069\pm0.00020$ 
is used to constrain $m_t$ and $\mh$ in addition to 
the present all electroweak data and the Tevatron $m_t$ data, 
$m_t=175\pm 6 \gev$. 
The inner and outer contours correspond to 
$\Delta\chi^2=1$ ($\sim$ 39\% CL), 
and $\Delta\chi^2=4.61$ ($\sim$ 90\%~CL), respectively. 
The minimum of $\chi^2$ is marked by the sign ``$\times$''. 
Thin dotted/solid lines show the SM predictions for $\sbar^2$ 
when $m_t$ and $\mh$ are given.
}
\label{fig:mtmh_future_sb2}
\end{figure}
 
Until the start-up of the B-factory the SLD Collaboration hopes to 
increase their statistics with polarized beams (${\rm P}_e \sim 77\%$) to 
500k $Z$-decays, which would allow them to reduce the uncertainty 
on $A_{LR}^0$ by a factor of two, without yet hitting the limit set 
by the systematic error\cite{rowson96}. 
Such a measurement would determine $\sbar^2(\mmz)$ to about 
$\pm 0.00023$, i.e. one single experiment is reaching then the same 
precision as presently all experiments together. 
With 1M $Z$-events, the error can be reduced to $\pm 0.00015$.   
It is clear that a reproduction of the existing mean value with 
a significantly reduced error would cause a conflict with the other 
measurements and would put 
in question the interpretation within the SM.
 
At hadron colliders the measurement of the lepton forward-backward 
asymmetries allows to derive also precise values of the weak angle. 
In the Snowmass'96 report Baur and Demarteau\cite{baur96} estimate 
that an uncertainty of 0.00013 can be expected for an integrated 
luminosity of 30 fb$^{-1}$ at Tevatron\cite{tev33,tev2000}.  
The LHC experiments may not improve this further\cite{baur96} 
without significantly 
extending the rapidity coverage of their lepton detector. 

The error of $\sbar^2(\mmz )$ may further be reduced at a future 
linear $\epem$ collider (LC) by measuring the beam polarization 
asymmetries on the $Z$ pole, if a significantly improved determination 
of the electron beam polarization is achieved\cite{jlc90}.  
It should be emphasised here that the present uncertainty of 
0.00023 in theoretical predictions of $\sbar^2(\mmz )$ due to 
the uncertainty in $\bar{\alpha}(\mmz )$ should not discourage 
further attempts to improve its measurement, because we 
anticipate a significant improvement in the $\bar{\alpha}(\mmz )$ 
estimate and also because it leads to a severe constraint on 
new physics 
independent of $\bar{\alpha}(\mmz)$. 
Within the SM, precise measurements of $\sbar^2(\mmz )$ and $\mw$ 
will reduce the allowed region of $\mt$ and $\mh$ even without 
improving the $\bar{\alpha}(\mmz )$ estimate, because they 
depend on different combinations of these parameters; 
see Eq.~(\ref{gzsb2mw_para}).

We show in Fig.~\ref{fig:mtmh_future_sb2} the impact of future 
improvement in $\sbar^2(\mmz)$ in the ($\protect\mh,\,m_t$) plane, 
for $\alpha_s=0.118$ and $\delta_\alpha=0.03$(a), 
$\delta_\alpha=0.12$(b). 
An assumed future 
value of
\bea 
\label{sb2_futuredata}
 \sbar^2(\mmz)=0.23069\pm0.00020
\eea
is used to constrain $m_t$ and $\mh$ in addition to 
the 
presently available data
and the 
present 
Tevatron $m_t$ data, 
$m_t=175\pm 6 \gev$. 
The inner and outer contours correspond to 
$\Delta\chi^2=1$ ($\sim$ 39\% CL), 
and $\Delta\chi^2=4.61$ ($\sim$ 90\%~CL), respectively. 
It is clearly seen that the allowed band in the ($\mh,\,\mt$) plane 
is significantly narrowed but the individual error of $\mh$ and $\mt$ 
is not reduced very much.
The sensitivity of the future constraints to $\bar{\alpha}(\mmz )$ 
can be 
judged 
by comparing the two figures.  

The assumed mean value of 0.23069 is chosen to retain the 
$\chi^2_{\rm min}$ point of the present data. 
The effect of
changing the average and dispersion of 
the assumed $\sbar^2(\mmz)$ data 
can be deduced from the two figures.

\subsection{$W$ mass}
%
\begin{figure}[t]
\begin{center}
 \leavevmode\psfig{file=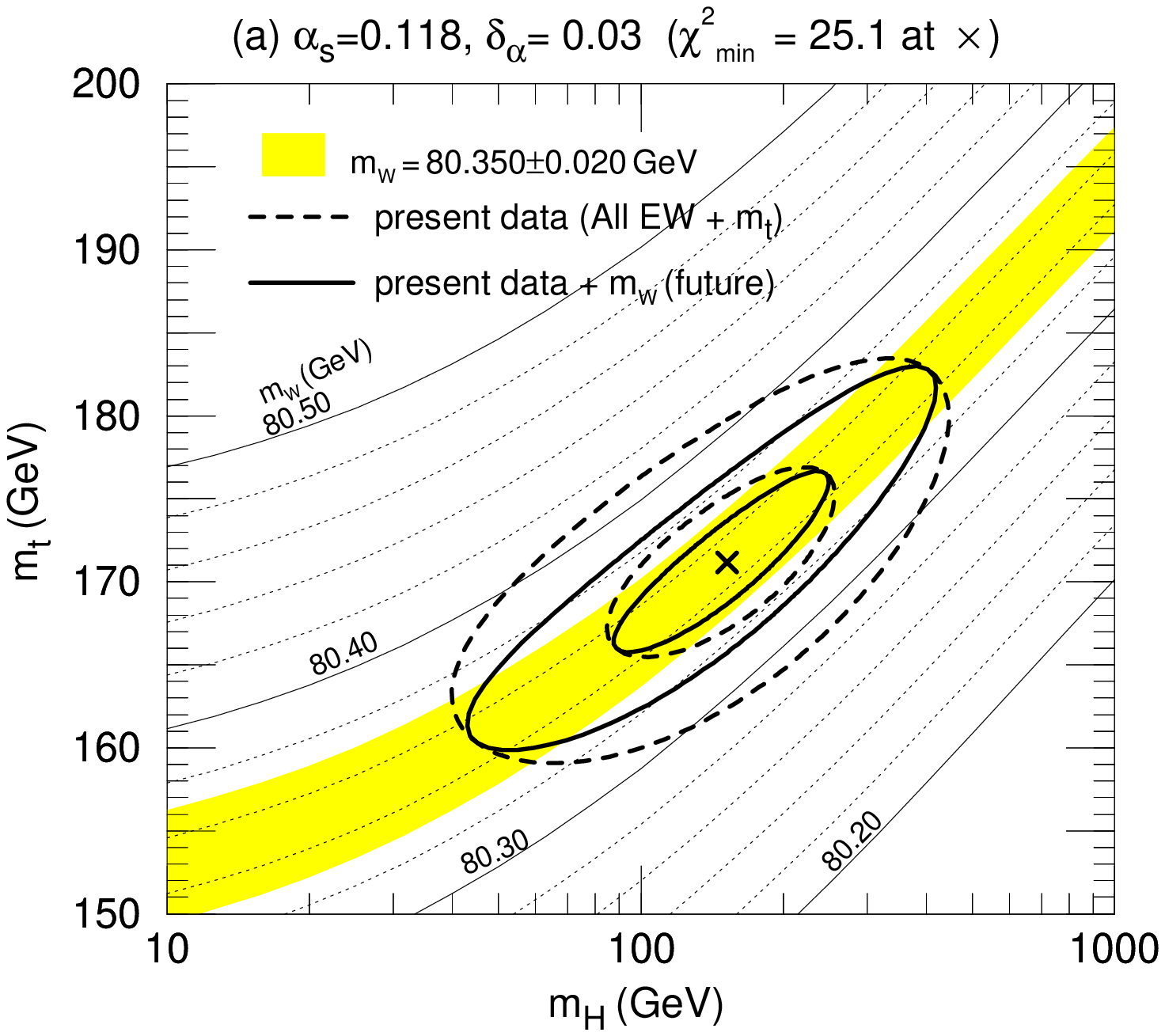,width=7cm,silent=0}
 \leavevmode\psfig{file=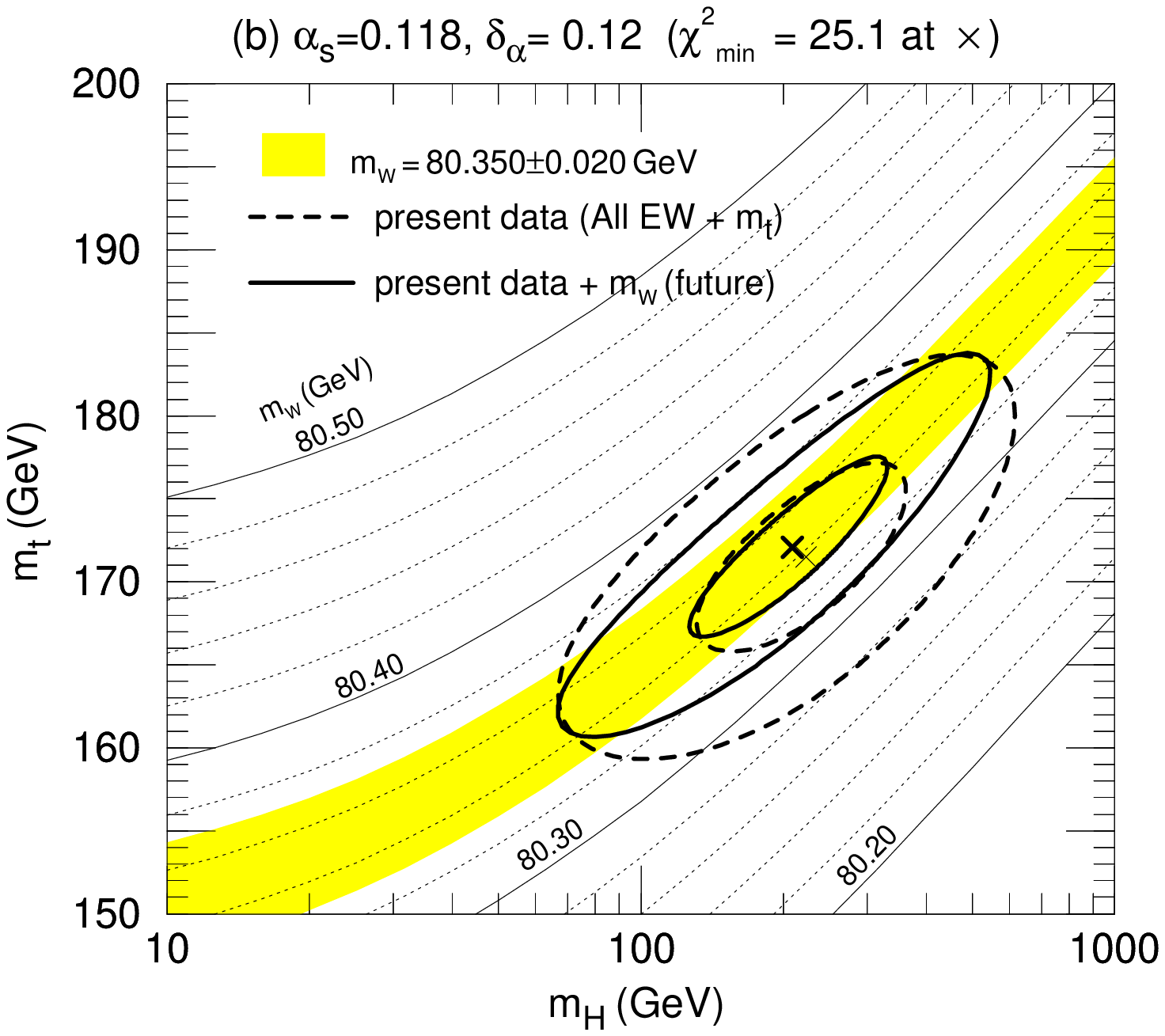,width=7cm,silent=0}
\end{center}
\caption{\protect\footnotesize\sl
Impact of future improvement in $W$ mass measurement
in the ($\protect\mh,\,m_t$) plane, for $\alpha_s=0.118$ 
and $\delta_\alpha=0.03$(a), $\delta_\alpha=0.12$(b). 
An assumed future data $\mw=80.350\pm 0.020$ 
is used to constrain $m_t$ and $\mh$ in addition to 
the present all electroweak data and the Tevatron $m_t$ data, 
$m_t=175\pm 6 \gev$. 
The inner and outer contours correspond to 
$\Delta\chi^2=1$ ($\sim$ 39\% CL), 
and $\Delta\chi^2=4.61$ ($\sim$ 90\%~CL), respectively. 
The minimum of $\chi^2$ is marked by the sign ``$\times$''. 
Thin dotted/solid lines show the SM predictions for $\mw$ 
when $m_t$ and $\mh$ are given.
}
\label{fig:mtmh_future_mw}
\end{figure}
%

Improved values on the $W$ mass are expected from CDF, D0 at the Tevatron,
from the HERA experiments and from the collaborations at LEP2. 
It is expected to obtain the $W$-mass to 31 MeV for a 1 fb$^{-1}$ run at
the Tevatron, which may be reduced to 11 MeV for 10 fb$^{-1}$\cite{baur96},
while at LEP2 in a 500 pb$^{-1}$-run 35 MeV\cite{wslep2}
is expected. 
In a high luminosity run at HERA a precision of 60 MeV is 
estimated\cite{wshera96}.
Further improved measurements on $\mw$ may be anticipated at 
a future linear $\epem$ collider\cite{mwatlc} or at a 
$\mu^+\mu^-$ collider\cite{eltwatmumu}.
Such measurements will provide a narrow band in the $(\mh\,,\mt)$-plane 
similar in width and orientation to the present asymmetry band and 
constitute a crucial piece of information in challenging the validity 
of the SM.
 
We show in Fig.~\ref{fig:mtmh_future_mw} the impact of future improvement 
in $W$ mass measurement in the ($\protect\mh,\,m_t$) plane, 
for $\alpha_s=0.118$ and $\delta_\alpha=0.03$(a), 
$\delta_\alpha=0.12$(b). 
An assumed future 
value of
\bea 
\label{mw_futuredata}
	\mw (\gev )=80.350\pm 0.020
\eea
is used to constrain $m_t$ and $\mh$ in addition to all the present 
electroweak data and the 
present 
Tevatron value for the top quark mass, 
$m_t=175\pm 6 \gev$. 
The allowed region in the ($\mh,\,\mt$) plane shrinks considerably,
but the individual errors of $\mh$ and $\mt$  remain essentially
unaltered as expected from Eq.~\ref{gzsb2mw_para}.   
By comparing the two figures, (a) and (b), the sensitivity of the 
future constraints to $\bar{\alpha}(\mmz )$ can be studied.

\begin{figure}[t]
\begin{center}
 \leavevmode\psfig{file=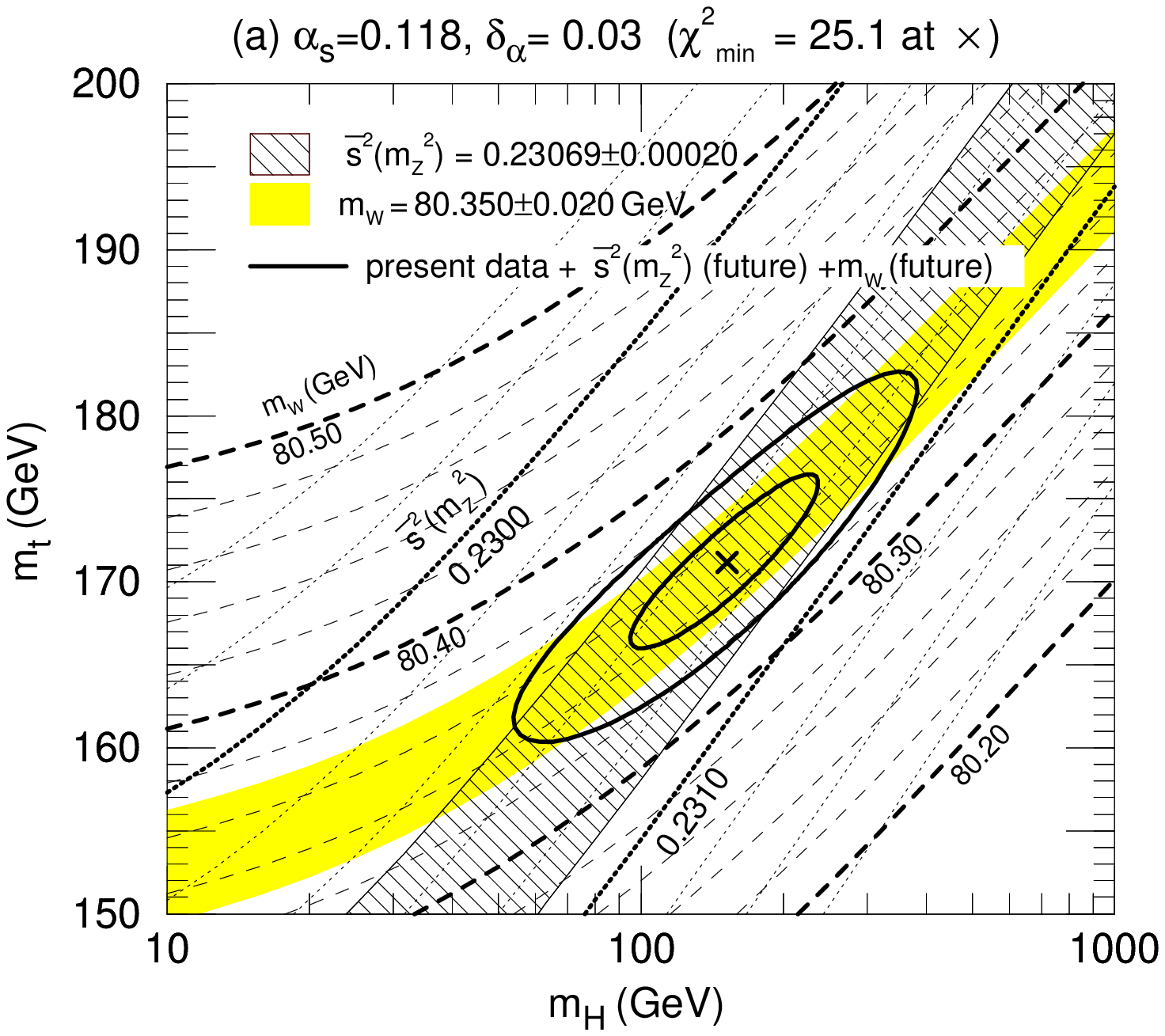,width=7cm,silent=0}
 \leavevmode\psfig{file=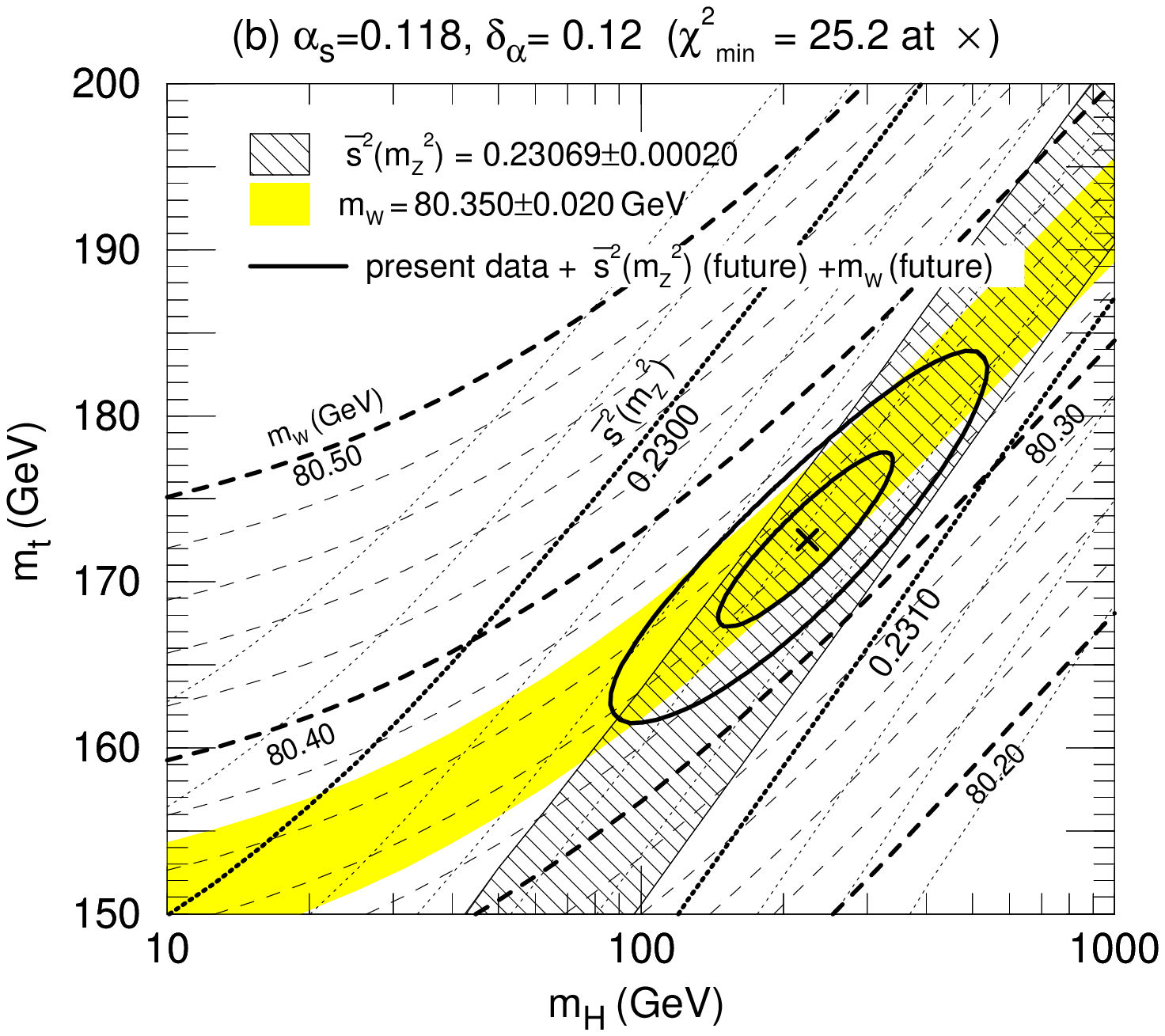,width=7cm,silent=0}
\end{center}
\caption{\protect\footnotesize\sl
Impact of future improvement in $\sbar^2(\mmz)$ and 
$W$ mass measurements in the ($\protect\mh,\,m_t$) plane,  
for $\alpha_s=0.118$ and $\delta_\alpha=0.03$(a), 
$\delta_\alpha=0.12$(b). 
Assumed future data $\sbar^2(\mmz)=0.23069\pm0.00020$ and 
$\mw=80.350\pm 0.020$ are used to constrain $m_t$ and $\mh$ 
in addition to the present all electroweak data and the 
Tevatron $m_t$ data, $m_t=175\pm 6 \gev$. 
The inner and outer contours correspond to $\Delta\chi^2=1$ 
($\sim$ 39\% CL), and $\Delta\chi^2=4.61$ ($\sim$ 90\%~CL), 
respectively. 
The minimum of $\chi^2$ is marked by the sign ``$\times$''. 
Thin dotted/solid lines show the SM predictions for $\mw$ 
when $m_t$ and $\mh$ are given.
}
\label{fig:mtmh_future_sb2andmw}
\end{figure}
%
The assumed mean value of $80.350~\gev$ is chosen to retain the 
$\chi^2_{\rm min}$ point of the present data.  
The effect of
changing the average and dispersion of 
the assumed $\mw$ data can be deduced from the two
figures.

The precise determinations of both $\sbar^2(\mmz )$ and $\mw$ 
provide independent constraints on $\mt$ and $\mh$, as can be clearly
seen by overlaying Fig.~\ref{fig:mtmh_future_sb2} and 
Fig.~\ref{fig:mtmh_future_mw}.  
Shown in Fig.~\ref{fig:mtmh_future_sb2andmw} is the impact of 
future improvement in $\sbar^2(\mmz)$ and $W$ mass measurements 
in the ($\mh,\,\mt$) plane, for $\alpha_s=0.118$ and 
$\delta_\alpha=0.03$(a), $\delta_\alpha=0.12$(b). 
Assumed future 
values 
$\sbar^2(\mmz)=0.23069\pm0.00020$ and 
$\mw=80.350\pm 0.020$ are shown again by shaded regions.  
Not only the reduction of the width of the allowed band in 
the ($\mh,\,\mt$) plane, but also the individual errors of 
$\mt$ and $\mh$ are now reduced 
considerably. 

In order to examine the future constraints on 
($\mt$, $\mh$, $\da$, $\alpha_s$) from the electroweak 
precision measurements, we repeat the four parameter fit with 
the present electroweak measurements plus the above two additional 
"data" on $\sbar^2(\mmz )$ (\ref{sb2_futuredata}) and $\mw$ 
(\ref{mw_futuredata}).  
We find 
from the electroweak data only 
%
\fitofmtxhalpsdafuture
By comparing with the present constraints (\ref{fitofmtxhalpsda}), 
we find that the error of $\mt$ can be 
reduced by about a factor of three, that of $\xh$, i.e. the logarithm of 
$\mh$ in units of $100\gev$, and $\da$ by a factor of two. 
It may be worth noting that $\mt$ can be predicted to $5~\gev$ 
accuracy even without assuming external knowledge on 
$\mh$, $\alpha_s$, and $\bar{\alpha}(\mmz )$.  

By imposing the present knowledge of $m_t$, $\alpha_s$ and $\da$, i.e. 
$\mt=175\pm 6 \gev$, $\alpha_s=0.118\pm 0.003$ and 
$\da=0.03\pm 0.009$, the fit 
(\ref{fitofmtxhalpsdafuture}) becomes 
%
\fitofmtxhalpsdafuturewithmtalpsda
This should be compared with the corresponding result in 
(\ref{fitofmtxhalpsdawithmtalpsdaej}). 
It is rather surprising to observe that none of the individual 
errors of the four fitted parameters reduces significantly from 
the present errors in 
(\ref{fitofmtxhalpsdawithmtalpsdaej}). 
The mean values stay the same because we chose the mean values 
of the future $\sbar^2(\mmz )$ and $\mw$ data at the present 
minimum of the global $\chi^2$ fit.  
What did change by adding the above two future "data" are the
correlations 
among the errors, in particular, that between $\mt$ and 
$\log\mh$ is now very large, 0.84, and the negative correlation 
between $\log\mh$ and $\da$ has also been strengthened.
Therefore, we can expect 
an important 
improvement on the $\mh$ 
constraint once $\mt$ and $\da$ are measured accurately. 

\subsection{Top-quark mass}
%
It is tantalizing that the present top mass value from Tevatron 
(\ref{mt_tevatron}) lies just on the boundary of the region allowed 
by the electroweak data.
 
The long-range program (TeV33\cite{tev33,tev2000}) at the Tevatron 
envisages an ultimate precision of the top mass of about 2~GeV
based on an anticipated 
yearly integrated luminosity 
of 10 fb$^{-1}$. 
In the future, the error can be reduced further to 200~MeV at an $\epem$ 
LC\cite{mtatlc} and possibly down to 70~MeV at a muon 
collider\cite{eltwatmumu} with precise beam energy resolution. 
Figure~\ref{fig:mhlimit} shows us that once the top quark mass is 
precisely determined, the major remaining uncertainty in electroweak 
fits is 
due to $\delta_\alpha$, the magnitude of the QED running coupling 
constant at the $\mz$ scale. 

Next 
we examine the effect of a future measurement 
$\mt =175\pm 2~\gev$ on the four parameter fit 
(\ref{fitofmtxhalpsdafuturewithmtalpsda}):
%
\fitofmtxhalpsdafuturewithfuturemtalpsda
The error of the logarithm of $\mh$ has been reduced from 
$\pm 0.60$ (\ref{fitofmtxhalpsdafuturewithmtalpsda})
to $\pm 0.35$, which is substantial, but not satisfactory. 
We find that this error 
cannot be reduced 
significantly by further reducing the error of $\mt$ 
down to 1~GeV. 
This may be inferred from the reduced correlation between 
$\mt$ and $\log\mh$ in 
(\ref{fitofmtxhalpsdafuturewithfuturemtalpsda}).
The strongest correlation among the four errors now appears 
between $\log\mh$ and $\da$.  
It is clear that further progress about $\mh$ in the SM, 
and also about physics beyond the SM from its quantum effects, 
will critically depend on an improved determination of 
$\bar{\alpha}(\mmz )$.

As a final example, we present the four parameter fit result 
with one further constraint, $\da=0.03\pm 0.03$, 
where the error is assumed to be 
1/3 of the conservative estimate\cite{eidjeg95}, 
or 1/2 of the other two estimates\cite{mz94,swartz95}.  
We find
%
\fitofmtxhalpsdafuturewithfuturemtalpsfutureda
The error in $\log \mh$ is now reduced to
about $\pm 0.25$. 

To conclude, 
we examine the constraint on $\mh$ from ultimate electroweak 
measurements by making use of the expressions (\ref{gzsb2mw_para}).  
With the top-quark mass determination of order 100~MeV at a LC or 
at a muon collider, its error can be safely neglected in 
(\ref{gzsb2mw_para}).   
Once the $\alpha_s$ value is measured to the 1~\% level\cite{alps96}, 
the LEP1 constraint from $\Gamma_Z$ 
becomes more 
effective 
through (\ref{gammaz_approx}). 
Nevertheless we find that $\mh$ will be constrained essentially by 
the future measurements of $\sbar^2(\mmz )$ and $\mw$ within the SM: 
\bea
\label{ultimatesb2}
\Delta\,[\,\xh -0.44\,x_\alpha\,] &=& 
		\pm \frac{\Delta\sbar^2(\mmz )}{0.0005} \,,
\\
\label{ultimatemw}
\Delta\,[\,\xh -0.20\,x_\alpha\,] &=& 
		\pm \frac{\Delta\mw(\gev )}{0.06} \,.
\eea
Combining
the above two constraints 
(\ref{ultimatesb2}) and (\ref{ultimatemw}) gives 
\bsub
\label{ultimate}
\bea
\Delta\,[\,\xh -A\,x_\alpha\,] &=& \pm \sigma \,,
\\
\Delta\,x_\alpha &=& \pm \sigma_\alpha \,,
\eea
\esub
with 
\bsub
\bea
\frac{1}{\sigma^2} &\approx& 
        (\frac{0.0005}{\Delta\sbar^2(\mmz )})^2 
       +(\frac{0.06}{\Delta\mw})^2 \,,
\\
\frac{A}{\sigma^2} &\approx& 
         0.44\,(\frac{0.0005}{\Delta\sbar^2(\mmz )})^2 
        +0.20\,(\frac{0.06}{\Delta\mw})^2 \,,
\\
\frac{1}{\sigma_\alpha^2} &\approx& 
         0.19\,(\frac{0.0005}{\Delta\sbar^2(\mmz )})^2 
        +0.04\,(\frac{0.06}{\Delta\mw})^2 
		-\frac{A^2}{\sigma^2} 
		+\frac{1}{(\Delta\,x_\alpha^{\rm (ext)})^2}\,,
\eea
\esub
where 
$\Delta\,x_\alpha^{\rm (ext)}=\Delta[1/\bar{\alpha}(\mmz )]/0.09$ is 
the external constraint on $\bar{\alpha}(\mmz )$.
For instance, with $\Delta\sbar^2(\mmz )=0.00010$, 
$\Delta\mw(\gev )=0.010$, and $\Delta\,x_\alpha^{\rm (ext)}=0.30$, 
we find 
$\Delta\,[\,\xh-0.30\,x_\alpha\,]\,=\, \pm 0.13$ with 
$\Delta\,x_\alpha \,=\, \pm 0.29$.  
Hence with the above ultimate assumptions, the error of the SM 
prediction to $\log \mh$ reduces to $\pm 0.15$.  
On the other hand, once the Higgs boson is found, its mass 
may be 
measured so accurately that its error can be neglected in the 
electroweak radiative effects.  
The electroweak data 
(\ref{ultimatesb2}) and 
(\ref{ultimatemw}) then constrain 
$\da$ to $\pm 0.036$, 
or about 
40\% 
its present error\cite{eidjeg95}.  
Evidence for new physics may then be looked for by comparing 
the direct and indirect measurements of $\bar{\alpha}(\mmz )$.  

\subsection{Future constraints on $S$, $T$, $U$ }

The impact on $S$, $T$, $U$ of the future
measurements of $\sbar^2(\mmz)$ (\ref{sb2_futuredata}) and 
$\mw$ (\ref{mw_futuredata}) is discussed briefly in this subsection. 
The analysis of section 4.4 can be repeated straightforwardly.  

It is again worth noting that only the following combinations of 
these parameters and $\da$, $\delg$ and $\alpha_s$ can be 
constrained by the three most accurately measurable quantities: 
%
%
%
\begin{subequations}\label{gzsb2mw_para_stu}
\begin{eqnarray}
  \Gamma_Z(\mev ) &\approx &
	2473.0 
	-9.5\,S' +25.0\,T' +1.7\,x'_s 
	-3.4\,\log[\, 1+(\frac{26\gev }{\mh})^2 \,] \,,
   \\ \label{sb2stu_para}
   \sbar^2(\mmz) &\approx&
	0.2334 +0.0036\,S' -0.0024\,T' 
   \\
    \mw (\gev ) &\approx&
	79.840 -0.291\,S' +0.417\,T' +0.332\,U''\,.
\end{eqnarray}
\end{subequations}
Here $x'_s = (\alpha'_s-0.118)/0.003 = 
[\alpha_s-0.118+1.54\,(\delb+0.00995)]/0.003$, and 
\begin{subequations}\label{stu_pp}
\begin{eqnarray}
S' &=& S -0.72\,\da \,, \\
T' &=& T +(0.0055-\delg)/\alpha \,, \\
U''&=& U' -0.87\,(T'-T) \nonumber \\
   &=& U -0.22\,\da +0.87\,(0.0055-\delg)/\alpha \,.
\end{eqnarray}
\end{subequations}
A linear combination of $S'$ and $T'$ 
will be better constrained by
future improvements in $\sbar^2(\mmz )$.  
Individual constraints will still be obtained from the LEP1 
$\Gamma_Z$ value, and hence they won't be improved significantly 
unless one can predict accurately the $\alpha_s'$ value including 
the $\zbb$ vertex factor.
The improved measurement on $\mw$ determines the combination $U''$.  
Therefore, we need to know $\delb$, $\delg$ and $\da$ 
accurately in order to constrain non-SM contributions to 
the $S$, $T$, $U$ parameters.  

As an example consider the result
of the three parameter fit 
with the new $\sbar^2(\mmz )$ and $\mw$ measurements of 
(\ref{sb2_futuredata}) and (\ref{mw_futuredata}), 
respectively :
\fitofstuwithfuturedata
As compared to the present result (\ref{fitofstu}), 
we find 
substantial 
reductions in the error of $U''$ but not in those 
of $S'$ and $T'$ individually.  
On the other hand, 
all correlations are stronger compared to those 
of Eq.~(\ref{fitofstu}). 
The most stringent constraint among the $S$, $T$, $U$ parameters 
now reads
\bea
T' -0.96 S' +0.45 U'' = 1.13\pm 0.036.
\label{stu_constraint_future}
\eea
When compared with the corresponding constraint 
(\ref{stu_constraint}) of the 
existing 
electroweak data, the allowed range of $T'$ for given $S'$ and $U'$ can
be reduced by a factor of two.


\section{Conclusions}
 
We have carried out a comprehensive analysis of the latest electroweak
data. 
The analysis updates our previous work (see Ref.\cite{hhkm}).
The total width $\Gamma_Z$, the hadronic width $\Gamma_h^0$ and 
the leptonic width $\Gamma_\ell$ agree well with the SM predictions
at the level of a few $10^{-3}$. 
The new measurement of $R_c$ is in agreement with the SM, and
also the new measurement of $R_b$, albeit within about two standard
deviations.
The asymmetry data determine the effective weak mixing parameter 
$\sin^2\theta_W$ to an accuracy of 0.1\% level, 
see Eq.~(\ref{fitofsb96withleptonuniversality}). 
Their average value agrees well with the SM, while their
dispersion is 
larger than statistically expected.
It is, however, fair to conclude 
that the progress both in precision and agreement of data with
SM expectation is impressive. 

  The ($S$, $T$) fit agrees well with the SM, whereas 
  the simple QCD-like Techni-Color (TC) model is ruled out at
  the 99\%CL.
The fitted $U$ parameter also agrees with the SM prediction.  
The fact
that all the $S$, $T$, $U$ parameters agree well with the SM 
prediction for the top quark mass as observed at the Tevatron 
and 
the Higgs boson mass below a few hundred GeV implies that 
any dynamical model of the electroweak symmetry breaking without 
a light Higgs boson 
should not only give a 
negative
$S_{\rm new}$, but also a $T_{\rm new}$-value which is constrained 
severely by the data for the given $S_{\rm new}$ and $U_{\rm new}$; 
see Eq.~(\ref{stu_constraint}).
The above conclusion remains valid even if 
the model contributes a sizeable amount to the $\zbb$
vertex, since 
the strong correlation between $S_{\rm new}$ and $T_{\rm new}$ 
comes from the accurate measurement of the effective weak 
mixing angle, $\sbar^2(\mmz )$, which is independent of 
$R_b$ or the assumed $\alpha_s$ value. 
For the $U$ parameter, $|U_{\rm new}|\simlt 0.4$ should be satisfied.
  The uncertainty in the running QED coupling constant at 
  the $\mz$ scale, $\bar{\alpha}(\mmz)$, is shown as {\it the}
  serious limiting factor for future improvements in the
  determination 
  of the $S$ parameter.

The global fit in the minimal SM in terms of ($m_t$, $\mh$) yields 
values for the top mass, $\mt = 153 \pm 10 \gev$ 
(\ref{fitofmtxhalpsdawithalpsdaej_mean}), 
or $\mt = 158 \pm 12 \gev$ 
(\ref{fitofmtxhalpsdawithalpsdaejwithoutrbrc_mean}) 
if we drop the present 
$R_b$ constraint, 
which agrees with the direct measurements from the Tevatron, 
$\mt = 175 \pm 6 \gev$\cite{mt96}. 
The corresponding allowed range in $\mh$ is 
$\mh = 50^{+50}_{-30} \gev$ 
(\ref{fitofmtxhalpsdawithalpsdaej_mean}) and 
$\mh = 60^{+100}_{-40} \gev$ 
(\ref{fitofmtxhalpsdawithalpsdaejwithoutrbrc_mean}) respectively.  
Once $m_t$ is accurately measured the present electroweak 
data will impose stringent limits
on the Higgs-boson mass which are not 
affected by the $R_b$ data (see Table~7 in section 5.2).
For instance the present electroweak data favor a light Higgs boson if 
$\mt \simlt 170\gev$ while a heavier Higgs boson is favored if 
$\mt \simgt 180\gev$:  
the 95\%~CL upper and lower mass bounds, 
$\mh<360 \gev$ for $m_t = 170 \gev$ and  
$\mh>130 \gev$ for $m_t = 180 \gev$ are obtained 
by 
using 
$\alpha_s=0.118\pm0.003$\cite{pdg96} and $\da=0.03\pm0.09$\cite{eidjeg95}. 
In order to further improve the constraint on $\mh$ not only precise 
measurement on $m_t$ are required, but also improved measurements 
on $\Delta\alpha_{\rm had}(\mmz )$ and $\alpha_s$.  

For the agreement of the SM predictions with precision experiments 
it is indispensable to include radiative effects due to  
`vertex-like' corrections which may be regarded as indirect evidence 
for the universal weak-boson self-couplings.
Their direct investigation will soon be carried out at LEP\,2.

Finally, we studied prospects of future improvements in the 
electroweak precision experiments. 
Major improvements are 
expected from further running and detector upgrades in
the determination of the 
mixing parameter $\sbar^2(\mmz )$ at SLC, Tevatron, 
and at a future linear $\epem$ collider (LC); $\mw$ will be 
measured more accurately at LEP2, Tevatron upgrades, LHC, LC 
and, perhaps, at a muon collider.
The error in the top-quark mass may be reduced to 2~GeV at Tevatron, 
200~MeV at LC, and even further down at a muon collider.  
These measurements will constrain physics beyond the SM very 
stringently, say in the $(S_{\rm new}, T_{\rm new}, U_{\rm new})$ 
parameter space, where not only $T_{\rm new}$ but also $U_{\rm new}$ 
will be constrained severely as function of $S_{\rm new}$, 
whose constraint can be improved with a better $\alpha(\mmz)$
knowledge. 
Within the SM, 
the constraint on the Higgs boson mass
will not 
improve significantly 
beyond $\pm 0.35$ for $\log \mh$, 
unless a substantial 
improvement in the $\alpha(\mmz )$ 
estimate is achieved also.
 
\section*{Acknowledgements}
We would like to thank
S.~Aoki, U.~Baur, B.K.~Bullock, D.~Charlton, 
M.~Drees, S.~Eidelman, S.~Erredi, G.L.~Fogli, 
W.~Hollik, R.~Jones, J.~Kanzaki, J.H.~K\"uhn, 
C.~Mariotti, A.D.~Martin, K.~McFarland, 
T.~Mori, M.~Morii, D.R.O.~Morrison, 
B.~Pietrzyk, P.B.~Renton, P.~Rowsen, 
M.H.~Shaevitz, D.~Schaile, M.~Swartz, R.~Szalapski, 
T.~Take\-uchi, P.~Vogel, P.~Wells and D.~Zeppenfeld
for discussions.


%


\begin{thebibliography}{99}
%
\bibitem{lepewwg96}
The LEP Collaborations ALEPH, DELPHI, L3, OPAL, 
the LEP Electroweak Working Group and the SLD Heavy Flavour Group, 
preprint CERN-PPE/96-183 (December 1996). 
 
\bibitem{hhkm}
K.~Hagiwara, D.~Haidt, C.S.~Kim and S.~Matsumoto, \ZP{C64}{1994}{559}; 
{\bf C68} (1995) 352(E).

\bibitem{sm95}
S.~Matsumoto, \MPL{A10}{1995}{2553}.

\bibitem{lepewwg95}
The LEP Collaborations ALEPH, DELPHI, L3, OPAL 
and the LEP Electroweak Working Group, 
preprint CERN-PPE/95-172 (1995).

\bibitem{stu}
M.E.~Peskin and T.~Takeuchi, \PRL{65}{1990}{964}; \PRD{46}{1992}{381}.

\bibitem{mw96}
M.~Rijssenbeek, talk at ICHEP96, Warsaw, 25-31 July 1996, 
Fermilab-Conf-96-365-E, to appear in the proceedings.  
 
\bibitem{ccfr95_kevin}
K.~McFarland, talk at the {\it
XV Workshop on Weak Interactions and Neutrinos},
Talloires, France, 4--8 Sep 1995.

\bibitem{mt96}
CDF Collaboration, J.~Lys, talk at ICHEP96, Warsaw, 25-31 July 1996, 
Fermilab-Conf-96-409-E, to appear in the proceedings;\\
D0 Collaboration, S.~Protopopescu, 
talk at ICHEP96, Warwaw, 25-31 July 1996,
to appear in the proceedings;\\
P.~Tipton, talk at ICHEP96, Warsaw, 25-31 July 1996, 
to appear in the proceedings.

\bibitem{khlp95}
K.~Hagiwara, Proceedings of 
{\it 17th International Symposium on Lepton-Photon Interactions}(LP95)
(Beijing, P.R.~China, 10-15 August 1995), p.63.

\bibitem{pinch}
J.M.~Cornwall and J.~Papavassiliou, \PRD{40}{1989}{3474};\\
J.~Papavassiliou and K.~Phillippides, \ibid {\bf 48}(1993)4225;
\ibid {\bf 51}(1995)6364;\\
J.~Papavassiliou, \ibid {\bf 50}(1994)5998.
 
\bibitem{pinch2}
G.~Degrassi and A.~Sirlin, \NPB{383}{1992}{73}; \PRD{46}{1992}{3104};
G.~Degrassi, B.~Kniehl and A.~Sirlin, \ibid {\bf 48}(1993)R3963.
 
\bibitem{kl89}
D.C.~Kennedy and B.W.~Lynn, \NPB{322}{1989}{1}.

\bibitem{hms95}
K.~Hagiwara, S.~Matsumoto and R.~Szalapski, \PLB{357}{1995}{411}; 
K.~Hagiwara, T.~Hatsukano, R.~Ishihara and R.~Szalapski, 
preprint KEK-TH-497, hep-ph/9612268, to appear in Nucl.Phys.B. 

\bibitem{prw96} 
J.~Papavassiliou, E.~de Rafael and N.J~Watson, 
Preprint CPT-96-P-3408, hep-ph/9612237. 

\bibitem{del_gf}
A.~Sirlin, \PRD{22}{1980}{971}.

\bibitem{mz94} 
A.D.~Martin and D.~Zeppenfeld, 
\PLB{345}{1995}{558}.
 
\bibitem{swartz95} 
M.L.~Swartz, 
\PRD{53}{1995}{5268}.

\bibitem{eidjeg95} 
S.~Eidelman and F.~Jegerlehner, \ZPC{67}{1995}{602}.

\bibitem{bp95}
H.~Burkhardt and B.~Pietrzyk, \PLB{356}{1995}{398}.

\bibitem{kniehl90}
B.A.~Kniehl, \NPB{347}{1990}{86}. 

\bibitem{burkhardt89}
H.~Burkhardt, F.~Jegerlehner, G.~Penso and C.~Verzegnassi, 
\ZPC{43}{1989}{497}.

\bibitem{jeg91}
F.~Jegerlehner, in {\it Testing the
Standard Model}, eds.~M.~Cveti\v{c} 
and P.~Langacker (World Scientific, 1991).

\bibitem{jeg92}
F.~Jegerlehner,\,cited by B.A.~Kniehl in 
{\it Proc.~Europhysics Marseille}\,1993,\,p.639.

\bibitem{3loopt}
K.G.~Chetyrkin, J.H.~K\"uhn and M.~Steinhauser, 
\PLB{351}{1995}{331}; \\
L.~Avdeev, J.~Fleischer, S.~Mikhailov and O.~Tarasov, 
\PLB{336}{1994}{560}; 
Erratum \ibid {\bf 349} 597 (1995).

\bibitem{3loopvv}
K.G.~Chetyrkin, J.H.~K\"uhn and M.~Steinhauser, 
\PRL{75}{1995}{3394}.

\bibitem{qcdeltw}
A.~Czarnecki and J.H.~K\"uhn, 
\PRL{77}{1996}{3955}.

\bibitem{renton}
P.B.~Renton, 
Proceedings of {\it 17th International Symposium on Lepton
Photon Interactions} (LP95) 
(Beijing, P.R. China, 10--15 August 1995), p35.

\bibitem {pdg96}
Particle Data Group, R.M.~Barnett \etal, 
\PRD{54}{1996}{1}.

\bibitem{rbrc_opal96}
The OPAL Collaboration, G.~Alexander \etal, 
\ZPC{72}{1996}{1}.

\bibitem{rb_opal96}
The OPAL Collaboration, K.~Ackerstaff \etal,
preprint CERN-PPE/96-167 (Nov 1996). 

\bibitem{rb_aleph96}
The ALEPH Collaboration, 
talk by J.Steinberger in CERN Seminar, : 8 Oct 1996;\\
The ALEPH Collaboration, R.~Barate \etal, Preprint CERN-PPE/97-017;\\
The ALEPH Collaboration, R.~Barate \etal, Preprint CERN-PPE/97-018.

\bibitem{fh88}
G.L.~Fogli and D.~Haidt, \ZPC{40}{1988}{379}.
 
\bibitem{langacker94}
P.~Langacker, in {\it Precision Tests of the Standard 
Electroweak Model}, ed.~by P.~Langacker 
(World Scientific, 1994).

\bibitem{appelquist93}
T.~Appelquist and J.~Terning, \PLB{315}{1993}{139};
\PRD{50}{1994}{2116}.

\bibitem{ellis95}
J.~Ellis, G.L.~Fogli and E.~Lisi, \PLB{343}{1995}{282}. 

\bibitem{altarelli96}
G.~Altarelli, talk at the {\it NATO Advanced Study Institute on Techniques 
and Concepts of High Energy Physics}, St.\ Croix, 10--23 July 1996, 
hep-ph/9611239.  

\bibitem{langacker97}
P.~Langacker and J.~Erler, presented at the 
{\it Ringberg Workshop on the Higgs Puzzle}, December 1996, 
hep-ph/9703428.  

\bibitem{rosner97}
J.L.~Rosner, hep-ph/9704331, submitted to Comments on Nuclear 
and Particle Physics.  

\bibitem{ellis96}
J.~Ellis, G.L.~Fogli and E.~Lisi, \PLB{389}{1996}{321}.  

\bibitem{boer96}
W.~de~Boer, A.~Dabelstein, W.~Hollik, W.~M\"osle and U.~Schwickerath, 
hep-ph/9609209 v4 (Nov, 1996). 

\bibitem{novikov93}
V.A.~Novikov, L.B.~Okun and M.I.~Vysotsky, 
\MPL{A8}{1993}{2529}; 
Erratum {\bf A8} 3301 (1993).

\bibitem{mhmssm}
Y.~Okada, M.~Yamaguchi, T.~Yanagida, 
\PTP{85}{1991}{1}; \PLB{262}{1991}{54}; 
H.~Haber, R.~Hempfling, 
\PRL{66}{1991}{1815}; \PLB{262}{1991}{54}; 
J.~Ellis, G.~Ridorfi, F.~Zwirner, 
\PLB{257}{1991}{83}; \PLB{262}{1991}{477}. 

\bibitem{altmodels}
J.D.~Bjorken, \PRD{19}{1978}{335};\
P.D.~Hung and J.J.~Sakurai, \NPB{143}{1978}{81}.
 
\bibitem{sirlin94}
P.~Gambio and A.~Sirlin, \PRL{73}{1994}{621}.
 
\bibitem{kh_ucla95}
K.~Hagiwara,  {\it Proceedings of the International Symposium 
on Vector Boson Self-Interactions}, 
eds. U.~Baur, S.~Errede and T.~M\"uller 
(Los Angeles, 1995, AIP Press), p.185.

\bibitem{novikov94}
V.A.~Novikov, L.B.~Okun, A.N.~Rozanov and M.I.~Vysotsky, 
\MPL{A9}{1994}{2641};
Z.~Hioki, \PLB{340}{1994}{181}.

\bibitem{rowson96}
P.~Rowson, private communication.

\bibitem{tev33}
D.~Amidei \etal, Preprint CDF/DOC/TOP/PUBLIC/3265 (Aug 1995).

\bibitem{baur96}
U.~Baur and M.~Demarteau, {\it Precision electroweak physics 
at future collider experiments}, hep-ph/9611334 v2, 
to be published in 
{\it Proceedings of the 1996 DPF/DPB Summer Study on 
New Directions for High-Energy Physics (Snowmass 96)}.  

\bibitem{tev2000} 
D.~Amidei and R.~Brock, {\it Future electroweak physics at the 
Fermilab Tevatron}: Report of the TeV2000 Study Group, 
Fermilab-Pub-96/082 (April 1996).

\bibitem{jlc90}
T.~Omori, {\it Proceedings of the 2'nd Workshop on JLC }, 
ed. by S.~Kawabata, KEK Proceedings 91-10, p.315.  

\bibitem{wslep2}
A.~Ballestrero \etal, {\it Proceedings of the Workshop
on Physics at LEP2}, G.~Altarelli, T.~Sj\"ostrand and 
F.~Zwirner (eds.), CERN Yellow Report CERN 96-01(1996), 
Vol.1, p.141.

\bibitem{wshera96}
R.J.~Cashmore \etal, MPI/PTh/96-105 and {\it Proceedings
of the Workshop on future Physics at HERA} 1996.

\bibitem{mwatlc}
A.~Miyamoto, {\it Physics and Experiments with Linear $\epem$ Colliders},
edited by F.A.~Harris \etal \,(World Scientific, 1993), p.141.

\bibitem{mtatlc}
L.H.~Orr, {\it Physics and Experiments with Linear $\epem$ Colliders},
eds.\ A.Miyamoto \etal \,(World Scientific, 1996), p.129. 

\bibitem{eltwatmumu}
V.~Barger, M.S.~Berger, J.F.~Gunion and T.~Han, 
"Precision $W$-boson and top-quark mass determination at a muon 
collider", hep-ph/9702334.

\bibitem{alps96}
P.N.~Burrows \etal, "Prospects for the precision measurement of 
$\alpha_s$", hep-ex/9612012, to be published in 
{\it Proceedings of the 1996 DPF/DPB Summer Study on 
New Directions for High-Energy Physics (Snowmass 96)}.  

\end{thebibliography}
\end{document}